\documentclass[onecolumn]{svjour3}

\journalname{PLoS Computational Biology}

\usepackage{graphicx}
\usepackage{subcaption}
\usepackage{amsmath}
\usepackage{amssymb}
\usepackage{psfrag}
\usepackage{color}
\usepackage[nice]{nicefrac}
\usepackage{mathtools}
\usepackage[linesnumbered,ruled,lined]{algorithm2e}
\usepackage[right]{lineno}

\usepackage{hyperref}
\usepackage{changebar,soul}
\usepackage{tikz-cd}

\renewenvironment{itemize}{
  \begin{list}{-}
    {\setlength{\parsep}{3pt}
      \setlength{\labelwidth}{24pt}
      \setlength{\itemsep}{1pt}
      \setlength{\topsep}{3pt}}}{\end{list}}

\usepackage[numbers,sort&compress]{natbib}

\newcommand{\heading}[1]{\smallskip\noindent\textit{\textbf{#1}}}
\newcommand{\lsyheadingsmall}[1]{\bigskip\noindent{\textbf{\small {#1}}\nopagebreak\smallskip\nopagebreak}}
\newcommand{\lsyheadingmed}[1]{\bigskip\noindent{\textbf{\small {#1}}\nopagebreak\medskip\nopagebreak}}

\usepackage{setspace}

\newcommand{\ddt}[1]{\frac{{\rm d}{#1}}{{\rm d}t}} % time derivative
\newcommand{\D}{{\rm d}} % differentiation
\newcommand{\ol}[1]{\overline{#1}} % mean value
\newcommand{\B}[1]{\vec{\bf #1}} % vector
\newcommand{\BM}{{\bf M}} % matrix M for L4
\newcommand{\BR}{{\bf R}} % matrix R for ref factor
\DeclarePairedDelimiterX{\expectarg}[1]{[}{]}{%
  \ifnum\currentgrouptype=16 \else\begingroup\fi
  \activatebar#1
  \ifnum\currentgrouptype=16 \else\endgroup\fi
}

\begin{document}

\title{A data-informed mean-field approach to mapping of cortical parameter
  landscapes}

\author{Zhuo-Cheng~Xiao$^{1}$ \and Kevin~K.~Lin$^{2}$ \and
  Lai-Sang~Young$^{1,3,\dagger}$}

\institute{  $^1$ Courant Institute of Mathematical Sciences, New York
  University, NY 10003, USA  \\
             $^2$ Department of Mathematics, University of Arizona, AZ
  85721, USA \\
             $^3$ Institute for Advanced Study, Princeton, NJ 08540, USA \\
             $^\dagger$ Corresponding author. 
             \email{lsy@cims.nyu.edu}
}

\date{~\\[-0.5in]Received: date / Accepted: date }

\maketitle

%%%%%%%%%%%%%%%%%%%%%%%%%%%%%%%%%%%%%%%%%%%%%%%%%%%%%%%%%%%%

%% PLOS guidelines at
%% https://journals.plos.org/ploscompbiol/s/submission-guidelines

%% Abstract

%% The Abstract comes after the title page in the manuscript file. The
%% abstract text is also entered in a separate field in the submission
%% system.

%% The Abstract should be succinct; it must not exceed 300
%% words. Authors should mention the techniques used without going into
%% methodological detail and should summarize the most important
%% results.

%% While the Abstract is conceptually divided into three sections
%% (Background, Methodology/Principal Findings, and
%% Conclusions/Significance), do not apply these distinct headings to
%% the Abstract within the article file.

%% Do not include any citations. Avoid specialist abbreviations.
\begin{abstract}
Constraining the many biological parameters that govern cortical
dynamics is computationally and conceptually difficult because of the
curse of dimensionality.  This paper addresses these challenges by
proposing (1) a novel data-informed mean-field (MF) approach to
efficiently map the parameter space of network models; and (2) an
organizing principle for studying parameter space that enables the
extraction biologically meaningful relations from this high-dimensional
data.  We illustrate these ideas using a large-scale network model of
the {\it Macaque} primary visual cortex.  Of the 10-20 model parameters,
we identify 7 that are especially poorly constrained, and use the MF
algorithm in (1) to discover the firing rate contours in this 7D
parameter cube.  Defining a ``biologically plausible" region to consist
of parameters that exhibit spontaneous Excitatory and Inhibitory firing
rates compatible with experimental values, we find that this region is a
slightly thickened codimension-1 submanifold.  An implication of this
finding is that while plausible regimes depend sensitively on
parameters, they are also robust and flexible provided one compensates
appropriately when parameters are varied.  Our organizing principle for
conceptualizing parameter dependence is to focus on certain 2D parameter
planes that govern lateral inhibition: Intersecting these planes with
the biologically plausible region leads to very simple geometric
structures which, when suitably scaled, have a universal character
independent of where the intersections are taken.  In addition to
elucidating the geometry of the plausible region, this invariance
suggests useful approximate scaling relations.  Our study offers, for
the first time, a complete characterization of the set of all
biologically plausible parameters for a detailed cortical model, which
has been out of reach due to the high dimensionality of parameter space.
\end{abstract}

%% Author Summary

%% We ask that all authors of research articles include a 150-200 word
%% non-technical summary of the work as part of the manuscript to
%% immediately follow the abstract. This text is subject to editorial
%% change, should be written in the first-person voice, and should be
%% distinct from the scientific abstract.

%% Aim to highlight where your work fits within a broader context;
%% present the significance or possible implications of your work simply
%% and objectively; and avoid the use of acronyms and complex
%% terminology wherever possible. The goal is to make your findings
%% accessible to a wide audience that includes both scientists and
%% non-scientists.

%% Authors may benefit from consulting with a science writer or press
%% officer to ensure they effectively communicate their findings to a
%% general audience.

\heading{Author Summary.}  Cortical circuits are characterized by a high
degree of structural and dynamical complexity, and this biological
reality is reflected in the large number of parameters in even
semi-realistic cortical models.  A fundamental task of computational
neuroscience is to understand how these parameters govern network
dynamics.  While some neuronal parameters can be measured {\em in vivo},
many remain poorly constrained due to limitations of available
experimental techniques.  Computational models can address this problem
by relating difficult-to-measure parameters to observable quantities,
but to do so one must overcome two challenges: (1) the computational
expense of mapping a high dimensional parameter space, and (2)
extracting biological insights from such a map.  This study aims to
address these challenges in the following ways: First, we propose a
parsimonious data-informed algorithm that efficiently predicts
spontaneous cortical activity, thereby speeding up the mapping of
parameter landscapes.  Second, we show that lateral inhibition provides
a basis for conceptualizing cortical parameter space, enabling us to
begin to make sense of its geometric structure and attendant scaling
relations.  We illustrate our approach on a biologically realistic model
of the monkey primary visual cortex.



%% Author contributions

%% Provide at minimum one contribution for each author in the submission
%% system. Use the CRediT taxonomy to describe each contribution. Read
%% the policy and the full list of roles.y

%% Contributions will be published with the final article, and they
%% should accurately reflect contributions to the work. The submitting
%% author is responsible for completing this information at submission,
%% and we expect that all authors will have reviewed, discussed, and
%% agreed to their individual contributions ahead of this time.

%% PLOS Computational Biology will contact all authors by email at
%% submission to ensure that they are aware of the submission.

%% \heading{Author Contributions.} \ktext{Categories are Conceptualization:
%%   Data Curation; Formal Analysis; Funding Acquisition; Investigation;
%%   Methodology; Project Administration; Resources; Software; Supervision;
%%   Validation; Visualization; and Writing.  See
%%   \href{https://journals.plos.org/ploscompbiol/s/authorship#loc-author-contributions}{this
%%     page}.}

\smallskip\noindent{\bf Author Contributions.}  Conceptualization,
Methodology, Analysis, Writing: ZCX, LSY, KL.  Data Curation and
Software: ZCX.

%%%%%%%%%%%%%%%%%%%%%%%%%%%%%%%%%%%%%%%%%%%%%%%%%%%%%%
\section*{\large Introduction}

From spatially and temporally homogeneous but sensitive resting states
to highly structured evoked responses, neuronal circuits in the cerebral
cortex exhibit an extremely broad range of dynamics in support of
information processing in the brain \cite{mitra2007observed,
  makeig2004mining, spiers2006thoughts, del2007brain, erb2013brain,
  schiff2014large, barttfeld2015signature, zalesky2014time}.
Accompanying this dynamical flexibility is a high degree of
morphological and physiological complexity \cite{braitenberg2013cortex,
  roth1994cell, van2001need, zeng2017neuronal, silver2010neuronal,
  im2012micrornas, debanne2011axon}.  As a result, any effort to
characterize cortical circuits necessarily involves a large number of
biological parameters \cite{wang2002probabilistic, friston2009cortical,
  wolf2014dynamical, peron2020recurrent, niell2021cortical,
  mochol2015stochastic}.  Understanding the range of parameters
compatible with biologically plausible cortical dynamics and how
individual parameters impact neural computation are, in our view, basic
questions in computational neuroscience.

Due to limitations of available experimental techniques, many neuronal
and network parameters are poorly constrained.  Biologically realistic
network models can bridge this gap by quantifying the dependence of
observable quantities like firing rates on parameters, thereby
constraining their values.  However, two challenges stand in the way of
efforts to map the parameter landscape of detailed cortical networks.
First, a direct approach, i.e., parameter sweeps using network models,
may be extremely costly or even infeasible.  This is because even a
single layer of a small piece of cortex consists of tens of thousands of
neurons, and the computational cost grows rapidly with the size of the
network.  This cost is compounded by the need for repeated model runs
during parameter sweeps, and by the ``curse of dimensionality,'' i.e.,
the exponential growth of parameter space volume with the number of
parameters.  Second, even after conducting parameter sweeps, one is
still faced with the daunting task of making sense of the high
dimensional data to identify {\em interpretable, biologically meaningful
  features}.

This paper addresses the twin challenges of computational cost and
interpretable cortical parameter mapping.  Starting from a biologically
realistic network model, we define as ``viable" those parameters that
yield predictions compatible with empirically observed firing rates, and
seek to identify the viable region.  To mitigate the computational cost
of parameter space scans, we propose a parsimonious, data-informed
mean-field (MF) algorithm.  MF methods replace rapidly-fluctuating
quantities like membrane potentials with their mean values; they have
been used a great deal in neuroscience \cite{wilson1972excitatory,
  wilson1973mathematical, amari1977dynamics, ermentrout1979mathematical,
  hopfield1982neural, cohen1983absolute, hopfield1984neurons,
  treves1993mean, ermentrout1998neural, brunel1999fast,
  bressloff2001geometric, gerstner2000population, coombes2005waves,
  mattia2002population, el2009master, faugeras2009constructive,
  parr2020modules}.  MF models of neuronal networks are all based on the
relevant biology to different degrees, but most rely on idealized
voltage-rate relations (so-called ``gain'' or ``activation'' function)
to make the system amenable to analysis (see, e.g.,
\cite{ermentrout2010mathematical, gerstner2014neuronal} and the
Discussion for more details).  In contrast, our MF equations are derived
from a biologically realistic network model: Instead of making
assumptions on gain functions, our MF equations follow closely the
anatomical and physiological information incorporated in the network,
hence reflecting its key features.  To stress this tight connection to a
realistic network model, we have described our method as a ``{\it
  data-informed} MF approach".  As we will show, the algorithm we
propose is capable of accurately predicting network firing rates at a
small fraction of the expense of direct network simulations.

We illustrate the power of this approach and how one might conceptualize
the mapping it produces using a biologically realistic model of Macaque
primary visual cortex (V1).  Focusing on spontaneous activity, our main
result is that the viable region is a thin neighborhood of a {\em
  codimension-1 manifold} in parameter space.  (A codimension-1 manifold
is an $(n-1)$-dimensional surface in an $n$-dimensional parameter
space.)  Being approximately codimension-1 implies that the viable
region is simultaneously sensitive and flexible to parameter changes:
sensitive in that a small perturbation can easily move a point off the
manifold; flexible in the sense that it allows for a great variety of
parameter combinations, consistent with the wide variability observed in
biology.  Our analysis of parameter dependence is based on the following
organizing principle: By restricting attention to certain 2D planes
associated with lateral inhibition, we discover geometric structures
that are remarkably similar across all such ``inhibition planes".  Our
findings suggest a number of simple {\em approximate scaling relations}
among neuronal parameters.

The Macaque V1 network model we use for illustration involves
$\gtrsim4\times10^4$ neurons in Layer 4C$\alpha$ of V1, and has been
carefully benchmarked to a large number of known features of V1 response
\cite{chariker2016orientation}.  In \cite{chariker2016orientation}, the
authors focused on a small parameter region which they had reason to
believe to be viable.  The present study produces a much more
comprehensive characterization of the set of all viable parameters
defined in terms of spontaneous activity.  The reason we have focused on
spontaneous activity is that it is a relatively simple, homogeneous
equilibrium steady state, and understanding it is necessary before
tackling more complex, evoked responses.  However, as all cortical
activity depends on a delicate balance between excitation and
inhibition, even background dynamics can be rather nontrivial.

Parameter search and tuning are problems common to all areas of
computational biology.  By significantly reducing the cost of mapping
parameter landscapes, we hope the computational strategy proposed in the
present paper will enable computational neuroscientists to construct
high-fidelity cortical models, and to use these models to shed light on
spontaneous and evoked dynamics in neural circuitry.  Moreover, reduced
models of the type proposed here may be useful as a basis for parameter
and state estimation on the basis of experimental data.

%%%%%%%%%%%%%%%%%%%%%%%%%%%%%%%%%%%%%
%%%%%%%%%%%%%%%%%%%%%%%%%%%%%%%%%%%%%
%%%%%%%%%%%%%%%%%%%%%%%%%%%%%%%%%%%%%%%
\section*{\large Results}

As explained in the Introduction, this work (1) proposes a novel
data-informed mean-field approach to facilitate efficient and systematic
parameter analysis of neuronal networks, which we validate using a
previously constructed model of the monkey visual cortex; and (2) we
develop ways to conceptualize and navigate the complexities of
high-dimensional parameter spaces of neuronal models by organizing
around certain relationships among parameters, notably those governing
lateral inhibition.

Sect.~\ref{sect:1} describes the network model of the visual cortex that
will be used both to challenge the MF algorithm and to assess its
efficacy, together with a brief introduction to the algorithm itself;
details are given in {\bf Methods}.  Sect.~\ref{sect:2} uses the
algorithm to explore the parameter landscape of the model.  Qualitative
analysis is offered along the way leading to a conceptual understanding
of parameter dependence.

%%%%%%%%%%%%%%%%%%%%%%%%%%%%%%%%%%%%%%%%%
%%%%%%%%%%%%%%%%%%%%%%%%%%%%%%%%%%%%%%%%
\section{Network model and parameter landscape}
\label{sect:1}

We use as starting point the large-scale network model in
\cite{chariker2016orientation}.  This is a mechanistic model of an input
layer to the primary visual cortex (V1) of the Macaque monkey, which has
vision very similar to that of humans~\cite{hubel1962receptive,
  callaway1998local, lund1988anatomical, douglas2004neuronal,
  angelucci2002circuits} Among existing neuronal network models,
\cite{chariker2016orientation} is at the very high end on the scale of
details and biological realism: It incorporates a good amount of known
neuroanatomy in its network architecture, capturing the dynamics of
individual neurons as well as their dynamical interaction.

In \cite{chariker2016orientation}, the authors located a small set of
potentially viable parameters, which they refined by benchmarking the
resulting regimes against multiple sets of experimental data.  No claims
were made on the uniqueness or optimality of the parameters considered.
Indeed, because of the intensity of the work involved in locating viable
parameters, little attempt was made to consider parameters farther from
the ones used.  This offers a natural testing ground for our novel
approach to parameter determination: We borrow certain aspects of the
model from \cite{chariker2016orientation}, including network
architecture, equations of neuronal dynamics and parameter structure,
but instead of using information on the parameters found, we will search
for viable parameter regions using the techniques developed here.

For a set of parameters to be viable, it must produce firing rates
similar to those of the real cortex, including {\it background} firing,
the spontaneous spiking produced when cortex is unstimulated.
Background activity provides a natural way to constrain parameters: It
is an especially simple state of equilibrium, one in which spiking is
statistically homogeneous in space and time
%
%% \ztext{?}\znote{My feeling is that the temporal dynamics still yield
%% weak gamma from time to time}\knote{I don't see any contradictions or
%% problems here -- ``homogeneous'' to me means statistically stationary
%% / homogeneous; added ``statistically'' to clarify}
%
and involves fewer features of cortical dynamics.  For example, synaptic
depression and facilitation are not known to play essential roles in
spontaneous activity.  A goal in this paper will be to systematically
identify all regions in parameter space with acceptable background
firing rates.

%%%%%%%%%%%%%%%%%%%%%%%%%%%%%%%%%%%%
\subsection{Network model of an input layer to primate visual cortex}
\label{sect:1.1}

The network model is that of a small patch of Macaque V1, layer
4C$\alpha$ (hereafter ``L4"), the input layer of the Magnocellular
pathway.  This layer receives input from the lateral geniculate nucleus
(LGN) and feedback from layer 6 (L6) of V1.  Model details are as in
\cite{chariker2016orientation}, except for omission of features not
involved in background activity.  We provide below a model description
that is sufficient for understanding -- at least on a conceptual level
-- the material in {\bf Results}.  Precise numerical values of the
various quantities are given in {\bf SI}.  For more detailed discussions
of the neurobiology behind the material in this subsection, we refer
interested readers to \cite{chariker2016orientation} and references
therein.

\lsyheadingsmall{Network architecture}

Of primary interest to us is L4, which is modeled as a 2D network of
point neurons.  Locations within this layer are identified with
locations in the retina via neuronal projections, and distances on the
retina are measured in degrees.  Cells\footnote{In this paper, ``cells"
and ``neurons" both refer to nerve cells in the primary visual cortex.}
are organized into hypercolumns of about 4000 neurons each, covering a
$0.25^\circ \times 0.25^\circ$ area.  The neurons are assumed to be of
two kinds: $75-80\%$ are Excitatory (E), and the rest are Inhibitory
(I).  The E-population is evenly placed in a lattice on the cortical
surface; the same is true for I-cells.  Projections to E- and I-cells
are assumed to be isotropic, with probabilities of connection described
by truncated Gaussians. E-neurons (which are assumed to be spiny stellate cells) have longer axons, about twice that of I-neurons (which are
assumed to be basket cells).  E-to-E coupling is relatively sparse, at
about $15\%$ at the peak, while E-to-I, I-to-E and I-to-I coupling is
denser, at about $60\%$.  Connections are drawn randomly subject to the
probabilities above.  On average, each E-neuron receives synaptic input
from slightly over 200 E-cells and about 100 I-cells, while each I-cell
receives input from $\sim 800$ E-cells and 100 I-cells.  Exact numbers
are given in {\bf SI}.

Cells in L4 also receive synaptic input from two external sources,
``external" in the sense that they originate from outside of this layer.
One source is LGN: Each L4 cell, E or I, is assumed to be connected to 4
LGN cells on average; each LGN cell is assumed to provide 20 spikes per
sec, a typical firing rate in background.  These spikes are assumed to
be delivered to L4 cells in a Poissonian manner, independently from cell
to cell.  L6 provides another source of synaptic input: We assume each
E-cell in L4 receives input from 50 E-cells from L6, consistent with
known neurobiology, and that each L6 E-cell fires on average 5 spikes
per sec in background.  For I-cells, the number of presynaptic L6 cells
is unknown; this is one of the free parameters we will explore.  Spike
times from L6 are also assumed to be Poissonian and independent from
cell to cell, a slight simplification of \cite{chariker2016orientation}.
All inputs from external sources are excitatory.  Finally, we lump
together all top-down modulatory influences on L4 not modeled into a
quantity we call ``ambient".  Again, see \textbf{SI} for all pertinent details.

\lsyheadingsmall{Equations of neuronal dynamics}

We model only the dynamics of neurons in L4.  Each neuron is modeled as
a conductance-based point neuron whose membrane potential $v$ evolves
according to the leaky integrate-and-fire (LIF) equation
\cite{lapicque1907excitation, dayan2001theoretical}
\begin{equation} \label{LIF}
\ddt{v} = - g^L (v-V_{\rm rest}) - g^E(t)(v-V^E) - g^I(t)(v-V^I)\ .
\end{equation}
Following \cite{chariker2016orientation}, we nondimensionalize $v$, with
resting potential $V_{\rm rest} = 0$ and spiking threshold $V_{\rm th}
=1$.  In Eq.~(\ref{LIF}), $v$ is driven toward $V_{\rm th}$ by the
Excitatory current $g^E(t)(v-V^E)$, and away from it by the leak term
$g^L v$ and the Inhibitory current $g^I(t)(v-V^I)$.  When $v$ reaches
$1$, a spike is fired, and $v$ is immediately reset to $0$, where it is
held for a refractory period of $2$ ms.  The membrane leakage times
$1/g^L=20$ ms for E-neurons and 16.7 ms for I-neurons, as well as the
reversal potentials $V^E= 14/3$ and $V^I=-2/3$, are standard
\cite{koch1999biophysics}.

The quantities $g^E(t)$ and $g^I(t)$, the Excitatory and Inhibitory
conductances, are defined as follows. First,
\begin{equation}
  \label{I-conductance}
  g^I(t) = S^{QI} \sum_{t^I_{\rm spike}} G_{I}(t-t^I_{\rm spike})\ .
\end{equation}
Here, the neuron whose dynamics are described in Eq.~(\ref{LIF}) is
assumed to be of type $Q$, $Q=\mbox{E}$ or I, and the constant $S^{QI}$ is the
synaptic coupling weight from I-neurons to neurons of type $Q$.  The
summation is taken over $t^I_{\rm spike}$, times at which a spike from a
presynaptic I-cell from within layer 4C$\alpha$ is received by the
neuron in question.  Upon the arrival of each spike, $g_I(t)$ is
elevated for 5-10 ms and $G_{I}(\cdot)$ describes the waveform in the
IPSC.  Second, the Excitatory conductance $g_E(t)$ is the sum of 4
terms, the first three of which are analogs of the right side of Eq.~(\ref{I-conductance}), with an EPSC lasting 3-5 ms: they represent
synaptic input from E-cells from within L4, from LGN and from L6. The
4th term is from ambient.

\medskip

This completes the main features of the network model. Details are given
in {\bf SI}.

%%%%%%%%%%%%%%%%%%%%%%%%%%%%%%%%%%%%
\subsection{Parameter space to be explored}
\label{sect:1.2}

Network dynamics can be very sensitive -- or relatively robust -- to
parameter changes, and dynamic regimes can change differently depending
on which parameter (or combination of parameters) is varied.  To
demonstrate the multiscale and anisotropic nature of the parameter
landscape, we study the effects of parameter perturbations on L4 firing
rates, using as reference point a set of biologically realistic
parameters \cite{chariker2016orientation}.  Specifically, we denote L4
E/I-firing rates at the reference point by $(f_{E}^0,f_{I}^0)$, and fix
a region $\mathcal{F}$ around $(f_{E}^0,f_{I}^0)$ consisting of firing
rates $(f_E, f_I)$ we are willing to tolerate.  We then vary network
parameters one at a time, changing it in small steps and computing
network firing rates $(f'_E,f'_I)$ until they reach the boundary of
$\mathcal{F}$, thereby determining the minimum perturbations needed to
force L4 firing rates out of $\mathcal{F}$.

%%%%%%%%%%%%%%%%%%%%%%%%%%%%%%%%
\begin{table}[htbp]
  \begin{center}
    \begin{tabular}{|c|c|l|l|l|}
      \hline
      Group              & Parameter        & Meaning                       & Value  & Bounds        \\ \hline
      within L4          & $S^{EE}$         & E-to-E  synaptic weight       & 0.024  & (-3\%, 1\%)   \\ 
                         & $S^{II}$         & I-to-I synaptic weight        & 0.120  & (-4\%, 1\%)   \\ 
                         & $S^{EI}$         & I-to-E synaptic weight        & 0.0362 & (-1\%, 3\%)   \\ 
                         & $S^{IE}$         & E-to-I synaptic weight        & 0.0176 & (-1\%, 3\%)   \\ \hline
      LGN to L4          & $S^{E{\rm lgn}}$ & lgn-to-E synaptic weight      & 0.048  & (-5\%, 3\%)   \\ 
                         & $S^{I{\rm lgn}}$ & lgn-to-I synaptic weight      & 0.096  & (-6\%, 9\%)   \\
                         & $F^{E{\rm lgn}}$ & total \# lgn spikes/s to E    & 80 Hz  & (-7\%, 4\%)   \\
                         & $F^{I{\rm lgn}}$ & total \# lgn spikes/s to I    & 80 Hz  & (-9\%, 11\%)  \\ \hline
      L6 to L4           & $S^{E{\rm L6}}$  & L6-to-E synaptic weight       & 0.008  & (-16\%, 11\%) \\ 
                         & $S^{I{\rm L6}}$  & L6-to-I synaptic weight       & 0.0058 & (-19\%, 30\%) \\
                         & $F^{E{\rm L6}}$  & total \# L6 spikes/s to E     & 250 Hz & (-16\%, 10\%) \\
                         & $F^{I{\rm L6}}$  & total \# L6 spikes/s to I     & 750 Hz & (-16\%, 29\%) \\\hline
      amb to L4          & $S^{{\rm amb}}$  & ambient-to-E/I synaptic wt.   & 0.01   & (-8\%, 6\%)   \\
                         & $F^{E{\rm amb}}$ & rate of ambient to E          & 500 Hz & (-7\%, 5\%)   \\
                         & $F^{I{\rm amb}}$ & rate of ambient to I          & 500 Hz & (-10\%, 27\%) \\\hline
    \end{tabular}
 \medskip
    \caption{Network parameters and response.  Using the parameters from
      \cite{chariker2016orientation} as a reference point, we set
      $(f_{E}^0,f_{I}^0) = (3.85, 13.32)$ Hz and $\mathcal{F} =
      \big\{(f_E,f_I)~\big|~\nicefrac{f_E}{f_E^0}\in(\nicefrac{2}{3},
      \nicefrac{4}{3}),\nicefrac{f_I}{f_E}\in(3, 4.5)\big\}$ and vary
      one parameter at a time.  We then compute the minimum perturbation
      needed to force the network firing rates out of $\mathcal{F}$.
      Values such as $F^{E\rm P}$, where $P\in\{{\rm lgn}, {\rm L6},
      {\rm amb}\}$, represent the total number of spikes per second
      received by an E-cell in L4 from source $P$.  For example, in the
      reference set, each L4 cell has 4 afferent LGN cells on average,
      the mean firing rate of each is assumed to be 20 spikes/s, so
      $F^{E{\rm lgn}}=$ 80 Hz.  Column 5 gives the lower and upper
      bounds of single-parameter variation (rounded to the nearest 1\%)
      from the reference point that yield firing rates within
      $\mathcal{F}$.}
    \label{Table1: Ref-point}
  \end{center}
\end{table}
%%%%%%%%%%%%%%%%%%%%%%%%%%%%%%%%

Table~\ref{Table1: Ref-point} shows the results.  We categorize the
parameters according to the aspect of network dynamics they govern.  As
can be seen, L4 firing rates show varying degrees of sensitivity to
perturbations in different parameter groups.  They are most sensitive to
perturbations to synaptic coupling weights within L4, where deviations
as small as 1\% can push the firing rates out of $\mathcal{F}$.  This
likely reflects the delicate balance between excitation and inhibition,
as well as the fact that the bulk of the synaptic input to a L4 neuron
comes from lateral interaction, facts consistent with earlier findings
\cite{chariker2016orientation}.  With respect to parameters governing inputs from external
sources, we find perturbing LGN parameters to have the most impact,
followed by amb and L6, consistent with their net influence on $g_{E,I}$
in background.  We also note that the parameters governing afferents to
I cells are more tolerant of perturbations than those for E cells.

These observations suggest that network dynamics depend in a complex and
subtle way on parameters; they underscore the challenges one faces when
attempting to tune parameters ``by hand."  We now identify the
parameters in the network model description in Sect.~\ref{sect:1.1} to
be treated as \textit{free parameters} in the study to follow.

\lsyheadingmed{Free parameters}

We consider a parameter ``free'' if it is hard to measure (or has not
yet been measured) directly in the laboratory, or when data offer
conflicting guidance.  When available data are sufficient to confidently
associate a value to a parameter, we consider it fixed.  Following this
principle, we designate the following 6 synaptic coupling weights
governing recurrent interactions within L4 and its thalamic inputs as
free parameters:
$$
S^{EE}, \quad S^{EI}, \quad S^{IE}, \quad S^{II}, \quad S^{E{\rm lgn}},
\quad S^{I{\rm lgn}}\ .
$$
As shown in Table~\ref{Table1: Ref-point}, these are also the parameters
to which network response rates are the most sensitive.  As for
$S^{E{\rm L6}}$ and $S^{I{\rm L6}}$, which govern synaptic coupling from
L6 to E- and I-neurons in L4, we assume
$$
S^{E{\rm L6}} = \frac13 \times S^{EE}, \qquad S^{I{\rm L6}} = \frac13 
\times S^{IE}\ 
$$
following \cite{stratford1996excitatory} (see also
\cite{chariker2016orientation}).  This means that in our study, these
quantities will vary, but they are indexed to $S^{EE}$ and $S^{IE}$ in a
fixed manner and we will not regard them as free parameters.

A second category of parameters govern external sources.  Here we regard
$F^{E{\rm lgn}}$, $F^{I{\rm lgn}}$ and $F^{E{\rm L6}}$ as fixed to the
values given in Table~\ref{Table1: Ref-point}.  L6 firing rates in
background have been measured, but we know of no estimates on the number
of presynaptic L6 cells to I-cells, so we treat $F^{I{\rm L6}}$ (which
combines the effects of both) as a free parameter.  The relation between
$S^{I{\rm L6}}$ and $S^{IE}$ assumed above is in fact unknown from
experiments.  On the other hand, we are assuming that errors in the
estimate of $S^{I{\rm L6}}$ can be absorbed into $F^{I{\rm L6}}$, which
we vary.  As for ``ambient", these inputs are thought to be significant,
though not enough to drive spikes on their own.  Since so little is
known about this category of inputs, we fix the values of $S^{\rm amb}$,
$F^{E{\rm amb}}$, and $F^{I{\rm amb}}$ to those given in
Table~\ref{Table1: Ref-point}, having checked that they meet the
conditions above.

\medskip 

As discussed earlier, we are interested in L4 firing rates under
background conditions.  Denoting E- and I- firing rates by $f_E$ and
$f_I$ respectively, the aim of our study can be summarized as follows:

\bigskip \noindent{\bf Aim:} {\it To produce maps of $f_E$ and $f_I$ as
  functions of the 7 parameters
\begin{equation}
  \label{freeparam}
  S^{EE}, \ S^{EI}, \ S^{IE}, \ S^{II}, \ S^{E{\rm lgn}}, \ S^{I{\rm
      lgn}}, \mbox{ and } \ F^{I{\rm L6}}\ ,
\end{equation}
to identify biologically relevant regions, and to provide a conceptual
understanding of the results. }

%%%%%%%%%%%%%%%%%%%%%%%%%%%%%%%%%
\subsection{A brief introduction to our proposed MF approach}
\label{sect:1.3}

The approach we take is a MF computation of firing rates augmented by
synthetic voltage data, a scheme we will refer to as ``MF+v".  To
motivate the ``+v" part of the scheme, we first write down the MF
equations obtained from Eq.~(\ref{LIF}) by balancing mean membrane
currents.  These MF equations will turn out to be incomplete.  We
discuss briefly how to secure the missing information; details are given
in {\bf Methods}.

\lsyheadingmed{MF equations}

Eq.~(\ref{LIF}) reflects instantaneous current balance across the cell
membrane of a L4 neuron.  Assuming that this neuron's firing rate
coincides with that of the L4 E/I-population to which it belongs and
neglecting (for now) refractory periods, we obtain a general relation
between firing rates and mean currents by integrating Eq.~(\ref{LIF}).
We will refer to the equations below as our ``MF equations".  They have
the general form
\begin{subequations}
  \label{MF}
  \begin{eqnarray} \label{MF1} 
    f_E & = & \mbox{ [4E $\to$ E] } + \mbox{ [4I $\to$ E] } + \mbox{ contributions from LGN, L6, amb
      and leak },\\ \label{MF2}
    f_I & = & \mbox{ [4E $\to$ I] } + \mbox{ [4I $\to$ I] } + \mbox{ contributions from LGN, L6, amb
      and leak } ,
  \end{eqnarray} 
\end{subequations}
where [4E $\to$ E] represents the integral of the current contribution
from E-cells in L4 to E-cells in L4, [4I $\to$ E] represents the
corresponding quantity from I-cells in L4 to E-cells in L4, and so on.
The contribution from lateral, intralaminar interactions can be further
decomposed into, e.g.,
\begin{eqnarray*}
\mbox{ [4E $\to$ E] } & = & N^{EE} \times f_E \times S^{EE} \times (V^E-\bar v_{E}), \\
\mbox{ [4I $\to$ E] } & = & N^{EI} \times f_I \times S^{EI} \times (V^E-\bar v_{E})\ .
\end{eqnarray*}
Here $N^{EE}$ and $N^{EI}$ are the mean numbers of presynaptic E- and
I-cells from within L4 to an E-neuron, $f_E, f_I, S^{EE}$ and $S^{EI}$
are as defined earlier, and $\bar v_{E}$ is the mean membrane potential
$v$ among E-neurons in L4. Other terms in Eq.~(\ref{MF1}) and in Eq.~(\ref{MF2})
are defined similarly; detailed derivation of the MF equations is given
in {\bf Methods}.

Network connectivity and parameters that are not considered ``free
parameters" are assumed to be fixed throughout. If additionally we fix a
set of the 7 free parameters in (\ref{freeparam}), then Eq.~(\ref{MF})
is linear in $f_E$ and $f_I$, and are easily solved --- except for two
undetermined quantities, $\bar v_E$ and $\bar v_I$.  For network
neurons, $\bar v_E$ and $\bar v_I$ are emergent quantities that cannot
be easily estimated from the equations of evolution or parameters
chosen.

\lsyheadingmed{Estimating mean voltages} 

We explain here the ideas that lead to the algorithm we use for
determining $\bar v_E$ and $\bar v_I$, leaving technical details to {\bf
  Methods}.

Our first observation is that the values of $f_E$ and $f_I$ computed
from Eq.~(\ref{MF}) depend delicately on $\bar v_E$ and $\bar v_I$;
they can vary wildly with small changes in $\bar v_E$ and $\bar v_I$.
This ruled out the use of (guessed) approximate values, and even called
into question the usefulness of the MF equations.  But as we demonstrate
in {\bf Methods}, if one collects mean voltages $\bar v_E$ and $\bar
v_I$ from network simulations and plug them into Eq.~(\ref{MF}) to solve
for $f_E$ and $f_I$, then one obtains results that agree very well with
actual network firing rates.  This suggests Eq.~(\ref{MF}) can be
useful, provided we correctly estimate $\bar v_E$ and $\bar v_I$.

As it defeats the purpose of an MF approach to use network simulations
to determine $\bar v_E$ and $\bar v_I$, we sought to use a pair of
LIF-neurons, one E and one I, to provide this information.  To do that,
we must create an environment for this pair of neurons that is similar
to that within the network, incorporating the biological features with
the LIF neurons.  For example, one must use the same parameters and give
them the same external drives, i.e., LGN, L6, and ambient.  But a good
fraction of the synaptic input to neurons in L4 are generated from
lateral interactions; to simulate {\em that,} we would have to first
learn what $f_E$ and $f_I$ are.  The problem has now come full circle:
what we need are {\it self-consistent} values of $f_E$ and $f_I$ for the
LIF-neurons, so that their input and output firing rates coincide.

These and other ideas to be explained (e.g., efficiency and stability)
go into the algorithm proposed.  In a nutshell, we use the aid of a pair
of LIF-neurons to help tie down $\bar v_E$ and $\bar v_I$, and use the
MF equations to compute $f_E$ and $f_I$.  This mean-field algorithm
aided by voltage closures (MF+v) is discussed in detail in {\bf
  Methods}.  We present next firing rate plots generated using this
algorithm.

%%%%%%%%%%%%%%%%%%%%%%%%%%%%%%%%%%%%%%
%%%%%%%%%%%%%%%%%%%%%%%%%%%%%%%%%%%%%%%
\section{Dependence of firing rates on system parameters}
\label{sect:2}

Even with a fast algorithm, so that many data points can be computed,
discovery and representation of functions depending on more than 3 or 4
variables can be a challenge, not to mention conceptualization of the
results.  In Sects.~\ref{sect:2.1}--\ref{sect:2.3}, we propose to
organize the 7D parameter space described in Sect.~\ref{sect:1.2} in
ways that take advantage of insights on how the parameters interact:
Instead of attempting to compute 6D level surfaces for $f_E$ and $f_I$
embedded in the 7D parameter space, we identify a biologically plausible
region of parameters called the ``viable region", and propose to study
parameter structures by slicing the 7D space with certain 2D planes
called ``inhibition planes". We will show that intersections of the
viable region and inhibition planes -- called ``good areas" -- possess
certain canonical geometric structures, and that these structures offer
a biologically interpretable landscape of parameter dependence. The
three terms, {\it viable regions}, {\it inhibition planes} and {\it good
  areas}, the precise definitions of which are given in
Sect.~\ref{sect:2.1}, are objects of interest throughout this section.
In Sect.~\ref{sect:2.4} we show comparison of firing rate computations
from our algorithm and from actual network simulations.

\subsection{Canonical structures in inhibition planes}
\label{sect:2.1}

We have found it revealing to slice the parameter space using 2D planes
defined by varying the parameters governing lateral inhibition, $S^{IE}$
and $S^{EI}$, with all other parameters fixed.  As we will show, these
planes contain very simple and stable geometric structures around which
we will organize our thinking about parameter space.  Fig.~\ref{Fig1:
  Canonical Set}A shows one such 2D slice.  We computed raw contour
curves for $f_E$ and $f_I$ on a $480\times480$ grid using the MF+v
algorithm, with red curves for $f_E$ and blue for $f_I$.

%%%%%%%%%%%%%%%%%%%%%%%%%%%%%%%%%%%%%%%%%%%%%%%%%%%%%%%%%%% Fig1
\begin{figure}%[htbp]
  \begin{center}
    %% \captionsetup{type=figure} 
  \begin{subfigure}{.678\textwidth}
  {\bf A}\\
  \includegraphics*[bb=0in 0.05in 5.9in 6in,width=\textwidth]{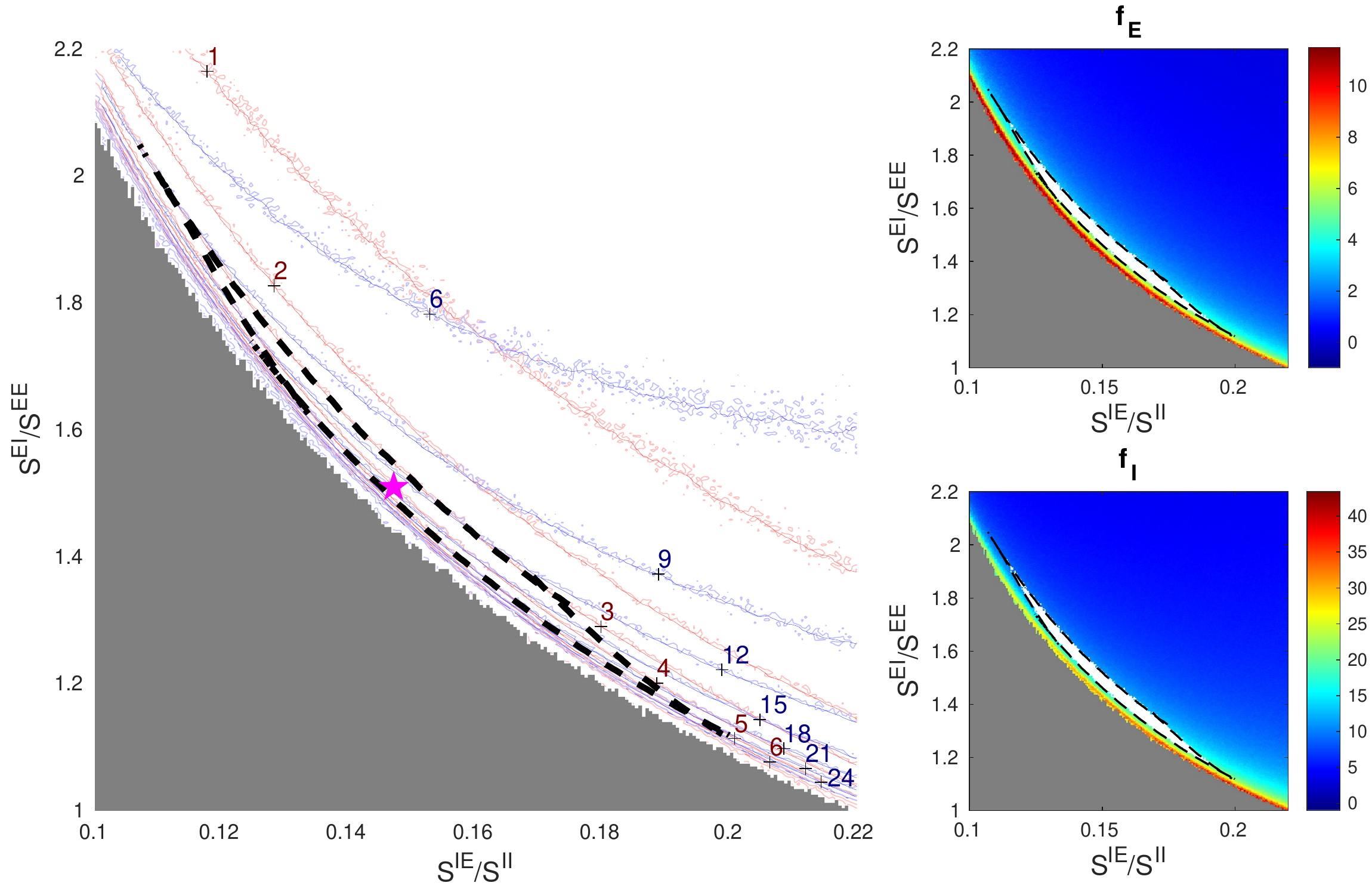}
  \end{subfigure}%
  \begin{subfigure}{.322\textwidth}
  {\bf B}\\
  \includegraphics*[bb=5.9in 0.05in 8.7in 6in,width=\textwidth]{{Figure1}.pdf}
  \end{subfigure}%  
    \caption{Canonical structures in an inhibition plane.  {\bf A.}
      Firing rates contours computed from firing rate maps on a 480$
      \times$480 grid, showing $f_E = 1-6$ Hz (red) and $f_I = 6-24$ Hz
      (blue).  The good area is indicated by the black dash lines ($f_E
      \in (3,5)$ Hz and $f_I/f_E \in (3,4.25)$); the reference point is
      indicated by the purple star ($\star$).  The MF+v method becomes
      unstable and fails when the inhibitory index $\mbox{SI}_E$ is too
      low (the gray region).  {\bf B.} Firing rate maps of $f_E$ (upper)
      and $f_I$ (lower), in which the good areas are indicated by white
      bands surrounded by black dash lines.  The other five free
      parameters are as in Table~\ref{Table1: Ref-point}: $S^{EE} =
      0.024$, $S^{II} = 0.12$, $S^{E{\rm lgn}} = 2\times S^{EE}$,
      $S^{I{\rm lgn}} = 2\times S^{E{\rm lgn}}$, and $F^{I{\rm L6}} =
      3\times F^{E{\rm L6}}$.}
    \label{Fig1: Canonical Set}
  \end{center}
\end{figure}
%%%%%%%%%%%%%%%%%%%%%%%%%%%%%%%%%%%%%%%%%%%%%%%%%%%%%%%%%%% Fig1

A striking feature of Fig.~\ref{Fig1: Canonical Set}A is that the level
curves are roughly hyperbolic in shape.  We argue that this is
necessarily so.  First, note that in Fig.~\ref{Fig1: Canonical Set}A, we
used ${S^{IE}}/{S^{II}}$ as $x$-axis and ${S^{EI}}/{S^{EE}}$ as $y$-axis.  The
reason for this choice is that ${S^{IE}}/{S^{II}}$ can be viewed as the
degree of cortical excitation of I-cells, and ${S^{EI}}/{S^{EE}}$ the
suppressive power of I-cells from the perspective of E-cells.  The
product
\begin{displaymath}
  \mbox{SI}_E := \frac{S^{EI}}{S^{EE}} \times \frac{S^{IE}}{S^{II}}
\end{displaymath}
can therefore be seen as a \emph{suppression index} for E-cells: the
larger this quantity, the smaller $f_E$.  This suggests that the
contours for $f_E$ should be of the form $xy = $ constant, i.e., they
should have the shape of hyperbolas. As E and I firing rates in local
populations are known to covary, these approximately hyperbolic shapes
are passed to contours of $f_I$.

A second feature of Fig.~\ref{Fig1: Canonical Set}A is that $f_I$
contours are less steep than those of $f_E$ at lower firing rates. That
I-firing covaries with E-firing is due in part to the fact that I-cells
receive a large portion of their excitatory input from E-cells through
lateral interaction, at least when E-firing is robust.  When $f_E$ is
low, $f_I$ is lowered as well as I-cells lose their supply of excitation
from E-cells, but the drop is less severe as I-cells also receive
excitatory input from external sources. This causes $f_I$ contours to
bend upwards relative to the $f_E$-hyperbolas at lower firing rates, a
fact quite evident from Fig.~\ref{Fig1: Canonical Set}.

We now define the {\it viable region}, the biologically plausible region
in our 7D parameter space, consisting of parameters that produce firing
rates we deem close enough to experimentally observed values.  For
definiteness, we take these to be~\cite{ringach2002orientation}
$$ 
f_E \in (3,5) \mbox{ Hz \quad and \quad} f_I/f_E \in (3, 4.25),
$$ 
and refer to the intersection of the viable region with the 2D slice
depicted in Fig.~\ref{Fig1: Canonical Set}A as the ``good area".  Here,
the good area is the crescent-shaped set bordered by dashed black lines.
For the parameters in Fig.~\ref{Fig1: Canonical Set}, it is bordered by
4 curves, two corresponding to the $f_E=3$ and $5$ Hz contours and the
other two are where $f_I/f_E = 3$ and $4.25$.  That such an area should
exist as a narrow strip of finite length, with unions of segments of
hyperbolas as boundaries, is a consequence of the fact that $f_E$ and
$f_I$-contours are roughly but not exactly parallel.  Fig.~\ref{Fig1:
  Canonical Set}B shows the good area (white) on firing rate maps for
$f_E$ and $f_I$.

Hereafter, we will refer to 2D planes parametrized by ${S^{IE}}/{S^{II}}$
and ${S^{EI}}/{S^{EE}}$, with all parameters other than $S^{IE}$ and
$S^{EI}$ fixed, as {\em inhibition planes}, and will proceed to
investigate the entire parameter space through these 2D slices and the
good areas they contain.  Though far from guaranteed, our aim is to show
that the structures in Fig.~\ref{Fig1: Canonical Set}A persist, and to
describe how they vary with the other 5 parameters.

Fig.~\ref{Fig1: Canonical Set} is for a particular set of parameters.
We presented it in high resolution to show our computational capacity
and to familiarize the reader with the picture.  As we vary parameters
in the rest of this paper, we will present only heat maps for $f_E$ for
each set of parameters studied.  The good area, if there is one, will
be marked in white, in analogy with the top panel of Fig.~\ref{Fig1:
  Canonical Set}B.

Finally, we remark that the MF+v algorithm does not always return
reasonable estimates of L4 firing rates.  MF+v tends to fail especially
for a low suppression index (gray area in Fig.~\ref{Fig1: Canonical
  Set}A), where the network simulation also exhibits explosive,
biologically unrealistic dynamics.  This issue is discussed in {\bf
  Methods} and {\bf SI}.

%%%%%%%%%%%%%%%%%%%%%%%%%%%%%%%%
\subsection{Dependence on external drives}
\label{sect:2.2}

There are two main sources of external input, LGN and L6 (while ambient
input is assumed fixed).  In both cases, it is their effect on E- {\it
  versus} I-cells in L4, and the variation thereof as LGN and L6
inputs are varied, that is of interest here.

%% In both cases, it is the variation of their effect on E- {\it versus}
%% I-cells in L4 that is of interest here.

\lsyheadingmed{LGN-to-E \emph{versus} LGN-to-I}

Results from \cite{stratford1996excitatory} suggest that the sizes of
EPSPs from LGN are $\sim2\times$ those from L4.  Based on this, we
consider the range $S^{E\rm lgn}/S^{EE}\in(1.5, 3.0)$ in our study.
Also, data show that LGN produces somewhat larger EPSCs in I-cells than
in E-cells~\cite{beierlein2003two}, though the relative coupling weights
to E and I-cells are not known.  Here, we index $S^{I\rm lgn}$ to
$S^{E\rm lgn}$, and consider $S^{I\rm lgn}/S^{E\rm lgn}\in(1.5, 3.0)$.

\begin{figure}%[htbp]
  \begin{center}
    %% \captionsetup{type=figure} 
    \includegraphics[width=\textwidth]{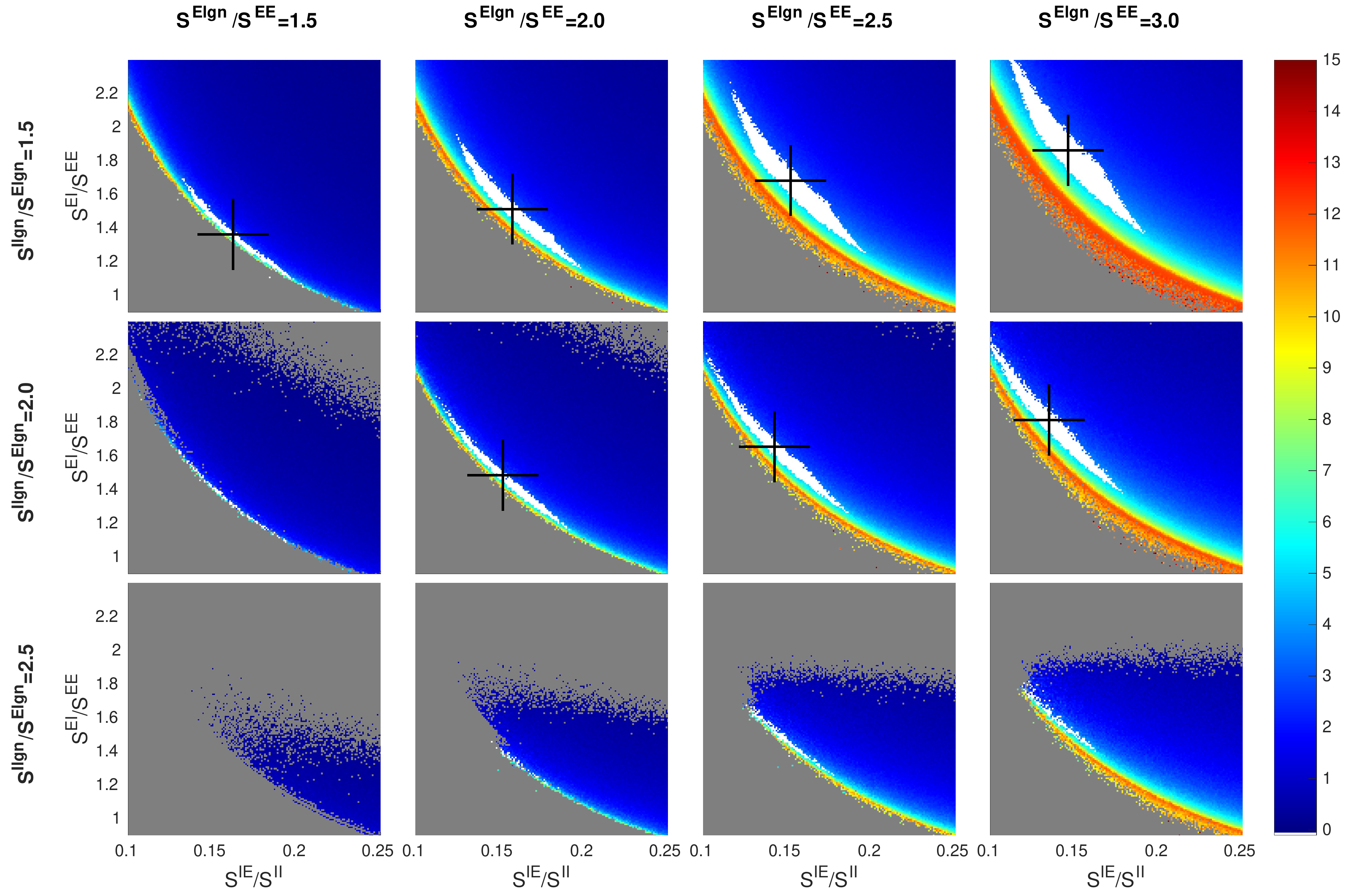}
    \caption{Dependence of firing rate and good area on LGN synaptic
      strengths.  A 3$\times$4 matrix of panels is shown: each row
      corresponds to a fixed value of $S^{I{\rm lgn}}/S^{E{\rm lgn}}$
      and each column a fixed value of ${S^{E{\rm lgn}}}/{S^{EE}}$. Each
      panel shows a heat map for $f_E$ on an inhibition plane (color bar
      on the right); $x$ and $y$-axes are as in Fig.~\ref{Fig1:
        Canonical Set}.  Good areas are in white, and their centers of
      mass marked by black crosses.  The picture for ${S^{E{\rm
            lgn}}}/{S^{EE}} =2$ and ${S^{I{\rm lgn}}}/{S^{E{\rm
            lgn}}}=2$ (row 2, column 2) corresponds to the $f_E$ rate
      map in Fig.~\ref{Fig1: Canonical Set}. Other free parameters are
      $S^{EE} = 0.024$, $S^{II} = 0.12$, and $F^{I{\rm L6}} = 3\times
      F^{E{\rm L6}}$.}
    \label{Fig2: LGN}
  \end{center}
\end{figure}

Fig.~\ref{Fig2: LGN} shows a $3 \times 4$ matrix of 2D panels, each one
of which is an inhibition plane (see Fig.~\ref{Fig1: Canonical Set}).
This is the language we will use here and in subsequent figures: we will
refer to the rows and columns of the matrix of panels, while $x$ and $y$
are reserved for the axes for each smaller panel.

%%%%%%%%%%%%%%%%%%%%%%%%%%%%%%%%%%%%%%%%%%%%%%%%%%%%%%%%%%% Fig2
%%%%%%%%%%%%%%%%%%%%%%%%%%%%%%%%%%%%%%%%%%%%%%%%%%%%%%%%%%% Fig2

We consider first the changes as we go from left to right in each row of
the matrix in Fig.~\ref{Fig2: LGN}.  With $S^{I{\rm lgn}}/S^{E{\rm
    lgn}}$ staying fixed, increasing $S^{E{\rm lgn}}/S^{EE}$ not only
increases LGN's input to E, but also increases LGN to I by the same
proportion.  It is evident that the rate maps in the subpanels are all
qualitatively similar, but with gradual changes in the location and the
shape of the good area.  Specifically, as LGN input increases, (i) the
center of mass of the good area (black cross) shifts upward and to the
left following the hyperbola, and (ii) the white region becomes wider.

To understand these trends, it is important to keep in mind that the
good area is characterized by having firing rates within a fairly narrow
range.  As LGN input to E increases, the amount of suppression must be
increased commensurably to maintain E-firing rate.  Within an inhibition
plane, this means an increase in $S^{EI}$, explaining the upward move of
black crosses as we go from left to right.  Likewise, the amount of
E-to-I input must be decreased by a suitable amount to maintain I-firing
at the same level, explaining the leftward move of the black crosses and
completing the argument for (i).  As for (ii), recall that the SI$_E$
measures the degree of suppression of E-cells {\it from within L4} (and
L6, the synaptic weights of which are indexed to those in L4).
Increased LGN input causes E-cells to be less suppressed than their
SI$_E$ index would indicate.  This has the effect of spreading the $f_E$
contours farther apart, stretching out the picture in a northeasterly
direction perpendicular to the hyperbolas and widening the good area.

Going down the columns of the matrix, we observe a compression of the
contours along the same northeasterly axis and a leftward shift of the
black crosses.  Recall that the only source of excitation of I-cells
counted in the SI$_E$-index is from L4 (hence also L6).  When LGN to E
is fixed and LGN to I is increased, the additional external drive to I
produces a larger amount of suppression than the SI$_E$-index would
indicate, hence the compression.  It also reduces the amount of $S^{IE}$
needed to produce the same I-firing rate, hence the leftward shift of
the good area.

The changes in LGN input to E and to I shown cover nearly all of the
biological range.  We did not show the row corresponding to $S^{I{\rm
    lgn}}/S^{E{\rm lgn}}=3$ because the same trend continues and there
are no viable regions.  Notice that even though we have only shown a $3
\times 4$ matrix of panels, the trends are clear and one can easily
interpolate between panels.

\lsyheadingmed{LGN-to-I \emph{versus} L6-to-I}

Next, we examine the relation between two sources of external drive: LGN
and L6.  In principle, this involves a total of 8 quantities: total
number of spikes and coupling weights, from LGN and from L6, received by
E and I-neurons in L4.  As discussed in Sect.~\ref{sect:1.2}, enough is
known about several of these quantities for us to treat them as
``fixed", leaving as free parameters the following three:
$S^{E\rm{lgn}}, S^{I\rm{lgn}}$ and $F^{I\rm{L6}}$.  Above, we focused on
$S^{I\rm{lgn}}/S^{E\rm{lgn}}$, the impact of LGN on I relative to E.  We
now compare $S^{I\rm{lgn}}/S^{E\rm{lgn}}$ to
$F^{I\rm{L6}}/F^{E\rm{L6}}$, the corresponding quantity with L6 in the
place of LGN.

As there is little experimental guidance with regard to the range of
$F^{I{\rm L6}}$, we will explore a relatively wide region of $F^{I{\rm
    L6}}$: Guided by the fact that
$$
(\# \mbox{ presynaptic E to an I-cell}) \ / \  (\# \mbox{ presynaptic E to an E-cell})
\ \ \sim \ \ 3.5-4\ ,
$$
in L4 and the hypothesis that similar ratios hold for inter-laminar
connections, we assume $F^{I{\rm L6}}/F^{E{\rm L6}} \in (1.5, 6)$.  We
have broadened the interval because it is somewhat controversial whether
the effect of L6 is net-Excitatory or net-Inhibitory: the modeling work
\cite{chariker2020contrast} on monkey found that it had to be at least
slightly net-Excitatory, while \cite{bortone2014translaminar} reported
that it was net-Inhibitory in mouse.

\begin{figure}%[htbp]
  \begin{center}
    %% \captionsetup{type=figure} 
    \includegraphics[width=\textwidth]{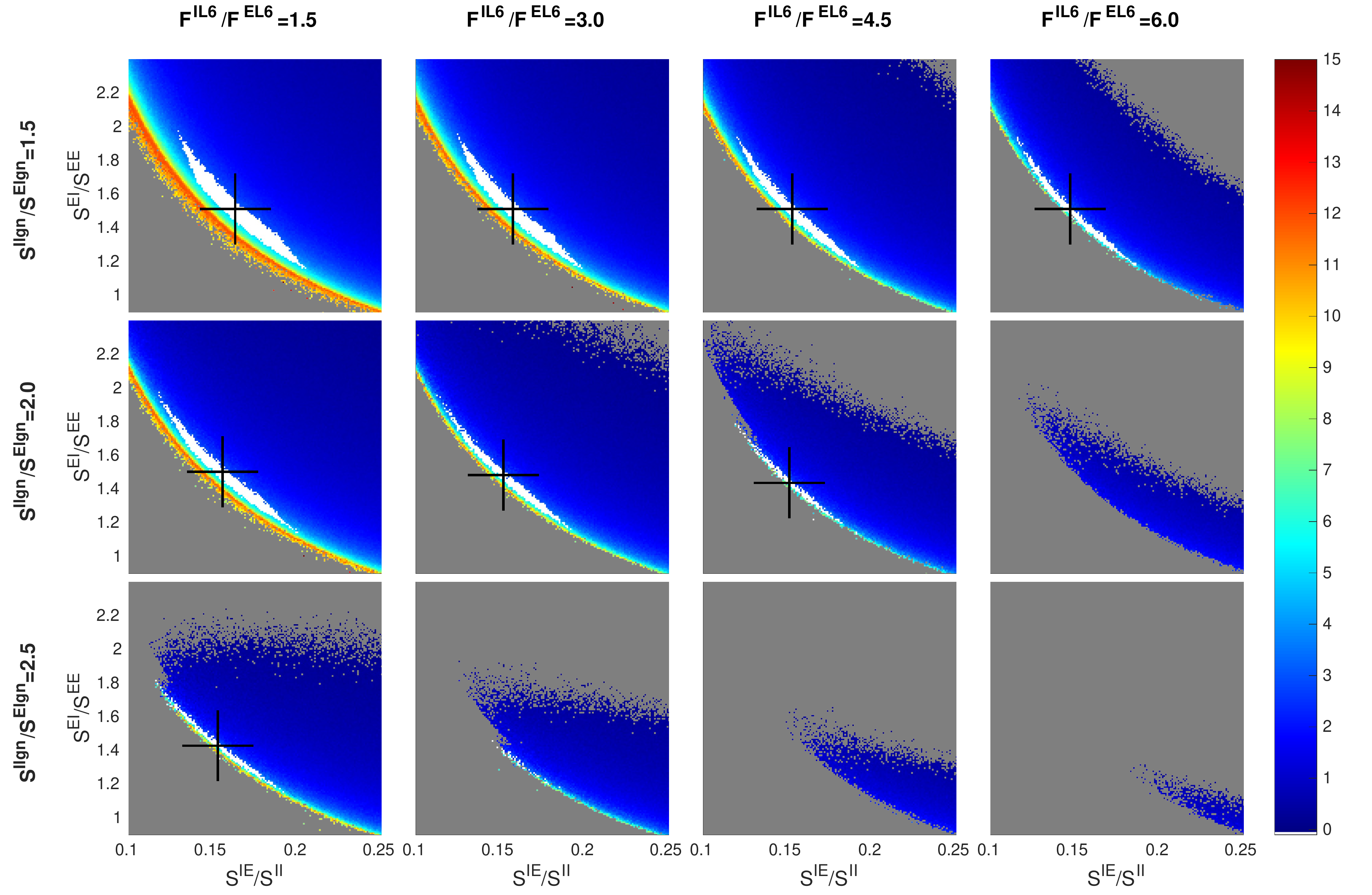}
    \caption{Dependence of firing rate and good area on LGN {\em versus}
      L6.  A 3$\times$4 matrix of panels is shown: each row corresponds
      to a fixed value of ${S^{I{\rm lgn}}}/{S^{E{\rm lgn}}}$
      and each column a fixed value of ${F^{I{\rm L6}}}/{F^{E{\rm L6}}}$. Smaller panels depict $f_E$ maps on
      inhibition planes and are as in Fig~\ref{Fig2: LGN}. The panel for
      ${F^{E{\rm L6}}}/{F^{I{\rm L6}}} =3$ and
      ${S^{I{\rm lgn}}}/{S^{E{\rm lgn}}}=2$ (row 2, column 2)
      corresponds to the $f_E$ rate map in Fig.~\ref{Fig1: Canonical
        Set}.  Other free parameters are $S^{EE} = 0.024$, $S^{II} =
      0.12$, and $S^{E{\rm lgn}} = 2\times S^{EE}$. }
    \label{Fig3: LGN-vs-L6}
  \end{center}
\end{figure}

Fig.~\ref{Fig3: LGN-vs-L6} shows, not surprisingly, that increasing LGN
and L6 inputs to I have very similar effects: As with
$S^{I\rm{lgn}}/S^{E\rm{lgn}}$, larger $F^{I{\rm L6}}/F^{E{\rm L6}}$
narrows the strip corresponding to the good area and shifts it
leftwards, that is, going from left to right in the matrix of panels has
a similar effect as going from top to bottom. Interpolating, one sees,
e.g., that the picture at $(S^{I{\rm lgn}}/S^{E{\rm lgn}}, F^{I{\rm
    L6}}/F^{E{\rm L6}})= (1.5, 4)$ is remarkably similar to that at $(2,
1.5)$. In background, changing the relative strengths of LGN to I vs to
E has a larger effect than the corresponding changes in L6, because LGN
input is a larger component of the Excitatory input than L6.  This
relation may not hold under drive, however, where L6 response is known
to increase significantly.

%%%%%%%%%%%%%%%%%%%%%%%%%%%%%%%%%%%%%%%%%%%%%%%%%%%%%%%%%%% Fig3
%%%%%%%%%%%%%%%%%%%%%%%%%%%%%%%%%%%%%%%%%%%%%%%%%%%%%%%%%%% Fig3

%%%%%%%%%%%%%%%%%%%%%%%%%%%%%%%%
\subsection{Scaling with $S^{EE}$ and $S^{II}$}
\label{sect:2.3}

\begin{figure}%[htbp]
  \begin{center}
    %% \captionsetup{type=figure} 
    \includegraphics[width=\textwidth]{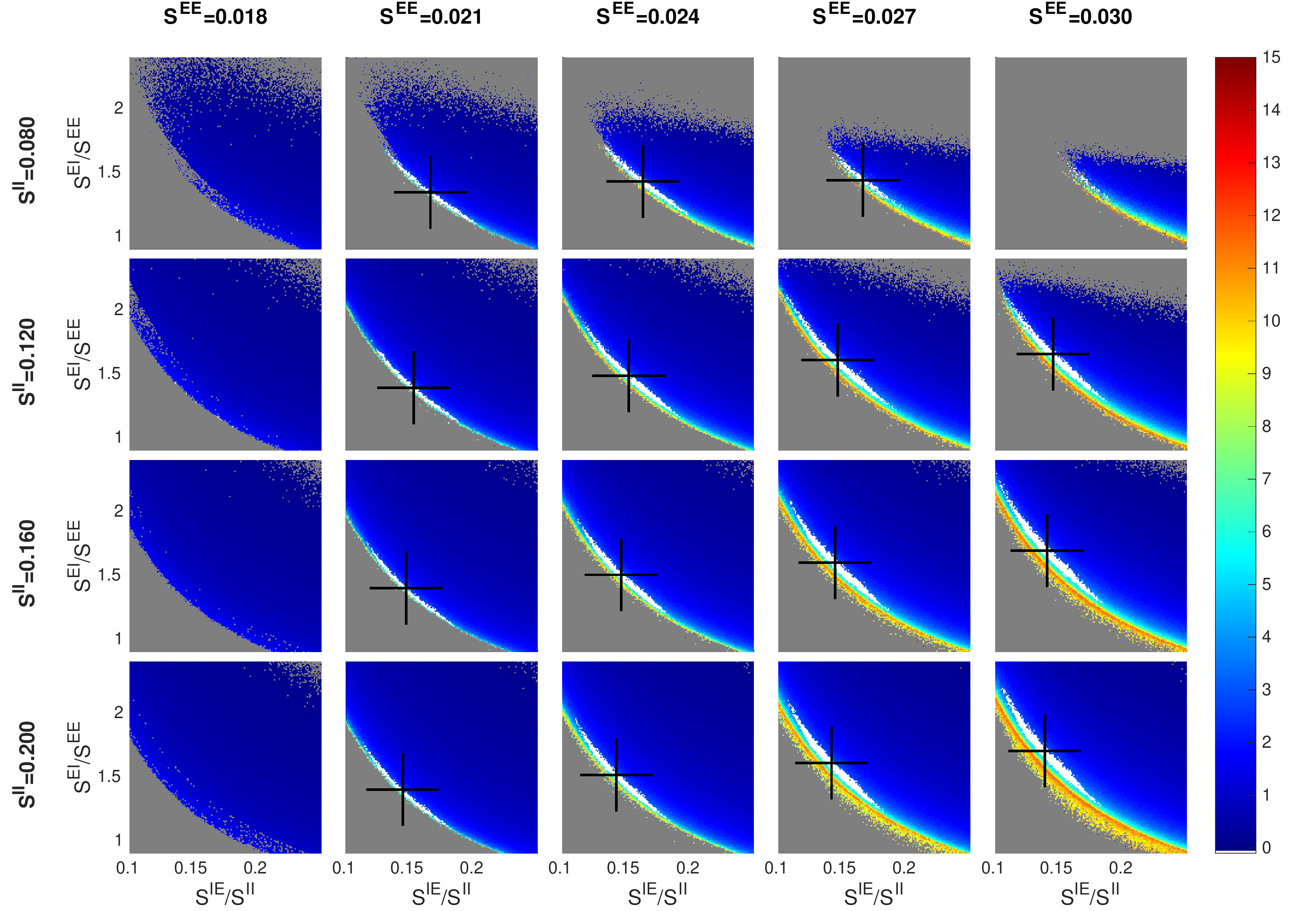}
    \caption{Dependence of firing rate and good area on $S^{EE}$ and
      $S^{II}$.  Smaller panels are as in Figures 2 and 3, with good
      areas (where visible) in white and black crosses denoting their
      centers of mass.  The picture for $S^{EE} =0.024$ and $S^{II}
      =0.120$ (row 2, column 3) corresponds to the $f_E$ rate map in
      Fig.~\ref{Fig1: Canonical Set}.  Other free parameters are
      $S^{E{\rm lgn}} = 2\times S^{EE}$, $S^{I{\rm lgn}} = 2\times
      S^{E{\rm lgn}}$, and $F^{I{\rm L6}} = 3\times F^{E{\rm L6}}$.}
    \label{Fig4: SEE-SII}
  \end{center}
\end{figure}

We have found $S^{EE}$ and $S^{II}$ to be the most ``basic" of the
parameters, and it has been productive indexing other parameters to
them.  In Fig.~\ref{Fig4: SEE-SII}, we vary these two parameters in the
matrix rows and columns, and examine changes in the inhibition planes.

We assume $S^{EE} \in (0.015, 0.03)$.  This follows from the
conventional wisdom~\cite{stratford1996excitatory,
  chariker2016orientation} that when an E-cell is stimulated {\it in
  vitro}, it takes 10-50 consecutive spikes in relatively quick
succession to produce a spike.  Numerical simulations of a biologically
realistic V1 model suggested $S^{EE}$ values lie well within the range
above \cite{chariker2015emergent}.  As for $S^{II}$, there is virtually
no direct information other than some experimental evidence to the
effect that EPSPs for I-cells are roughly comparable in size to those
for E-cells; see \cite{levy2012spatial} and also {\bf SI}.  We arrived
at the range we use as follows: With $S^{II} \in (0.08, 0.2)$ and
${S^{IE}}/{S^{II}} \in (0.1, 0.25)$, we are effectively searching
through a range of $S^{IE} \in (0.008, 0.05)$. As this interval extends
quite a bit beyond the biological range for $S^{EE}$, we hope to have
cast a wide enough net given the roughly comparable EPSPs for E and
I-cells.

%%%%%%%%%%%%%%%%%%%%%%%%%%%%%%%%%%%%%%%%%%%%%%%%%%%%%%%%%%% Fig4
%%%%%%%%%%%%%%%%%%%%%%%%%%%%%%%%%%%%%%%%%%%%%%%%%%%%%%%%%%% Fig4

Fig.~\ref{Fig4: SEE-SII} shows a matrix of panels with $S^{EE}$ and
$S^{II}$ in these ranges and the three ratios $S^{E{\rm
    lgn}}/S^{EE}=2.5$, $S^{I{\rm lgn}}/S^{E{\rm lgn}}=2$ and $F^{I{\rm
    L6}}/F^{E{\rm L6}} = 3$. As before, each of the smaller panels shows
an inhibition plane.  Good areas with characteristics similar to those
seen in earlier figures varying from panel to panel are clearly visible.

A closer examination reveals that (i) going along each row of the matrix
(from left to right), the center of mass of the good area (black cross)
shifts upward as $S^{EE}$ increases, and (ii) going down each column,
the black cross shifts slightly to the left as $S^{II}$ increases,
two phenomena we now explain.  Again, it is important to remember that
firing rates are roughly constant on the good areas.  

To understand (i),
consider the currents that flow into an E-cell, decomposing according to
source as follows: Let [4E], [6E], [LGN] and [amb] denote the total
current into an E-cell from E-cells in L4 and L6, from LGN and ambient,
and let [I] denote the magnitude of the I-current. As $f_E$ is
determined by the difference (or gap) between the Excitatory and
Inhibitory currents, we have
$$
\mbox{Gap} \ = \ \{[4E] + [6E] + [\mbox{LGN}] - [I]\} + [\mbox{amb}]\ .
$$
It is an empirical fact that the quantity in curly brackets is strictly
positive (recall that ambient alone does not produce spikes).  An
increase of $x \%$ in $S^{EE}$ will cause not only [4E] to increase but
also [6E] and [LGN], both of which are indexed to $S^{EE}$, to increase
by the same percentage. If $[I]$ also increases by $x \%$, then the
quantity inside curly brackets will increase by $x \%$ resulting in a
larger current gap. To maintain E-firing rate hence current gap size,
[I] must increase by more than $x \%$. Since I-firing rate is unchanged,
this can only come about through an increase in the ratio
${S^{EI}}/{S^{EE}}$, hence the upward movement of the black crosses.

To understand (ii), we consider currents into an I-cell, and let
[$\cdots$] have the same meanings as before (except that they flow into
an I-cell). Writing
$$
\mbox{Gap} \ = \ \{[4E] + [6E] - [I]\} + [\mbox{LGN}] + [\mbox{amb}] \ ,
$$
we observe empirically that the quantity inside curly brackets is
slightly positive.  Now we increase $S^{II}$ by $x \%$ and ask how
$S^{IE}$ should vary to maintain the current gap.  Since [LGN] and [amb]
are unchanged, we argue as above that $S^{IE}$ must increase by $< x\%$
(note that 6E is also indexed to $S^{IE}$).  This means
${S^{IE}}/{S^{II}}$ has to decrease, proving (ii).

\bigskip
We have suggested that the inhibition plane picture we have seen many
times is {\it canonical}, or {\it universal}, in the sense that through
any point in the designated 7D parameter cube, if one takes a 2D slice
as proposed, pictures qualitatively similar to those in Figures 2--4
will appear.  To confirm this hypothesis, we have computed a number of
slices taken at different values of the free
parameters (see {\bf SI}), at
$$
S^{EE} \in \{0.021,~0.024,~0.027\}, \qquad S^{II} \in \{0.12,~0.16,~0.2\}, \qquad 
F^{I{\rm L6}}/F^{E{\rm L6}} \in \{3.0,~4.5\}\ .
$$
For all $18 = 3\times 3\times 2$ combinations of these parameters, we
reproduce the $3 \times 4$ panel matrix in Fig.~\ref{Fig2: LGN}, i.e.,
the 4D slices of $(S^{E{\rm lgn}}/S^{EE},S^{I{\rm
    lgn}}/S^{II})\times({S^{EI}}/{S^{EE}},{S^{IE}}/{S^{II}})$, the first pair
corresponding to rows and columns of the matrix, and the second to the
$xy$-axes of each inhibition plane plot.  All 18 plots confirm the
trends observed above. Interpolating between them, we see that the
contours and geometric shapes on inhibition planes are indeed
universal, and taken together they offer a systematic, interpretable
view of the 7D parameter space.

%%%%%%%%%%%%%%%%%%%%%%%%%%%%%%%%%%%%%%%%%%%%%%%%%%%%%%%%%%% Fig5
\begin{figure}%[htbp]
  \begin{center}
  \begin{subfigure}{.5\textwidth}
  {\bf A}\\
  \includegraphics*[bb=0in 0in 4.6in 4.6in,width=\textwidth]{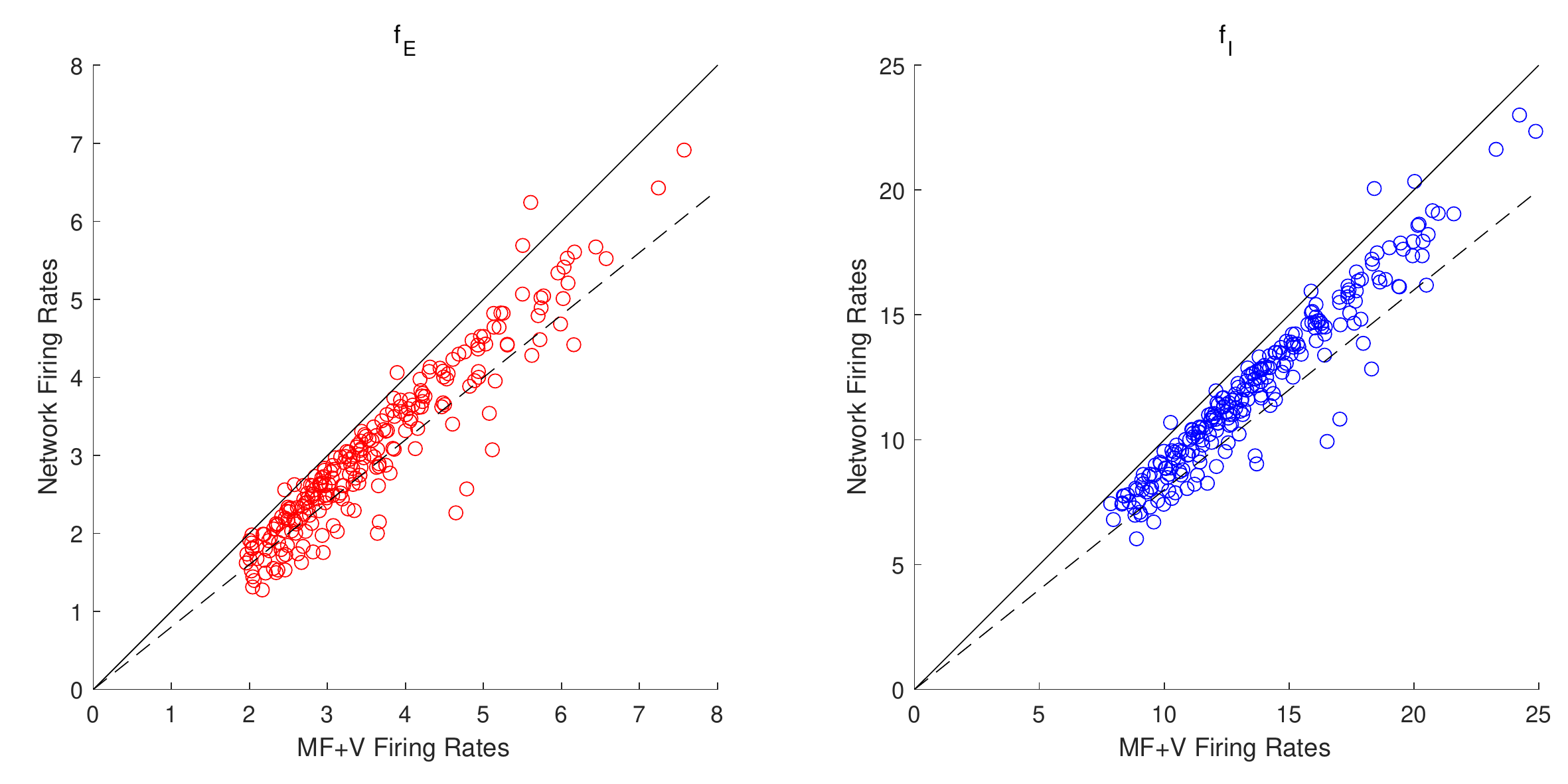}
  \end{subfigure}%
  \begin{subfigure}{.5\textwidth}
  {\bf B}\\
  \includegraphics*[bb=4.8in 0in 9.4in 4.6in,width=\textwidth]{{Figure5}.pdf}
  \end{subfigure}%
    \caption{Comparison of firing rates computed using the MF+v
      algorithm to those from direct network simulations.  The
      scatter plots show results for 128 sets of parameters randomly
      chosen in or near the good areas in Figs.~\ref{Fig1: Canonical
        Set}--\ref{Fig4: SEE-SII}.  \textbf{A.} Comparison of $f_E$.  \textbf{B.}
      Comparison of $f_I$.  Solid lines: $y = x$.  Dashed lines: $y =
      0.8x$.  A majority of data points fall in the range of 20\%
      accuracy.}
    \label{Fig5: SpotCheck}
  \end{center}
\end{figure}
%%%%%%%%%%%%%%%%%%%%%%%%%%%%%%%%%%%%%%%%%%%%%%%%%%%%%%%%%%% Fig5

\subsection{Comparing network simulations and the MF+v algorithm}
\label{sect:2.4}

Figs.~\ref{Fig1: Canonical Set}-\ref{Fig4: SEE-SII} were generated using
the MF+v algorithm introduced in Sect.~\ref{sect:1.3} and discussed in
detail in {\bf Methods}.  Indeed, the same analysis would not be
feasible using direct network simulations. But how accurately does the
MF+v algorithm reproduce network firing rates? To answer that question,
we randomly selected 128 sets of parameters in or near the good areas in
Figs.~\ref{Fig1: Canonical Set}-\ref{Fig4: SEE-SII}, and compared the
values of $f_E$ and $f_I$ computed from MF+v to results of direct
network simulations.  The results are presented in Fig.~\ref{Fig5:
  SpotCheck}.  They show that in almost all cases, MF+v overestimated
network firing rates by a little, with $<20\%$ error for $\sim 80\%$ of
the parameters tested.  In view of the natural variability of neuronal
parameters, both within a single individual under different conditions
and across a population, we view of this level of accuracy as sufficient
for all practical purposes.  Most of the larger errors are associated
with network E-firing rates that are lower than empirically observed (at
about 2 spikes/sec).

%%%%%%%%%%%%%%%%%%%%%%%%%%%%%%%%%%%%%%%%%%%%%%%%%%%%%%

\section{Other views of the viable region}

We have shown in Sect.~\ref{sect:2} that a systematic and efficient way
to explore parameter dependence is to slice the viable region using
inhibition planes with rescaled coordinate axes, but there are many
other ways to view the 6D manifold that approximates the viable region.
Here are some examples.

%%%%%%%%%%%%%%%%%%%%%%%%%%%%%%%%%%%%%%%%%%%%%%%%%%%%%%%%%%% Fig6
\begin{figure}%[htbp]
  \begin{center}
     \begin{subfigure}{.509\textwidth}
     {\bf A}\\
     \includegraphics*[bb=0.1in 0in 5.8in 4.7in,width=\textwidth]{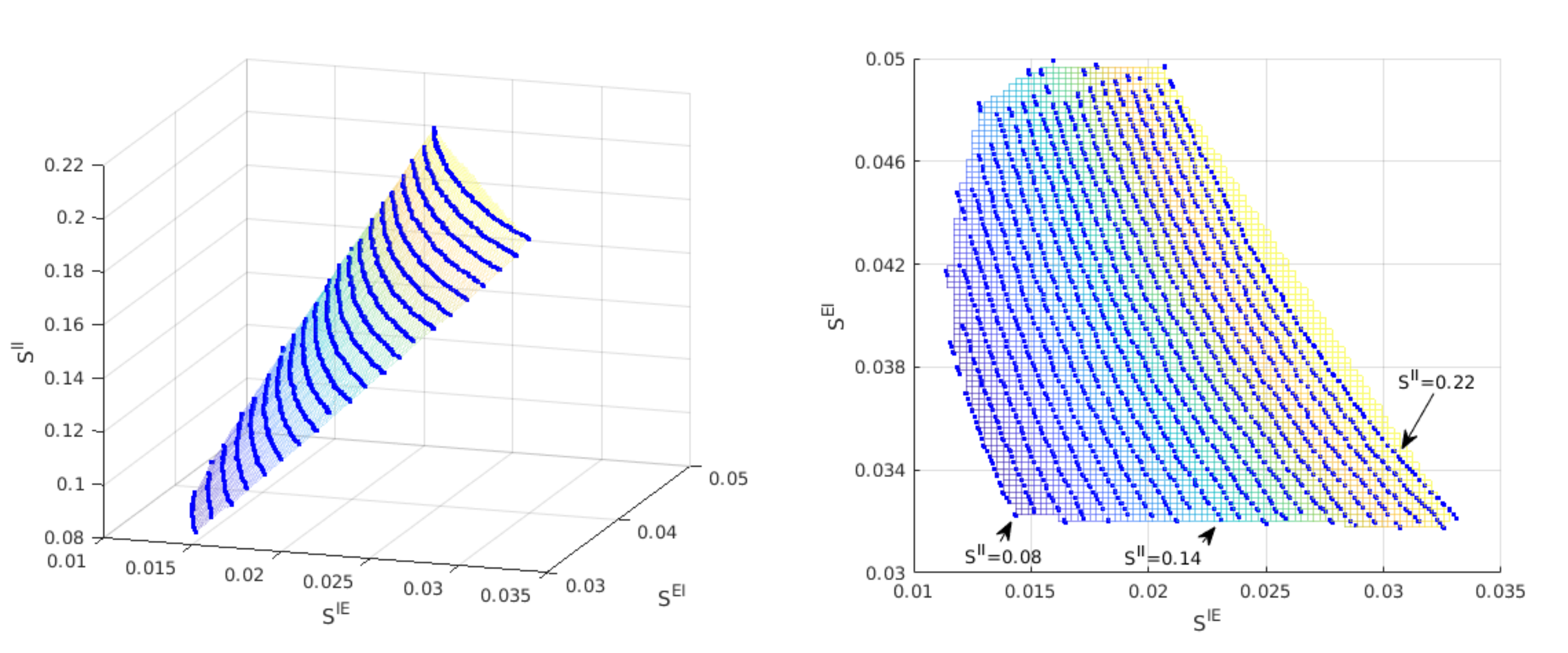}
     \end{subfigure}%
     \begin{subfigure}{.491\textwidth}
     {\bf B}\\
     \includegraphics*[bb=6.1in 0in 11.6in 4.7in,width=\textwidth]{{Figure6}.pdf}
     \end{subfigure}%

    \caption{Other views of the viable regions. \textbf{A.} A 2D surface 
    approximating the viable region projected to the 3D-space defined
    by $S^{IE} \times S^{EI} \times S^{II}$.
    Blue lines are projections of contours for $f_E= 4$ Hz 
    intersected with the good areas, computed on 21 different inhibition planes. Parameters are fixed at $S^{EE}$ =0.024, $S^{E \rm{lgn}}/S^{EE}$= 2, 
    $S^{I \rm{lgn}}/S^{E\rm{lgn}}$= 2, and $F^{I \rm{L6}} / F^{E \rm{L6}}$ = 2;
    $S^{II}$ varies from 0.22 for the top contour to 0.08 for the bottom contour. 
    \textbf{B.} A bird's eye view of the left panel, with colors indicating 
   corresponding locations of the projected E-contours.}
    \label{Fig6: SIE_SEI_SII}
  \end{center}
\end{figure}
%%%%%%%%%%%%%%%%%%%%%%%%%%%%%%%%%%%%%%%%%%%%%%%%%%%%%%%%%%% Fig6

Fig.~\ref{Fig6: SIE_SEI_SII} shows two views of the viable region
projected to two different low dimension subspaces.  The left panel
shows the viable region as a surface parametrized by hyperbolas with
varying aspect ratios.  This is how it looks in unscaled coordinates,
compared to the panels in, e.g., Fig.~\ref{Fig4: SEE-SII}, where we have
uniformized the aspect ratios of the hyperbolas by plotting against
${S^{IE}}/{S^{II}}$ instead of $S^{IE}$.  The right panel of
Fig.~\ref{Fig6: SIE_SEI_SII} shows a bird's-eye view of the same plot,
with the $\{f_E=4\}$-contours in the (unscaled) inhibition plane,
giving another view of the $S^{II}$ dependence.

\begin{figure}%[htbp]
  \begin{center}
    %% \captionsetup{type=figure} 
    \resizebox{5.2in}{!}{\includegraphics{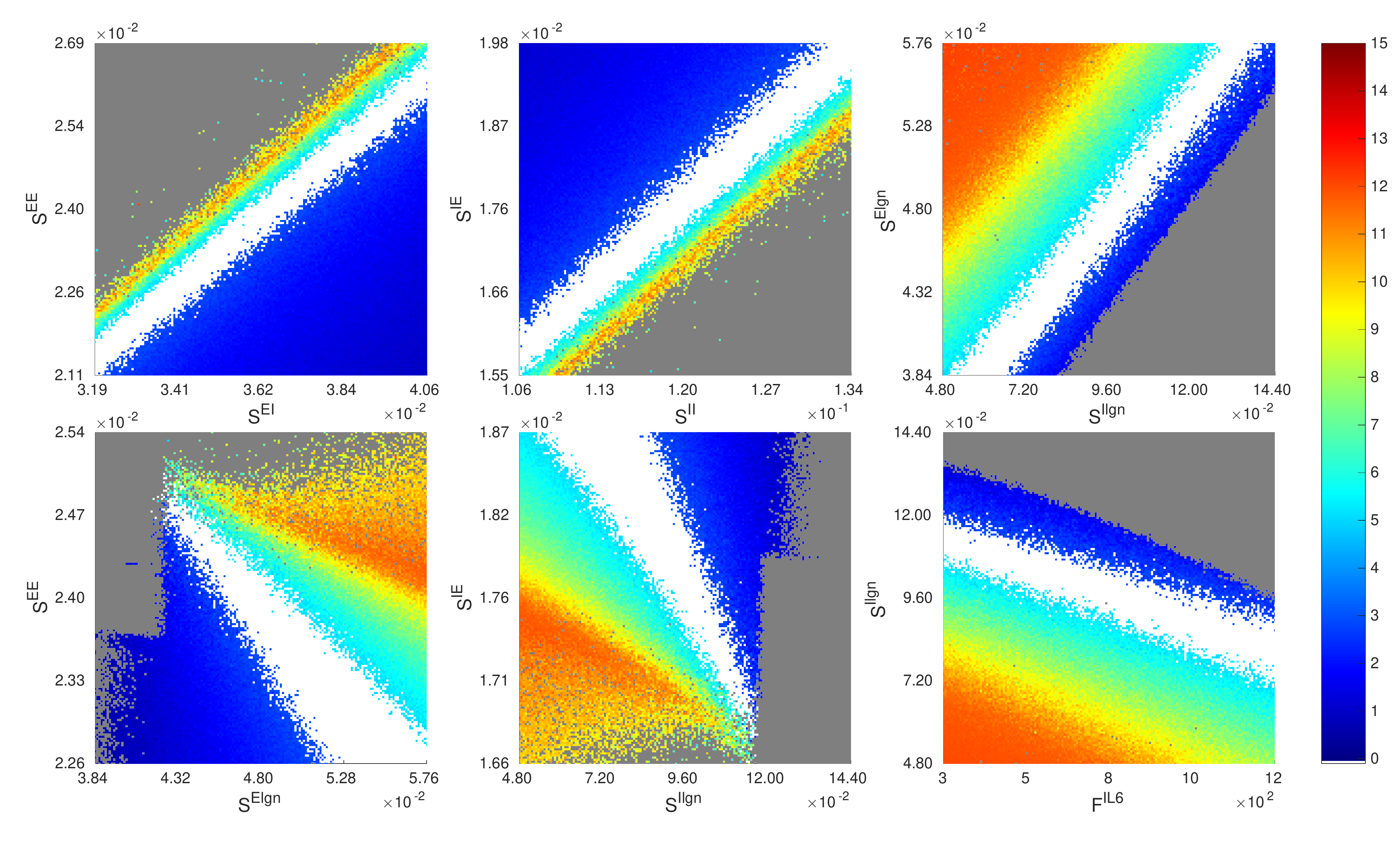}}
    \caption{Slicing the 7D parameter space from other directions
      through the reference point. We compute the rate maps of $f_E$ and
      indicate viable regions on six other 2D slices, namely, the planes
      of $S^{EE}\times S^{EI}$, $S^{IE}\times S^{II}$, $S^{E{\rm
          lgn}}\times S^{I{\rm lgn}}$, $S^{EE}\times S^{E{\rm lgn}}$,
      $S^{IE}\times S^{I{\rm lgn}}$, and $S^{I{\rm lgn}}\times F^{I{\rm
          L6}}$. Firing rates are as in the color bar, and as usual
          the good area is in white. }
    \label{Fig7: OtherSlices}
  \end{center}
\end{figure}

Table~\ref{Table1: Ref-point} gives a sense of how the viable region near
the reference parameter point looks when we cut through the 7D parameter
cube with 1D lines.  In Fig.~\ref{Fig7: OtherSlices}, we show several
heat maps for firing rates obtained by slicing the viable region with
various 2D planes though the same reference point.  In the top row, we
have chosen pairs of parameters that covary (positively), meaning to
stay in the viable region, these pairs of parameters need to be increased
or decreased simultaneously by roughly constant proportions.  The idea
behind these plots is that to maintain constant firing rates, increased
coupling strength from the E-population must be compensated by a
commensurate increase in coupling strength from the I-population (left
and middle panels), and increased drive to E must be compensated by a
commensurate increase in drive to I (right panel).  In the second row,
we have selected pairs of parameters that {\it covary negatively}, i.e.,
their {\it sums} need to be conserved to stay in the viable region.  The
rationale here is that to maintain constant firing rates, total
excitation from cortex and from LGN should be conserved (left and middle
panels), as should total drive from L6 and from LGN (right panel).

Thus, together with the results in Sect.~\ref{sect:2}, we have seen
three different ways in which pairs of parameters can relate: (i) they
can covary, or (ii) their sums can be conserved, or, (iii) as in the
case of inhibition planes, it is the {\it product} of the two parameters
that needs to be conserved. Like (iii), which we have shown to hold
ubiquitously and not just through this one parameter point, the
relations in (i) and (ii) are also valid quite generally.

%%%%%%%%%%%%%%%%%%%%%%%%%%%%%%%%%%%%%%%%%%%%%%%%%%%%%%%%%%% Fig7
%%%%%%%%%%%%%%%%%%%%%%%%%%%%%%%%%%%%%%%%%%%%%%%%%%%%%%%%%%% Fig7

\section*{\large Discussion}

We began with some very rough {\it a priori} bounds for the 7 free
parameters identified in Sect.~\ref{sect:1.2}, basing our choices on
physiological data when available and casting a wide net when they are
not.  We also identified a biologically plausible region (referred to as
``the viable region") defined to be the set of parameters that lead to
spontaneous E- and I-firing rates compatible with experimental values,
and sought to understand the geometry of this region of parameter space.
Our most basic finding is that the viable region as defined is a
slightly thickened 6D submanifold -- the amount of thickening varies
from location to location, and is so thin in places that for all
practical purposes the submanifold vanishes.  This is consistent with
Table~\ref{Table1: Ref-point}, which shows that varying certain
parameters by as little as $1\%$ can take us out of the viable region.
One can think of directions that show greater sensitivity in
Table~\ref{Table1: Ref-point} as more ``perpendicular" to the slightly
thickened 6D surface, while those that are more robust make a smaller
angle with its tangent plane.  The codimension-1 property is largely a
consequence of the E-I balance and has a number of biological
implications, the most important of which is that the parameters giving
rise to biologically plausible regimes are robust --- provided one {\it
  compensates} appropriately when varying parameters.  Such compensation
can come about from a variety of sources {\em in vivo}, e.g., synaptic
depression of I-neurons \cite{galarreta1999network,gibson1999two};
increased thresholds for potential generation of E-spikes due to $K_v$
currents \cite{yuan2005functional}; and a host of other homeostatic
mechanisms \cite{turrigiano2012homeostatic}.  To a first approximation,
one can view these mechanisms as regulating synaptic weights, and our
findings may be pertinent to anyone wishing to study homeostatic
mechanisms governing neuronal activity.

Our analysis offers a great deal more information on the structure of
the viable region beyond its being a thickened 6D manifold.  We have
found it profitable to slice the 7D parameter cube with {\it inhibition
  planes}, 2D planes containing the parameter axes $S^{EI}$ and
$S^{IE}$.  Each inhibition plane intersects the viable region in a
narrow strip surrounding a segment of a hyperbola (noted as ``good area"
above).  Moreover, in rescaled variables ${S^{EI}}/{S^{EE}}$ and
${S^{IE}}/{S^{II}}$, these hyperbolas are not only remarkably alike in
appearance but their exact coordinate locations and aspect ratios vary
little as we move from inhibition plane to inhibition plane, suggesting
{\it approximate scaling relations} for firing rates as functions of
parameters.

Summarizing, we found that most points in the viable region have $S^{EE}
\in [0.02, 0.03]$ and $S^{II} \in [0.1, 0.2]$; the lower limits of these
ranges were found in our parameter analysis and the upper limits were
{\it a priori} bounds.  Our parameter exploration also shows that
${S^{EI}}/{S^{EE}} \in (1,2)$, and ${S^{IE}}/{S^{II}} \in (0.1, 0.15)$,
$S^{E \rm{lgn}}/S^{EE} \in (1.5, 3), S^{I \rm{lgn}}/S^{E \rm{lgn}} \in
(1.5, 2)$, and $F^{I \rm{L6}}/F^{I \rm{L6}} \in (3,4.5)$.  We have
further observed a strong correlation between degeneration of good areas
and external inputs to I being too large in relation to that of E.  For
example, when $S^{E{\rm lgn}}/S^{EE}$ is too low, or when $S^{I{\rm
    lgn}}/S^{E{\rm lgn}}$ or $F^{I{\rm L6}}/F^{E{\rm L6}}$ is too high,
the hyperbolic strips in inhibition planes narrow, possibly vanishing
altogether.  (See also Sect.~\ref{sect:MF solutions}.)  We have offered
explanations in terms of a {\em suppression index} for
E-cells.

%%  \ztext{[insert: Bio implications of the manifold being
%% codimensional-1...]}

\medskip \noindent {\bf \small Relation to previous work on MF}

\smallskip
Since the pioneering work of Wilson, Cowan, Amari, and others
\cite{wilson1972excitatory, wilson1973mathematical, amari1977dynamics,
  ermentrout1979mathematical, hopfield1982neural, cohen1983absolute,
  hopfield1984neurons, treves1993mean, ermentrout1998neural,
  brunel1999fast, bressloff2001geometric, gerstner2000population,
  coombes2005waves, mattia2002population, el2009master,
  faugeras2009constructive, bressloff2010stochastic}, MF ideas have been
used to justify the use of firing rate-based models to model networks of
spiking neurons.  The basic idea underlying MF is to start with a
relation between averaged quantities, e.g., an equation similar or
analogous to Eq.~(\ref{MF}), and supplement it with an ``activation'' or
``gain'' function relating incoming spike rates and the firing rate of
the postsynaptic neuron, thus arriving at a closed governing equation
for firing rates.  MF and related ideas have yielded valuable
mathematical insights into a wide range of phenomena and mechanisms,
including pattern formation in slices \cite{ermentrout1979mathematical,
  ermentrout1998neural}, synaptic plasticity and memory formation
\cite{lim2015inferring, brunel2001effects, renart2004mean,
  barbieri2007irregular, vignoud2018interplay}, stability of attractor
networks \cite{knierim1995model, samsonovich1997path,
  laing2001stationary, guo2005existence, moreno2007noise,
  barbieri2008can, kropff2008emergence}, and many other features of
network dynamics involved in neuronal computation \cite{ben1995theory,
  stetter2000mean, zerlaut2018modeling, johnson2009dynamic,
  curto2009simple, van2009mean1, van2009mean2, markounikau2010dynamic,
  steyn2012gap, bressloff2014neural, carroll2016phase,
  levenstein2018excitable, bressloff2001geometric, brunel1999fast,
  coombes2003dynamics, coombes2005waves, kilpatrick2010effects,
  coombes2012interface, baladron2012mean, touboul2014propagation,
  el2009master, bressloff2010stochastic, nykamp2000population,
  cai2004effective, cai2006kinetic, shao2020dimensional,
  zhang2019coarse, chow2015path, ocker2017linking, bahri2020statistical,
  bressloff2015path, ocker2017statistics}.  However, as far as we are
aware, MF has not been used to systematically map out cortical parameter
landscape.

Another distinction between our approach and most previous MF models has
to do with intended use.  In most MF models, the form of the gain
function is {\em assumed}, usually given by a simple analytical
expression; see, e.g., \cite{ermentrout2010mathematical}.  In settings
where the goal is a general theoretical understanding and the relevant
dynamical features are insensitive to the details of the gain function,
MF theory enables mathematical analysis and can be quite informative.
Our goals are different: our MF models are computationally efficient
surrogates for realistic biological network models, models that are
typically highly complex, incorporating the anatomy and physiology of
the biological system in question.  For such purposes, it is essential
that our MF equations capture quantitative details of the corresponding
network model with sufficient accuracy.  In particular, we are not free
to design gain functions; they are dictated by the connectivity
statistics, types of afferents and overall structure of the network
model.  We have termed our approach ``data-informed MF" to stress these
differences with the usual MF theories.

We have tried to minimize the imposition of additional hypotheses beyond
the basic MF assumption of a closed model in terms of rates.  As
summarized in Results and discussed in depth in Methods and SI, we
sought to build an MF equation assuming only that the dynamics of
individual neurons are governed by leaky integrate-and-fire (LIF)
equations with inputs from lateral and external sources, and when
information on mean voltage was needed to close the MF equation, we
secured that from synthetic data using a pair of LIF neurons driven by
the same inputs as network neurons.  The resulting algorithm, which we
have called the ``MF+v" algorithm, is to our knowledge novel and is
faithful to the idea of data-informed MF modeling.

As we have shown, our simple and flexible approach produces accurate
firing rate estimates, capturing cortical parameter landscape at a
fraction of the cost of realistic network simulations.  It is also
apparent that its scope goes beyond background activity, and can be
readily generalized to other settings, e.g., to study evoked responses.

\medskip \noindent 
{\bf \small Contribution to a model of primate visual cortex}

\smallskip

Our starting point was \cite{chariker2016orientation}, which contains a
mechanistic model of the input layer to the monkey V1 cortex.  This
model was an ideal proving ground for our data-informed MF ideas: it is
a large-scale network model of interacting neurons built to conform to
anatomy.  For this network, the authors of
\cite{chariker2016orientation} located a small patch of parameters with
which they reproduced many visual responses both spontaneous and evoked.
Their aim was to show that such parameters existed; parameters away from
this patch were not considered --- and this is where they left off and
where we began: Our MF+v algorithm, coupled with techniques for
conceptualizing parameter space, made it possible to fully examine a
large 7D parameter cube.  In this paper, we identified {\it all} the
parameters in this cube for which spontaneous firing rates lie within
certain acceptable ranges.  The region we found includes the parameters
in \cite{chariker2016orientation} and is many times larger; it is a
slightly thickened 6D manifold that is nontrivial in size.  Which subset
of this 6D manifold will produce acceptable behavior when the model is
stimulated remains to be determined, but since all viable parameters --
viable in the sense of both background and evoked responses -- must lie
in this set, knowledge of its coordinates should provide a head-start in
future modeling work.

\smallskip
\medskip \noindent 
{\bf \small Taking stock and moving forward}

\smallskip

In a study such as the one conducted here, had we not used basic
biological insight and other simplifications (such as inhibition planes,
rescaled parameters, viable regions, and good areas) to focus the
exploration of the 7D parameter space, the number of parameter points to
be explored would have been $N^7$, where $N$ is the number of grid
points per parameter, and the observations in Table~\ref{Table1:
  Ref-point} suggest that $N=\mathcal O(100)$ may be the order of
magnitude needed.  Obtaining this many data points from numerical
simulation of the entire network would have been out of the question.
Even after pruning out large subsets of the 7D parameter cube and
leveraging the insights and scaling relations as we have done, producing
the figures in this paper involved computing firing rates for $\sim10^7$
distinct parameters.  That would still have required significant effort
and resources to implement and execute using direct network simulations.
In contrast, using the proposed MF+v algorithm, each example shown in
this paper can be implemented with moderate programming effort and
computed in a matter of hours on a modern computing cluster.

We have focused on background or spontaneous activity because its
spatially and temporally homogeneous dynamics provide a natural testing
ground for the MF+v algorithm.  Having tested the capabilities of MF+v,
our next challenge is to proceed to evoked activity, where visual
stimulation typically produces firing rates with inhomogeneous spatial
patterns across the cortical sheet.  The methods developed in this paper
continue to be relevant in such studies: evoked activity is often
locally constant in space (as well as in time), so our methods apply to
local populations, the dynamics of which form building blocks of
cortical responses to stimuli with different spatiotemporal structure.

%% As evoked activity is {\em locally} constant --- in space and in time
%% --- under a wide range of conditions, one possible approach is to use
%% MF+v to model local populations, then piece them together to model
%% cortical response to stimuli with different spatiotemporal structure.

Finally, we emphasize that while MF+v provides a tremendous reduction to
the cost of estimating firing rates given biological parameters, the
computational cost of a parameter grid search remains exponential in the
number of parameters (``curse of dimensionality'').  Nevertheless, we
expect our MF+v-based strategy, in combination with more efficient
representations of data in high dimensional spaces (e.g., sparse
grids~\cite{bungartz2004sparse}) and the leveraging of biological
insight, can scale to systems with many more degrees of complexity.

%% , can scale to systems with many times % more parameters.

%% \newpage
%%%%%%%%%%%%%%%%%%%%%%%%%%%% Methods %%%%%%%%%%%%%%%%%%%%%%%%%%%%%%%%%%%%%%

\section*{\large Methods}
\label{Mthd}
\renewcommand{\thesubsection}{M\arabic{subsection}}
\setcounter{subsection}{0}

As explained in {\bf Introduction} and {\bf Results}, we seek
parsimonious phenomenological models that are (i) simple and efficient;
(ii) flexible enough to accommodate key biological features and
constraints; and (iii) able to faithfully capture mean firing rates and
voltages of network models across a wide range of parameters.  We use
for illustration a model of the monkey primary visual cortex, treating
as ``ground truth" the network model described in Sect.~\ref{sect:1.1}
of {\bf Results}.  Here we elaborate on the MF+v scheme outlined in
Sect.~\ref{sect:1.3}, applying it to study firing rates in Layer
4C$\alpha$ (L4), an input layer to V1 in the network model.

\subsection{Mean-field rate-voltage relation}
\label{Mthd-Deri}

We begin by stating precisely and deriving the MF equations~(\ref{MF})
alluded to in Sect.~\ref{sect:1.3} of Results.  Consider an LIF model
(Eq.~(\ref{LIF})) for neuron $i$ in L4 of the network model.  We set
$V_{\rm rest} = 0$, $V_{\rm th} = 1$, and let $t_1,t_2,...t_n$ be the
spiking times of neuron $i$ on the time interval $[0,T]$ for some large
$T$.  Integrating Eq.~(\ref{LIF}) in time, we obtain
\begin{align}
  \label{Mthd:LIF-Intgl}
  N_i(T) = g^L \int_{\mathbb T} - v_i(t)~\D t + \int_{\mathbb T} g_i^E(t)(V^E-v_i(t))~\D t + \int_{\mathbb T} g_i^I(t)(V^I-v_i(t))~\D t\ ,
\end{align}
where $\mathbb{T} = [0,T]\setminus \mathcal{R}$, i.e., the time interval
$[0,T]$ minus the union $\mathcal{R}= \cup_{j=1}^n [t_j,t_j+\tau_{\rm
    ref}]$ of all refractory periods.  Let $f_i = \lim_{T\to\infty}
N_i(T)/T$ denote the mean firing rate of the $i$th neuron.  We then have
\begin{align}
    \label{Mthd:LIF-Corr}
    f_{i} \approx (1-\underbrace{f_i\cdot\tau_{\rm
        ref}}_{(\star)})\times\left[-g_R \ol{v}_i +
      \ol{g}_i^E(V^E-\ol{v}_i) + \ol{g}_i^I(V^I-\ol{v}_i)\right],
\end{align}
where $\ol{x} = \lim_{T\to\infty}\frac1{|{\mathbb T}|}\int_{\mathbb
  T}x(t)~dt$ and $|{\mathbb T}|$ is the total length of ${\mathbb T}$.
The term ($\star$) is the fraction of time the $i$th neuron is in
refractory, and $\ol{x}$ is the conditional expectation of the quantity
$x(t)$ given the cell is \emph{not} refractory at time $t$.  We have
neglected correlations between conductances and voltages, as is
typically done in mean-field (MF) theories.  See, e.g.,
\cite{treves1993mean}.

The long-time averages $\ol{g}_i^E$ and $\ol{g}_i^I$ reflect the numbers
and sizes of E/I-kicks received by neuron $i$.  In our network model
(see {\bf SI} for details), the only source of inhibition comes from
I-cells in L4, while excitatory inputs come from LGN, layer 6 (L6),
ambient inputs (amb), and recurrent excitation from E-cells in L4.  Mean
conductances can thus be decomposed into:
\begin{subequations}
\label{Mthd:Condc}
\begin{align}
  \ol{g}_i^E &= S^{i,\rm lgn}\times F^{i,\rm lgn} ~~+~~ S^{i,\rm
    L6}\times F^{i,\rm L6} ~~+~~ S^{i,\rm amb}\times F^{i,\rm amb} ~~+~~
  S^{i,E}\times F^{i,E}, \\
  \ol{g}_i^I &= S^{i,I}\times F^{i,I}\ ,
\end{align}
\end{subequations}
where for $P\in\{{\rm lgn}$, L6, amb, E, I$\}$, $S^{i,P}$ is the
synaptic coupling weight from cells in $P$ to neuron $i$, and $F^{i,P}$
is the total number of spikes per second neuron $i$ receives from source
$P$, i.e., from all of its presynaptic cells in $P$ combined.  Here and
in the rest of {\bf Methods}, ``E'' and ``I" refer to L4, the primary
focus of the present study, so that $F^{i,E}$, for example, is the total
number of spikes neuron $i$ receives from E-cells from within L4.

As discussed in the main text, we are interested in background or
spontaneous activity. During spontaneous activity, we may assume under
the MF limit that all E-cells in L4 receive statistically identical
inputs, i.e., for each $P$, $(S^{i,P}, F^{i,P})$ is identical for all
E-cells $i$ in L4. We denote their common values by $(S^{EP}, F^{EP})$,
and call the common firing rate of all E-cells $f_E$.  Corresponding
quantities for I-cells are denoted $(S^{IP}, F^{IP})$ and $f_I$.
Combining Eqs.~\ref{Mthd:LIF-Corr} and \ref{Mthd:Condc}, we obtain the MF
equations for E/I-cells:
\begin{subequations}
\label{Mthd:MF}
\begin{align}
  \nonumber
  f_{E} =  & \left\{\left[S^{E\rm lgn}F^{E\rm lgn} + S^{E\rm L6} F^{E\rm
      L6} + S^{E\rm amb} F^{E\rm amb} + S^{EE} N^{EE} f_E\cdot(1-p_{\rm
      fail})\right]\cdot(V^E-\ol{v}_E) \right. \\
  & +\left. \left(S^{EI}N^{EI} f_{I}\right)\cdot(V^I-\ol{v}_E)
  -g_{E}^L\cdot \ol{v}_E \right\} \times (1-f_E\cdot\tau_{\rm ref}), \\
    \nonumber
    f_{I} = & \left\{\left[S^{I\rm lgn}F^{I\rm lgn} + S^{I\rm L6}
      F^{I\rm L6} + S^{I\rm amb} F^{I\rm amb} + S^{IE} N^{IE}
      f_E\right]\cdot(V^E-\ol{v}_I) \right.\\
    & + \left. \left(S^{II}N^{II} f_{I}\right)\cdot(V^I-\ol{v}_I)
    -g_{I}^L\cdot \ol{v}_I \right\} \times (1-f_I\cdot\tau_{\rm ref})\ .
\end{align}
\end{subequations}
In Eq.~(\ref{Mthd:MF}), $g_{E}^L$ and $g_{I}^L$ are leakage conductances;
$N^{Q'Q}$ represents the average number of type-$Q$ neurons in L4
presynaptic to a type-$Q'$ neuron in L4.  These four numbers follow
estimations of neuron density and connection probability of Layer
4C$\alpha$ of the monkey primary visual cortex. Refractory periods are
$\tau_{\rm ref}$, and E-to-E synapses are assumed to have a synaptic
failure rate $p_{\rm fail}$, also fixed. Details are discussed in {\bf
  SI}.

%%%%%%%%%%%%%%%%%%%%%%%%%%%%%%%%%%%%%%%%%%%%%%%%%%%%%%%%%%% Fig5
\begin{figure}%[htbp]
  %% \captionsetup{type=figure} 
  {\bf A}\\
  \includegraphics*[bb=0in 1.8in 8.7in
    5.5in,width=\textwidth]{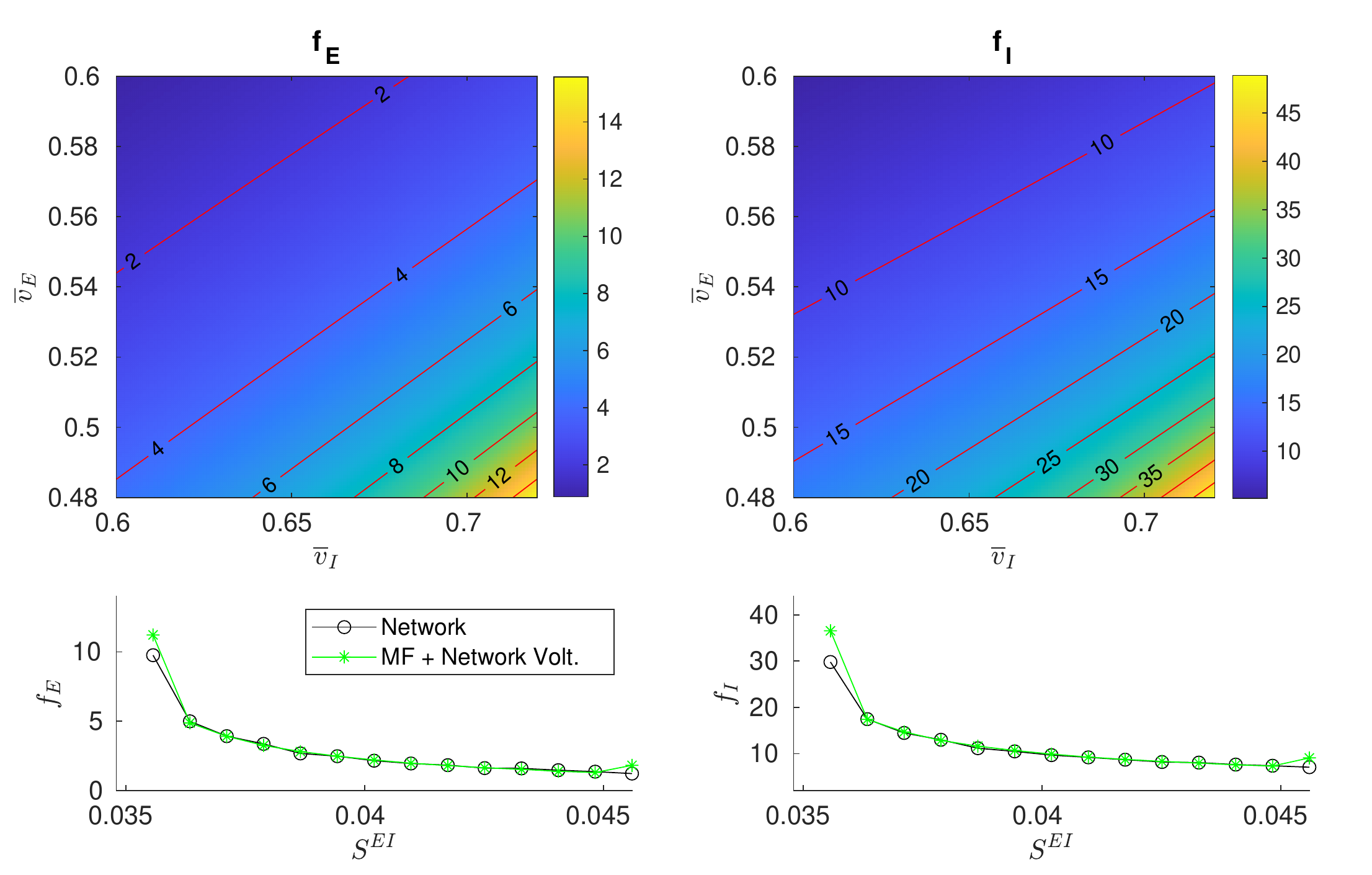}\\[1ex]
  {\bf B}\\
  \includegraphics*[bb=0in 0in 8.7in
    1.8in,width=\textwidth]{{Figure8}.pdf}\\
  \caption{MF approximations.  {\bf A.} Contours of E/I firing rates as
    functions of $(\ol{v}_E,\ol{v}_I)$.  {\bf B.} Comparison of network
    firing rates (black) and $(f_E, f_I)$ computed from
    Eq.~(\ref{Mthd:MF}) using network-computed mean voltages (green).}
  \label{FigM}
\end{figure}

%%%%%%%%%%%%%%%%%%%%%%%%%%%%%%%%%%%%%%%%%%%%%%%%%%%%%%%%%%% Fig5

We seek to solve Eq.~(\ref{Mthd:MF}) for $(f_E, f_I)$ given network
connections, synaptic coupling weights and external inputs.  That is, we
assume all the quantities that appear in Eq.~(\ref{Mthd:MF}) are fixed,
except for $f_E, f_I, \ol{v}_E$ and $\ol{v}_I$. The latter two, the mean
voltages $\ol{v}_E$ and $\ol{v}_I$, cannot be prescribed freely as they
describe the sub-threshold activity of L4 neurons once the other
parameters are specified.  Thus what we have is a system that is not
closed: there are four unknowns, and only two equations.

A second observation is that once $\ol{v}_E$ and $\ol{v}_I$ are
determined, Eq.~(\ref{Mthd:MF}) has a very simple structure.  To
highlight this near-linear structure, we rewrite Eq.~(\ref{Mthd:MF}) in
matrix form as
\begin{equation}
  \label{Mthd:MFMat}
  \B{f} = \BR(\B{f}) \times \left[\BM(\B{v})\cdot\B{f} + \B{s}(\B{v})\right].
\end{equation}
Here $\B{f}=(f_E, f_I)$, $\B{v}= (\ol{v}_E, \ol{v}_I)$, $\BM(\B{v})$ is
a (voltage-dependent) linear operator acting on L4 E/I firing rates,
$\B{s}$ includes inputs from external sources and leakage currents, and
$\BR$ accounts for refractory periods.  (See Sect.~\ref{sect:MF
  solutions}.))  Neglecting refractory periods, Eq.~(\ref{Mthd:MFMat})
is linear in $\B{f}$ assuming $\BM(\B{v})$ and $\B{s}(\B{v})$ are known.
At typical cortical firing rates in background, the refractory factor
$\BR$ contributes a small nonlinear correction.

To understand the dependence on $\B{v}$, we show in Fig.~\ref{FigM}A the
level curves of $f_E$ and $f_I$ as functions of $(\ol{v}_E,\ol{v}_I)$
from the mapping defined by Eq.~(\ref{Mthd:MF}).  As expected,
$(f_E,f_I)$ vary with $(\ol{v}_E,\ol{v}_I)$, the nearly straight
contours reflecting the near-linear structure of
Eq.~(\ref{Mthd:MFMat}). One sees both $f_E$ and $f_I$ increase steadily
(in a nonlinear fashion) as we decrease $\ol{v}_E$ and increase
$\ol{v}_I$, the dependence on $(\ol{v}_E,\ol{v}_I)$ being quite
sensitive in the lower right part of the panels. The sensitive
dependence of $f_E$ and $f_I$ on $(\ol{v}_E,\ol{v}_I)$ rules out
arbitrary choices on the latter in a MF theory that aims to reproduce
network dynamics. How to obtain reasonable information on
$(\ol{v}_E,\ol{v}_I)$ is the main issue we need to overcome.

We propose in this paper to augment Eq.~(\ref{Mthd:MF}) with values of
$(\ol{v}_E,\ol{v}_I)$ informed by (synthetic) data.  To gauge the
viability of this idea, we first perform the most basic of all tests: We
collect firing rates and mean voltages (averaged over time and over
neurons) computed directly from network simulations, and compare the
firing rates to $f_E$ and $f_I$ computed from Eq.~(\ref{Mthd:MF}) with
$(\ol{v}_E,\ol{v}_I)$ set to network-computed mean voltages. The results
for a range of synaptic coupling constants are shown in
Fig.~\ref{FigM}B, and the agreement is excellent except when firing
rates are very low or very high.

\subsection{The MF+v algorithm}
\label{Mthd-MV}

Simulating the entire network to obtain $(\ol{v}_E,\ol{v}_I)$ defeats
the purpose of MF approaches, but the results in Fig.~\ref{FigM}B
suggest that we might try using single LIF neurons to represent typical
network neurons, and use them to estimate mean voltages.

The idea is as follows: Consider a pair of LIF neurons, one E and one I,
and fix a set of parameters and external drives.  In order for this pair
to produce $(\ol{v}_E,\ol{v}_I)$ similar to the mean voltages in network
simulations, we must provide these cells with surrogate inputs that
mimic what they would receive if they were operating within a network.
However, the bulk of the input into L4 cells are from other L4 cells.
This means that in addition to surrogate LGN, L6, and ambient inputs, we
need to provide our LIF neurons \emph{surrogate L4 inputs} (both E and
I) commensurate with those received by network cells.  Arrival time
statistics will have to be presumed (here we use Poisson), but firing
rates should be those of L4 cells -- the very quantities we are seeking
from our MF model.  Thus there is the following consistency condition
that must be fulfilled: For suitable parameters and external inputs, we
look for values $\ol{v}_E$, $\ol{v}_I$, $f_E$, and $f_I$ such that
\begin{itemize}
\item given $\ol{v}_E$ and $\ol{v}_I$, Eq.~(\ref{Mthd:MF}) returns $f_E$
  and $f_I$ as firing rates; and
\item when L4 firing rates of $f_E$ and $f_I$ are presented to the LIF
  pair along with the stipulated parameters and external inputs, direct
  simulations of this LIF pair return values of $\ol{v}_E$ and
  $\ol{v}_I$.
\end{itemize}
If we are able to locate values of $\ol{v}_E$, $\ol{v}_I$, $f_E$, and
$f_I$ that satisfy the consistency relations above, it will follow that
the LIF pair, acting as surrogate for the E and I-populations in the
network, provide mean voltage data that enable us to determine mean
network firing rates in a self-consistent fashion.

%%%%%%%%%%%%%%%%%%%%%%%%%%%%%%%%
\begin{figure}
  {\bf A}\\
  \begin{large}
    \begin{tikzcd}
      &\boxed{\mbox{Single LIF (E)}}\arrow[bend right=8,"\ol{v}_E^{(p+1)}~~~" below]{rdd}\\
      \mbox{\small LGN}\arrow[dashed]{ru}\arrow[dashed]{rddd}\\
      \mbox{\small L6}\arrow[dashed]{ruu}\arrow[dashed]{rdd}&&\boxed{\mbox{MF Eq.~(\ref{Mthd:MF})}}\arrow[bend right=8,dashed,"~~~~~~~~f_{\{E,I\}}^{(p)}\cdots" above]{luu}\arrow[bend left=8,dashed,"f_{\{E,I\}}^{(p)}\cdots"]{ldd}\\
      \mbox{\small Ambient}\arrow[dashed]{rd}\arrow[dashed]{ruuu}\\
      &\boxed{\mbox{Single LIF (I)}}\arrow[bend left=8,"\ol{v}_I^{(p+1)}~~~" above]{ruu}\\
    \end{tikzcd}
  \end{large}\\[1ex]
  {\bf B}\\
  \includegraphics[width=\textwidth]{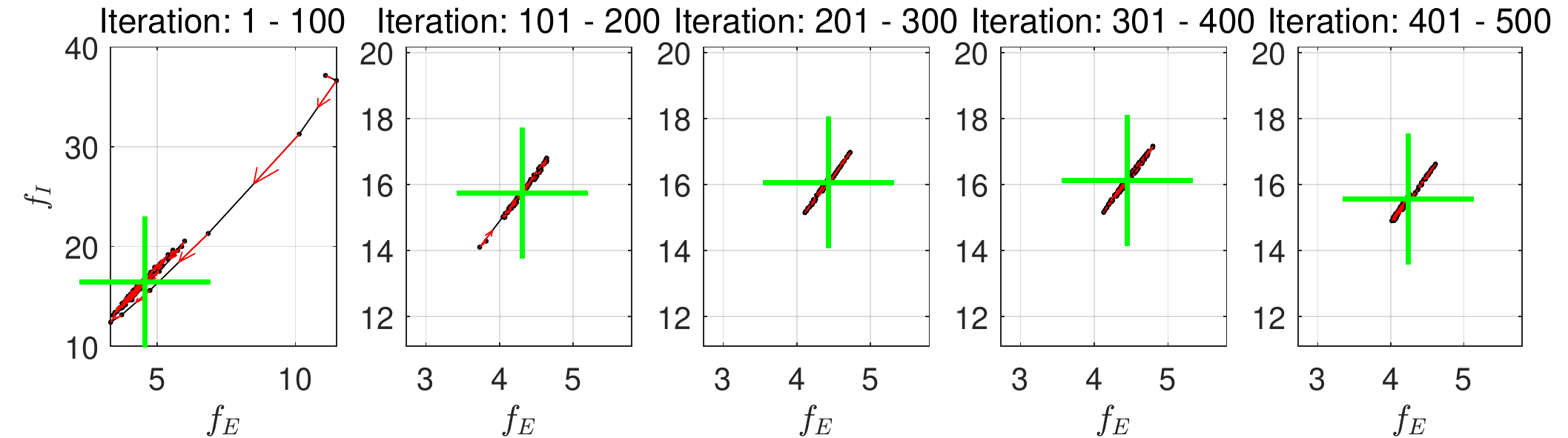}\\[1ex]
  {\bf C}\\
  \includegraphics[width=\textwidth]{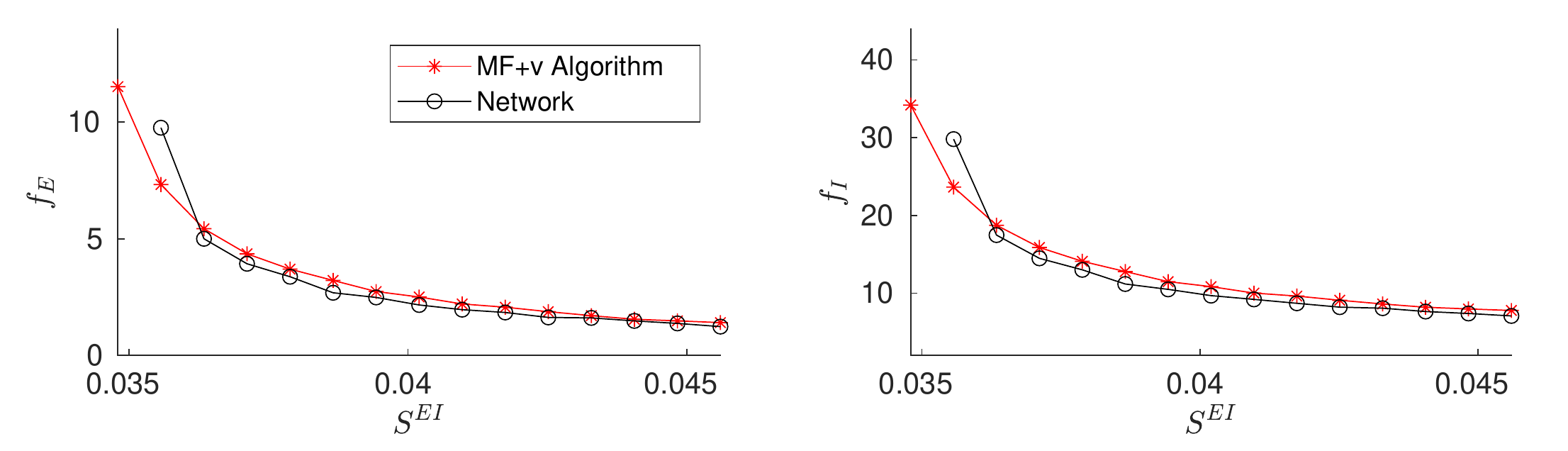}\\[1ex]
  \caption{{\bf The MF+v algorithm.}  {\bf A.} Schematic of the
    algorithm.  We begin by choosing initial values $f^0_E$ and $f^0_I$.
    These values are used to drive a pair of LIF neurons for 20 seconds
    (biological time).  The resulting membrane voltages are then fed
    into the MF equation~(\ref{Mthd:MF}), which gives us firing rates
    for the next iteration.  This is repeated until certain convergence
    criteria are met (see text).  In the above, all dashed lines are
    modeled by Poisson processes.  {\bf B.} 500 training iterations of
    the MF+v.  The means of every 100 iterations are indicated by green
    crosses, stability properties of which are quite evident. {\bf C.}
    Comparison of network (black) and MF+v computed (red) firing rates.}
  \label{fig:flowchart}
\end{figure} 

%%%%%%%%%%%%%%%%%%%%%%%%%%%%%%%%

A natural approach to finding self-consistent firing rates is to
alternate between estimating mean voltages using LIF neurons receiving
surrogate network inputs --- including L4 firing rates from the previous
iteration --- and using the MF formula (Eq.~(\ref{MF})) to estimate L4
firing rates using voltage values from the previous step.  A schematic
representation of this iterative method is shown in
Fig.~\ref{fig:flowchart}A.  In more detail, let $\B{v}_p$ and $\B{f}_p$
be the mean voltage and firing rate estimates obtained from the $p$th
iteration of the cycle above.  In the next iteration, we first
simulate the LIF cells for a prespecified duration $t^{LIF}$,
with L4 firing rate set to $\B{f}_p$, to obtain an estimate $\B{v}_{p+1}
= {\rm LIF}(\B{f}_p, t^{LIF})$ of the mean voltage.  We then update the
rate estimate by solving
\begin{equation}
  \B{f}_{p+1} = \BR(\B{f}_p) \times \left[\BM(\B{v}_{p+1})\cdot\B{f}_{p+1} +
    \B{s}(\B{v}_{p+1})\right]
\end{equation}
for $\B{f}_{p+1}$, leading to
\begin{equation}
  \B{f}_{p+1} ~~=~~{\rm MF}(\B{v}_{p+1},\B{f}_{p})~~:=~~\left[I -
    \BR(\B{f}_p) \BM(\B{v}_{p+1})\right]^{-1} \times \BR(\B{f}_p)
  \B{s}(\B{v}_{p+1}).
  \label{eq:MF-iterate}
\end{equation}
When the iteration converges, i.e., when $\B{f}_{p}=\B{f}_{p+1}=\B{f}$
and $\B{v}_p=\B{v}_{p+1}=\B{v}$, we have a solution $(\B{f},\B{v})$ of
Eq.~(\ref{Mthd:MFMat}).

Fig.~\ref{fig:flowchart}B shows 500 iterations of this scheme in
$\B{f}_p$-space.  Shown are iterations and running means (green
crosses).  Observe that after transients, the $\B{f}_p$ settle down to
what appears to be a narrow band of finite length, and wanders back and
forth along this band without appearing to converge.  We have
experimented with doubling the integration time $t^{LIF}$ and other
measures to increase the accuracy of the voltage estimates $\B{f}_p$,
but they have had no appreciable impact on the amplitudes of these
oscillations.  A likely explanation is that the contours of the E/I
firing rates (Fig.~\ref{FigM}A) are close to but not exactly parallel:
Had they been exactly parallel, a long line of $\B{v}$ would produce the
same $\B{f}$, implying the MF equations (\ref{Mthd:MF}) do not have
unique solutions.  The fact that they appear to be nearly parallel then
suggests a large number of near-solutions, explaining why our attempt at
fixed point iteration cannot be expected to converge in a reasonable
amount of time.

However, the oscillations shown in Fig.~\ref{fig:flowchart}B are very
stable and well defined, suggesting a pragmatic way forward: After the
iterations settle to this narrow band, we can run a large number of
iterations and average the $\B{f}_p$ to produce a single estimate.
Specifically, we first carry out a number of ``training'' iterations, and
when the firing rate estimates settle to a steady state by a heuristic
criterion, we compute a long-time average and output the result.
Combining this with the MF formula~(\ref{MF}) yields the MF+v algorithm.
See \textbf{SI} for more details.

Fig.~\ref{fig:flowchart}C compares MF+v predictions with network
simulations.  As one would expect, averaging significantly reduces
variance.  The results show strong agreement between MF+v predictions
and their target values given by direct network simulations.
  
Finally, we remark that a natural alternative to our MF+v algorithm
might be to forgo the MF equation altogether, and construct a
self-consistent model using single LIF neurons.  We found that such an
LIF-only method is much less stable than MF+v.  (See {\bf SI}.)  This is
in part because firing rates require more data to estimate accurately:
For each second of simulated activity, each LIF neuron fires only 3-4
spikes, whereas we can collect a great deal more data on voltages over
the same time interval.

\subsection{Issues related to implementation of MF+v}
\label{Mthd-Appl}

The hybrid approach of MF+v, which combines the use of the MF
formula~(\ref{MF}) with voltage estimates from direct simulations of
single neurons, has enabled us to seamlessly incorporate the variety of
kick sizes and rates from L6, LGN, and ambient inputs while benefiting
from the efficiency and simplicity of Eq.~(\ref{MF}).  Nevertheless
there are some technical issues one should be aware of.

\heading{Failure to generate a meaningful result.}  We have found that
MF+v iterations can fail to give meaningful answers in the following two
situations.  First, in some parameter regimes, the linear operator
$\BM(\B{v})$ can become singular, which can result in unreasonably low
or even negative values of $\B{f}_p$ in the MF+v algorithm (see {\bf
  SI}). Second, when firing rates are too low (e.g., $f_E<0.1$ Hz), the
low rate of L4 kicks to the LIF neurons in the MF+v model can result in
large fluctuations of $\B{v}_p$, which can destabilize the computed
$\B{f}_p$ unless the integration time $t^{\rm LIF}$ is sufficiently
large.

In {\bf Results}, whenever MF+v fails, we exclude the parameter set and
label it with a gray pixel in the canonical picture; see, e.g.,
Fig.~\ref{Fig1: Canonical Set}.

\heading{Computational cost.} The majority of computation in MF+v is
spent on collecting the mean voltages $\B{v}_p$ in each iteration, which
can be time-consuming depending on the time-scale and accuracy of
LIF-neuron simulation.  By repeating for $M$ iterations, the
computational cost of MF+v is $O\big(Mt^{\rm LIF}\big)$, where a general
choice of $t^{\rm LIF}\in(5,40)$s.  If we simulate each neuron for
$t^{\rm LIF}=20{\rm s}$ per iteration for up to $M=100$ iterations, we
typically obtain a firing rate estimate within $\sim20\%$
accuracy of the network firing rate in $\sim1.5$s.  In contrast,
the cost of simulating a large network with $N$ neurons typically grows
at least as fast as $N$, and may even grow nonlinearly in $N$.  With the
parameters in this paper, a typical network simulation using
$N\approx4\times10^4$ cells may require up to $\sim60$ seconds, which is
substantial when used to map a 7-dimensional parameter grid.  The MF+v
algorithm thus represents an $\sim40$-fold speedup over the
corresponding large network simulation in contemporary computing
environments.\footnote{Both network simulations and MF+v computations
are implemented using MATLAB\textsuperscript{\textregistered} 2020B with
Intel Xeon Platinum 8268 24Core 2.9GHz processors.}

It is possible to further reduce the computational cost of MF+v.  First,
instead of the simple iteration scheme we used in MF+v (i.e., Picard
iteration), one can use a stochastic variant of a higher-order method
(see, e.g., \cite{kushner2003stochastic}).  Second, one need not make an
independent computation of the mean voltages $\B{v}_p$ for each
iteration.  Instead, we can precompute the mean voltages for a coarse
grid in $(f_E,f_I)$ space, then, by interpolation and smoothing,
construct a table of values of mean voltages as functions of L4 rates.
This approximation can then be used as a surrogate for the Monte Carlo
simulation presently used in MF+v.  In certain regimes, it may also be
possible to compute $\B{v}_p$ in a more analytical manner, e.g., via the
Fokker-Planck equation for LIF neurons.

\begin{acknowledgement}
  ZCX was supported by the Swartz Foundation.  KL was supported in part
  by NSF grant DMS-182128.  LSY was supported in part by NSF grants
  SMA-1734854 and DMS-1901009.
\end{acknowledgement}

\bibliographystyle{plos2015}
\bibliography{main}

%%%%%%%%%%%%%%%%%%%%%%%%%%%% Supplemental %%%%%%%%%%%%%%%%%%%%%%%%%%%%%%%%%%%%%%%%%%%%
%% \newpage
\section*{Supplementary Information}
\renewcommand{\thesubsection}{S\arabic{subsection}} 
\setcounter{subsection}{0}
\renewcommand{\thefigure}{S\arabic{figure}} 
\setcounter{figure}{0}

\subsection{Details of network model: network architecture and governing equations}
\label{SI-NW}

In this section and the next, we describe the V1 network model we view
as ``ground truth''.  Except for minor simplifications, we follow
\cite{chariker2016orientation}; interested readers are directed to
\cite{chariker2016orientation} and references therein for more details.

\heading{``External" input to layer 4C$\alpha$ neurons.}  We model a
small piece of layer 4C$\alpha$ of the Macaque primary visual cortex
(V1), focusing on the background regime, i.e., spontaneous network
activity.  Layer 4C$\alpha$ neurons deliver excitatory and inhibitory
signals to each other.  In addition, each Layer 4C$\alpha$ neuron
receives excitatory input from three categories of external (meaning
external to 4C$\alpha$) sources, which we label {\it LGN}, {\it Layer
  6}, and {\it ambient}.  ``LGN'' refers to input from the lateral
geniculate nucleus, which carries visual signals from the retina;
``Layer 6'' represents recurrent excitation from Layer 6 of V1;
``ambient'' is an amalgamation of weak cortical neuromodulatory signals.
Since our primary goal is to investigate background firing rates (when
external input rates are low), we model signals from all three sources
as statistically independent, approximately Poisson (Bernoulli)
processes, i.e., to generate a point process of rate $R$ with timestep
$\Delta t$, we place 0 or 1 spike in each time bin with probability
$R\Delta t$.

\heading{Architecture of layer 4C$\alpha$ network in V1.}  Consider a
part of a 2D cortical sheet representing $3\times3$ hypercolumns, each
occupying $0.5\times0.5$ mm$^2$ in layer 4C$\alpha$.  This region
contains 26,244 E-cells and 8,649 I-cells, which we assume are uniformly
distributed, resulting in 3,877 cells (2,916 E, 961 I) per hypercolumn.
In our model, E-cells are assumed to be spiny stellate cells and I-cells PV
basket cells.  The strength of connectivity between cells depends on the
cell types (E or I, to be discussed later in \textit{Equations} and
Sect.~\ref{SI-Para}), while the probability of connection between any
two cells in a local circuit is determined by a truncated Gaussian
function of cell-to-cell distances. (In layer 4C$\alpha$ there are no
long-range connections.)  For E-to-Q and I-to-Q connections, the
standard deviations of the Gaussian functions are $0.2/\sqrt{2}$ mm and
$0.125/\sqrt{2}$ mm, respectively, reflecting the different reach of E
and I cells.  The peak connection probability for E$\to$E is 0.15, while
the numbers for E$\to$I, I$\to$E, and I$\to$I are set as 0.6, due to the
much denser I-cell connections.  We truncate all the Gaussian functions
at $X^o = 0.36$ mm.  Specifically, for a pair of neurons $(i,j)$ which
are $x$ mm away from each other, the projection probability from $j$ to
$i$ is
\begin{align*}
    P^{EE}(x) &= 0.15\times \exp\left\{-\left(\frac{x}{0.2}\right)^2\right\}, \\
    P^{EI}(x) &= 0.60\times \exp\left\{-\left(\frac{x}{0.125}\right)^2\right\}, \\
    P^{IE}(x) &= 0.60\times \exp\left\{-\left(\frac{x}{0.2}\right)^2\right\}, \\
    P^{II}(x) &= 0.60\times \exp\left\{-\left(\frac{x}{0.125}\right)^2\right\},
\end{align*}
and
\begin{align*}
    P^{ij}(x) = 0 \quad \mbox{for } x>X^o;\ i,j\in\{E,I\}.
\end{align*}
The connection probabilities and cell densities result in that, on
average, each E-cell has $\sim$210 presynaptic E-cells and $\sim$110 I
presynaptic cells, while each I -cell has $\sim$840 presynaptic E-cells
and $\sim$110 presynaptic I-cells.  We leave the neurons close to the
boundary as is and choose not to compensate for the missed presynaptic
neurons.

\heading{Equations.}  We use conductance-based leaky-integrate-and-fire
(LIF) models for neuronal dynamics within the Layer 4C$\alpha$ network.
For a neuron $i$ with cell type $Q \in \{E,I\}$, its membrane potential
$v_i$ advances by
\begin{equation} 
\label{SI:LIF}
\ddt{v_i} = - g_{Q}^L (v_i-V_{\rm rest}) - g_i^E(t)(v_i-V^E) - g_i^I(t)(v_i-V^I)\ .
\end{equation}
We nondimensionalize voltages, setting resting potential to $V_{\rm
  rest} = 0$ and spiking threshold $V_{\rm th} =1$.  Every time $v_i$
reaches $V_{\rm th}$, a spiking event occurs at neuron $i$ and the
signal is sent to all its postsynaptic neurons.  Afterwards, $v_i$
enters a refractory period for $\tau_{\rm ref}= 2$ ms right away, then
reset to $V_{\rm rest}$.

With the selections of the reversal potentials $V^E= 14/3$ and
$V^I=-2/3$ \cite{koch1999biophysics}, the membrane potential $v_i$ is
driven by three current terms in Eq.~(\ref{SI:LIF}):
\begin{enumerate}

\item Towards $V_{\rm rest} = 0$ due to the leaky current $- g_{Q}^L
  (v_i-V_{\rm rest})$, where $g_{Q}^L$ stands for membrane leakage
  conductance of cell type $Q$.

\item Towards $V^I=-2/3$ due to the inhibitory current $-
  g_i^I(t)(v_i-V^I)$, where the inhibitory conductance $g_i^I(t)$ is
  determined by the spiking series generated by inhibitory cells
  presynaptic to neuron $i$, i.e.,
  \begin{align}
    \label{SI:I-conductance}
    g_i^I(t) = S^{QI} \sum_{j\in N_{{\rm 4C},I}(i)} \sum_{k = 1}^\infty G_{\rm gaba}(t-t^j(k))\ .
  \end{align}
  Here, $S^{QI}$ stands for the connectivity from an I-cell to a
  type-$Q$ cell. Cell $j$ belongs to $N_{{\rm 4C},I}(i)$, the collection
  of all presynaptic I-cells to neuron $i$, generating a spiking time
  series $\{t^j\}$.  In addition, $G_{\rm gaba}(t)$ is a Green's
  function modeling the temporal increment of inhibitory conductances
  induced by each I-spike through GABA receptors (details provided in
  Sect.~\ref{SI-Para}).

\item Towards $V^E= 14/3$ due to the excitation current $-
  g_i^E(t)(v_i-V^E)$.  The excitatory conductance of neuron $i$,
  $g_i^E(t)$, consists of four components:
  \begin{align}
    \label{SI:E-conductance}
    \nonumber g_i^E(t) & = \underbrace{S^{Q\rm lgn} \sum_{k =
        1}^\infty G_{\rm ampa}(t-t^{i,{\rm lgn}}(k))}_{\mbox{(I) LGN}}
    \ +\ \underbrace{S^{Q\rm amb} \sum_{k = 1}^\infty G_{\rm
        ampa}(t-t^{i,{\rm amb}}(k))}_{\mbox{(II) ambient}} \\
    & + \underbrace{S^{Q\rm L6} \sum_{k = 1}^\infty \left[\rho_{\rm
          ampa}^QG_{\rm ampa}(t-t^{i,\rm L6}(k)) + \rho_{\rm
          nmda}^QG_{\rm nmda}(t-t^{i,\rm L6}(k)) \right]}_{\mbox{(III)
        Layer 6}} \\
    \nonumber & + \underbrace{S^{QE} \sum_{j\in N_{{\rm 4C},E}(i)}
      \sum_{k = 1}^\infty \left[\rho_{\rm ampa}^QG_{\rm
          ampa}(t-t^{j}(k)) + \rho_{\rm nmda}^QG_{\rm
          nmda}(t-t^j(k)) \right]}_{\mbox{(IV) Layer\ 4}}\ .
  \end{align}
  Terms I-IV represent synaptic conductances induced by LGN, ambient,
  Layer 6 input, and Layer 4 recurrent excitation, respectively.  For
  each neuron $i$, the spiking series in terms I-III are modeled by
  Poisson processes as described above.  For E-spikes from Layer 4 and
  6, two different types of excitatory synapses (AMPA and NMDA) induce different
  temporal increment of $g_i^E(t)$ (term III and IV), while only AMPA
  synapse is involved for LGN and ambient input (term I and II).  Here,
  $\rho_{\rm amda, nmda}^Q$ stand for the fractions of synaptic input
  received by AMPA and NMDA receptors in a type-$Q$ neuron; $(S^{Q\rm
    lgn}$, $S^{Q\rm amb}$, and $S^{Q\rm L6},S^{QE})$ denote the
  respective synaptic coupling weights of these sources towards type-$Q$
  cells.

\end{enumerate}
For the E-to-E input in Layer 4, two additional biological details are
incorporated in the model.  First, we consider a possibility for
synaptic failure $p_{\rm fail}=20\%$, i.e., whether the $k$-th spike
from neuron $j$ successfully induces a change in $g_i^E(t)$ depends on
an independent coin-toss with $p=0.8$.  Second, if a spike is
``successful", a random delay is added to $t^j(k)$ to model the fact that
E-neurons project to the dendrites of other E-cells, instead of the
soma.  In all, when neuron $i$ is an E-neuron, the term IV in
Eq.~(\ref{SI:E-conductance}) is replaced by
$$
S^{EE} \sum_{j\in N_{{\rm 4C},E}(i)} \sum_{k = 1}^\infty \sigma^j(k)
\left[\rho_{\rm ampa}^EG_{\rm ampa}(t-(t^j(k) + \tau^j(k))) + \rho_{\rm
    nmda}^EG_{\rm nmda}(t-(t^j(k) + \tau^j(k))) \right],
$$
where $\sigma^j(k)$ stands for the coin-toss, and the $\tau^j(k)$ are
independent, identically distributed random delay times uniformly
distributed on $[0,1]$ ms.

\subsection{Parameters}
\label{SI-Para}

We now list the specific parameter values used.  We remark that while
EPSC and IPSCs have been measured in the laboratory and so can be
assumed to be known, the coupling weights --- which involve how one
neuron affects another --- cannot be measured directly.  This is the
main reason we mostly regard the coupling weights as free parameters to
be investigated in this paper, in spite of experimental evidence that
may provide certain ranges for them.

\heading{Neuronal parameters.}  The following parameters are used for L4 neurons, 
and are fixed throughout the paper.
\begin{enumerate}
\item Reversal potentials: $V^I=-2/3$, $V^E=14/3$
\item Leakage conductances: $g_{E}^L = (20\ {\rm ms})^{-1}$, $g_{I}^L =
  (16.7\ {\rm ms})^{-1}$
\item Postsynaptic conductances:
  \begin{align*}
    G_s(t) = \frac{1}{\tau^{\rm decay}_s - \tau^{\rm rise}_s}\left(e^{-\frac{t}{\tau^{\rm rise}_s}} - e^{-\frac{t}{\tau^{\rm decay}_s}}\right),
  \end{align*}
  where $(\tau^{\rm rise}_s,\tau^{\rm decay}_s)$ stand for the time
  scales of activation/inactivation of synapse type $s = {\rm ampa,
    nmda, gaba}$; the time constants used here are
  \begin{itemize}
  \item $(\tau^{\rm rise}_{\rm ampa},\tau^{\rm decay}_{\rm ampa}) = (0.5,3)$ ms
  \item $(\tau^{\rm rise}_{\rm nmda},\tau^{\rm decay}_{\rm nmda}) = (2,80)$ ms
  \item $(\tau^{\rm rise}_{\rm gaba},\tau^{\rm decay}_{\rm gaba}) = (0.5,5)$ ms
  \end{itemize}
\item Fraction of AMPA and NMDA receptors activated by E-spikes:
  $(\rho^{E}_{\rm ampa},\rho^{E}_{\rm nmda}) = (0.8,0.2)$, and
  $(\rho^{E}_{\rm ampa},\rho^{E}_{\rm nmda}) = (0.67,0.33)$
\item Synaptic failure: $\sigma^j(k) = 1$ with $p = 0.8$, and
  $\sigma^j(k) = 0$ with $p = 0.2$
\item Synaptic delays: the $\tau^j(k)$ are uniformly distributed between
  $[0,1]$ ms
\item Refractory period: $\tau_{\rm ref} = 2$ ms
\end{enumerate}
For biophysical constants, see, e.g., \cite{koch1999biophysics}.  We
follow~\cite{chariker2016orientation, chariker2018rhythm} for all other
parameters.

\heading{Network parameters.} These include all synaptic coupling
weights and rates of external input, making up the parameter space in
which we compute the landscape of E- and I-firing rates.  We specify
below our choices of them.  Those parameters that are specified with
ranges form the 7D space of \textit{free parameters} we investigate in
this paper.  These \textit{a priori} constraints of free parameters are
discussed later in this section. Generally, we first choose the values of $S^{EE}$ and $S^{II}$
(independently of other parameters), then index some of the other
parameters to $S^{EE}$ and $S^{II}$.
\begin{enumerate}
    \item Synaptic coupling weights chosen independently: $S^{EE}\in(0.018,\ 0.030)$, $S^{II}\in(0.08,\ 0.20)$
    \item Synaptic coupling weights depending on $S^{EE}$:
    \begin{itemize}
        \item $S^{E\rm lgn} \in (1.5,\ 3)\times S^{EE}$, $S^{I\rm lgn} \in (1.5,\ 3)\times S^{E\rm lgn}$
        \item $S^{E\rm L6}  = \frac13\times S^{EE}$
        \item $S^{EI} \in (0.9,\ 2.4)\times S^{EE}$
    \end{itemize}
    \item Synaptic coupling weights depending on $S^{II}$:
    \begin{itemize}
        \item $S^{IE} \in (0.1,\ 0.25)\times S^{II}$
        \item $S^{I\rm L6}  = \frac13\times S^{IE}$
    \end{itemize}
    \item LGN input rates: $F^{E\rm lgn} = F^{I\rm lgn} = 80$ Hz
    \item Layer 6 input rates: $F^{E\rm L6} = 250$ Hz, $F^{I\rm L6} \in 
(1.5,\ 6)\times F^{E\rm L6}$
    \item Ambient input: 
    \begin{itemize}
        \item $F^{E\rm amb} = F^{I\rm amb} = 500$ Hz
        \item $S^{E\rm amb} = S^{I\rm amb} = 0.01$
    \end{itemize}
\end{enumerate}

\heading{Prior biological constraints and scaling conventions.} We
justify our choices of the ranges of \textit{free parameters} above.
Following (often indirect) suggestions from experimental observations
and heuristic reasoning, one can arrive at some bounds on them.  The
ranges we impose are broader than those suggested by available data; the
greater the uncertainty, the wider the net we cast.  Specifically:
\begin{itemize}
\item $S^{EE} \in (0.015, 0.03)$: This follows from the conventional
  wisdom that when an E-cell is stimulated {\it in vitro}, it takes
  10-50 consecutive spikes in relatively quick succession to produce a
  spike.  Numerical simulation suggests the assumed order of magnitude
  for $S^{EE}$ is reasonable~\cite{chariker2015emergent}.
\item $S^{EI} \in (0.9,\ 2.4)\times S^{EE}$ and $S^{IE} \in
  (0.1,\ 0.25)\times S^{II}$: In the absence of experimental guidance,
  we located these ranges numerically as follows: We examined firing
  rate maps for wider ranges than these, and found that, for the most
  part, the geometry on inhibition planes forces the good areas to lie
  within these ranges.

\item $S^{II} \in (0.08,\ 0.20)$: There is no direct empirical
  information on this parameter; however, there is evidence that EPSPs
  for I-cells are similar in size to those for
  E-cells~\cite{levy2012spatial}.  We choose the range for $S^{II}$ by
  following the logistic that $S^{II} \in (0.08,\ 0.20)$ and $S^{IE} \in
  (0.1,\ 0.25)\times S^{II}$ means $S^{IE} \in (0.008,\ 0.05)$, which
  contains and is significantly larger than the range of $S^{EE}$ above.
\item $S^{E\rm lgn}\in(1.5, 3.0)\times S^{EE}$: Results from
  \cite{stratford1996excitatory} suggest that the sizes of EPSPs from
  LGN are $\sim2\times$ those from L4.  We therefore assume a range
  around 2.
\item $S^{I\rm lgn}\in(1.5, 3.0)\times S^{E\rm lgn}$: we assume $S^{I\rm
  lgn} > S^{E\rm lgn}$ because it has been reported that LGN produces
  larger EPSCs in I-cells \cite{beierlein2003two}.

\item $F^{I{\rm L6}} \in (1.5, 6)\times F^{E{\rm L6}}$: Within L4, an
  I-cell has 3.5-4 times as many presynaptic E-neurons as an E-cell. If
  we hypothesize a similar ratio between L6 and L4, it would follow that
  $F^{I{\rm L6}} \in (3.5, 4)\times F^{E{\rm L6}}$.  We relax the
  interval to $(1.5, 6)$ because of uncertainty surrounding L6: whether
  the effect of L6 on L4 is net-excitatory or net-inhibitory is an issue
  that is currently unresolved for the real cortex.  A wider range also
  serves to absorb potential errors in the assumption that $S^{I{\rm
      L6}} = \frac13\times S^{IE}$.
\end{itemize}

\subsection{Closer look at MF+v: solvability of MF equation and implementation details}
\label{sect:MF solutions}

\heading{Biologically meaningful solutions of MF equation.}  In some
situations, the MF equation~(\ref{MF}) yields negative firing rates when
given valid mean voltages; we have indicated such parameters by gray in
all parameter plots.  Here, we discuss some of the reasons underlying
these failures.

First, recall that in Methods, we had asserted that the MF equation
Eq.~(\ref{Mthd:MF}) can be written in matrix form as
\begin{displaymath}
  \B{f} = \BR(\B{f}) \times \left[\BM(\B{v})\cdot\B{f} +
    \B{s}(\B{v})\right].
\end{displaymath}
(This was Eq.~(\ref{Mthd:MFMat}) in Methods.)  This can be seen by
defining
\begin{align}
  \BR(\B{f}) &= \begin{bmatrix} 
    1-f_E\tau_{\rm ref} & 0 \\ 
    0 & 1-f_I\tau_{\rm ref}\\
  \end{bmatrix}\nonumber\\[1ex]
  \BM(\B{v}) &= \begin{bmatrix}
    S^{EE} N^{EE} (1-p_{\rm fail})(V^E-\ol{v}_E) & S^{EI}N^{EI} (V^I-\ol{v}_E) \\[1ex]
    S^{IE} N^{IE}(V^E-\ol{v}_I) & S^{II}N^{II}(V^I-\ol{v}_I)\\
  \end{bmatrix}\\[1ex]
  \B{s}(\B{v}) &= \begin{bmatrix}
    \left(S^{E\rm lgn}F^{E\rm lgn} + S^{E\rm L6} F^{E\rm L6} + S^{E\rm amb} F^{E\rm amb} \right)(V^E-\ol{v}_E) - g_{E}^L\ol{v}_E \\[1ex]
    \left(S^{I\rm lgn}F^{I\rm lgn} + S^{I\rm L6} F^{I\rm L6} + S^{I\rm amb} F^{I\rm amb} \right)(V^E-\ol{v}_I) - g_{I}^L\ol{v}_I\\
  \end{bmatrix}\nonumber
\end{align}
and verifying directly.  Our interest is in finding nonnegative
solutions $\B{f}$ of this equation given mean voltages $\B{v}$.  The
solvability of Eq.~(\ref{Mthd:MFMat}) depends on the properties of the
matrix $\BM(\B{v})$, which in turn depends on L4 connectivity and the
mean voltages $\B{v}$.  Note that the entries of $\B{s}(\B{v})$
represent the mean currents into E and I cells, respectively, and must
be positive for cells to fire with positive rates.

Among the scenarios in which these equations fail to give meaningful
firing rates, by far the simplest is when the equations are (nearly)
singular.  Ignoring the refractory factor $\BR(\B{f})$ (whose effect is
perturbative), Eq.~(\ref{Mthd:MFMat}) is equivalent to
\begin{equation}
  \B{f} = \left[I - \BM(\B{v})\right]^{-1} \times \B{s}(\B{v})\ ,
\end{equation}
where 
\begin{align*}
  \left[I - \BM(\B{v})\right] = \begin{bmatrix} 1- \rm [4E\to E] &
    -\rm [4I\to E] \\ -\rm [4E\to I] & 1-\rm [4I\to I] \end{bmatrix} =
  \left[\begin{array}{c|c} \B\alpha_1 &
      \B\alpha_2 \end{array}\right]\ ,
\end{align*}
provided $I-\BM(\B{v})$ is nonsingular.  In the above, $\rm [P \to Q]$
indicate the corresponding entry of matrix $M(\B{v})$, i.e., the net
contribution to an E/I-cell from one E/I-kick.  When $\det(I -
\BM(\B{v}))=0$, the linearized equation above may not have a solution,
suggesting MF+v iteration is likely to fail when $I - \BM(\B{v})$ is
(nearly) singular.

%%%%%%%%%%%%%%%%%%%%%%%%%%%%%%%%%%%%%%%%%%%%%%%%%%%%%%%%%%% FigS3
\begin{figure}%[htbp]
  \begin{center}
  \begin{subfigure}{.33\textwidth}
  {\bf A}\\
  \includegraphics*[bb=0in 0.4in 3.2in 3.65in,width=\textwidth]{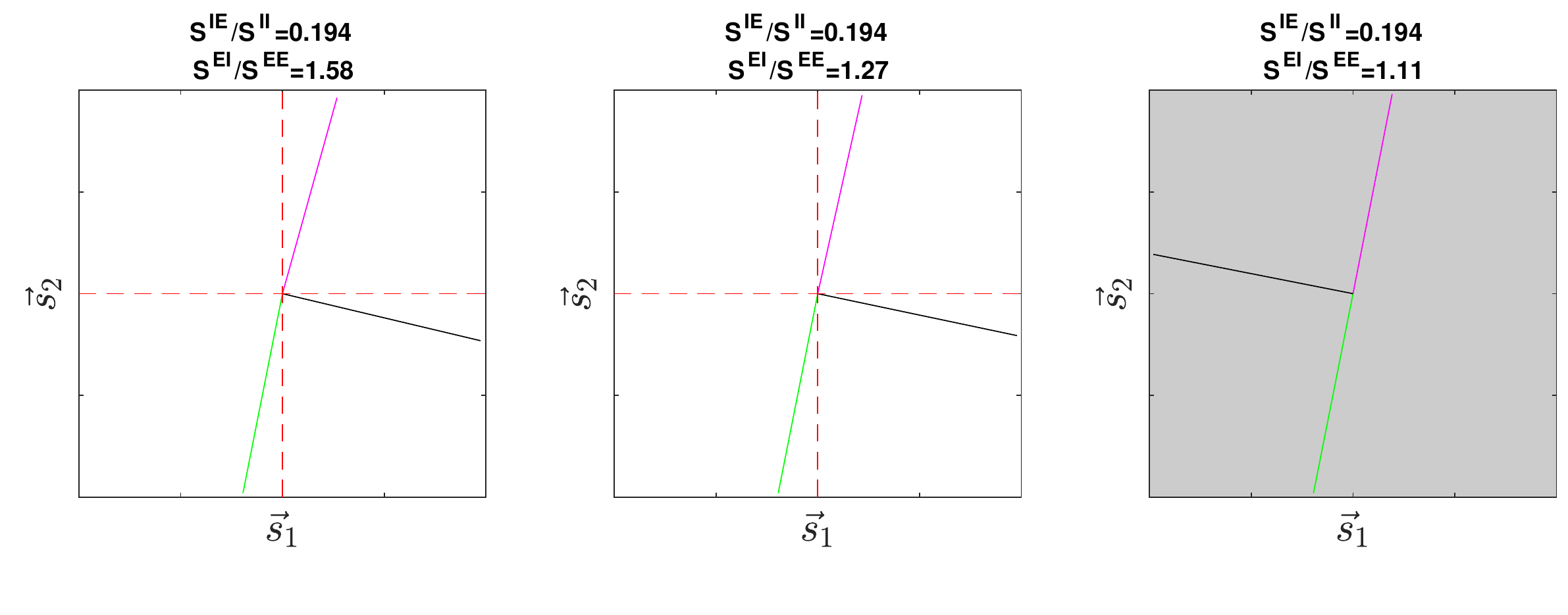}
  \end{subfigure}%
  \begin{subfigure}{.33\textwidth}
  {\bf B}\\
  \includegraphics*[bb=3.2in 0.4in 6.4in 3.65in,width=\textwidth]{{FigureS5_Singular2}.pdf}
  \end{subfigure}
  \begin{subfigure}{.33\textwidth}
  {\bf C}\\
  \includegraphics*[bb=6.4in 0.4in 9.6in 3.65in,width=\textwidth]{{FigureS5_Singular2}.pdf}
  \end{subfigure}
    
    \caption{A geometric view of the solvability of MF equations.  Each
      panel contains two normalized column vectors of $I-\BM(\B{v})$ ,
      i.e., $\B\alpha_1$ (green vector in the third quadrant) and
      $\B\alpha_2$ (purple vector in the first quadrant).  The black
      line segment marks the region $S = \left\{f_1\B\alpha_1 +
      f_2\B\alpha_2~\mid~f_1,f_2\geq0\right\}$.  \textbf{A.}
      Eq.~(\ref{Mthd:MFMat}) is nonsingular and well-behaved.
      \textbf{B.}  Eq.~(\ref{Mthd:MFMat}) is near-singular and its
      column vectors are nearly collinear.  \textbf{C.} ${\rm
        det}(I-\BM(\B{v}))$ changes sign and MF+v gives negative firing
      rates.}
    \label{FigS2:Singularity_2}
  \end{center}
\end{figure}
%%%%%%%%%%%%%%%%%%%%%%%%%%%%%%%%%%%%%%%%%%%%%%%%%%%%%%%%%%% FigS3

One can in fact take a more geometric view of the solvability of
Eq.~(\ref{Mthd:MFMat}), one that makes questions surrounding solvability
more transparent.  Let $\B\alpha_1$ and $\B\alpha_2$ be the columns of
$I-\B{M}(\B{v})$.  Eq.~(\ref{Mthd:MFMat}) yields nonnegative firing
rates precisely when there exist $f_1,f_2\geq0$ such that $f_1\B\alpha_1
+ f_2\B\alpha_2 = \B{s(v)}$.  This can be visualized by defining $S =
\left\{f_1\B\alpha_1 + f_2\B\alpha_2~\mid~f_1,f_2\geq0\right\}$ and
noting that the intersection of $S$ with the first quadrant
$\{s_1,s_2\geq0\}$ is precisely the set of all $\B{s}$ such that $\B{f}
= (I-\B{M}(\B{v}))^{-1}\B{s}$ results in nonnegative firing rates.
Fig.~\ref{FigS2:Singularity_2}A shows an example.  Here, the two
(normalized) column vectors $\B\alpha_1$ and $\B\alpha_2$ are linearly
independent, and the set $S$ (the region bounded by ${\rm
  span}(\B{\alpha}_1)$ and ${\rm span}(\B{\alpha}_2)$ and contains the
black line) has a large intersection with the first quadrant, so that
{\em most} nonnegative values of $\B{s}$ lead to nonnegative rates.
Note that these parameters lie well above the good area; {\it cf.}
Figs.~\ref{Fig1: Canonical Set} and \ref{FigS2:Singularity_2}.

To see what else might happen, we now move along a line in the
inhibition plane defined by ${S^{IE}}/{S^{II}} = 0.194$,
starting from the value ${S^{EI}}/{S^{EE}} = 1.58$ used in
Fig.~\ref{FigS2:Singularity_2}A and moving down.
Fig.~\ref{FigS2:Singularity_2}B shows what happens for
${S^{EI}}/{S^{EE}} = 1.27$, which lies within the good area:
$\B\alpha_1$ and $\B\alpha_2$ become more nearly collinear, though there
is still a sizable intersection between $S$ and the first quadrant, so
that most values of $\B{s}$ lead to positive firing rates.  However, as
$S^{EI}$ decreases even further, $\det(I-\BM(\B{v}))$ changes sign, and
the set $S$ abruptly flips to the other side of the dividing line,
leading Eq.~(\ref{Mthd:MFMat}) to produce negative firing rates for many
values of $\B{s}$.

%%%%%%%%%%%%%%%%%%%%%%%%%%%%%%%%%%%%%%%%%%%%%%%%%%%%%%%%%%% FigS4
\begin{figure}%[htbp]
  \begin{center}
    \includegraphics[width=\textwidth]{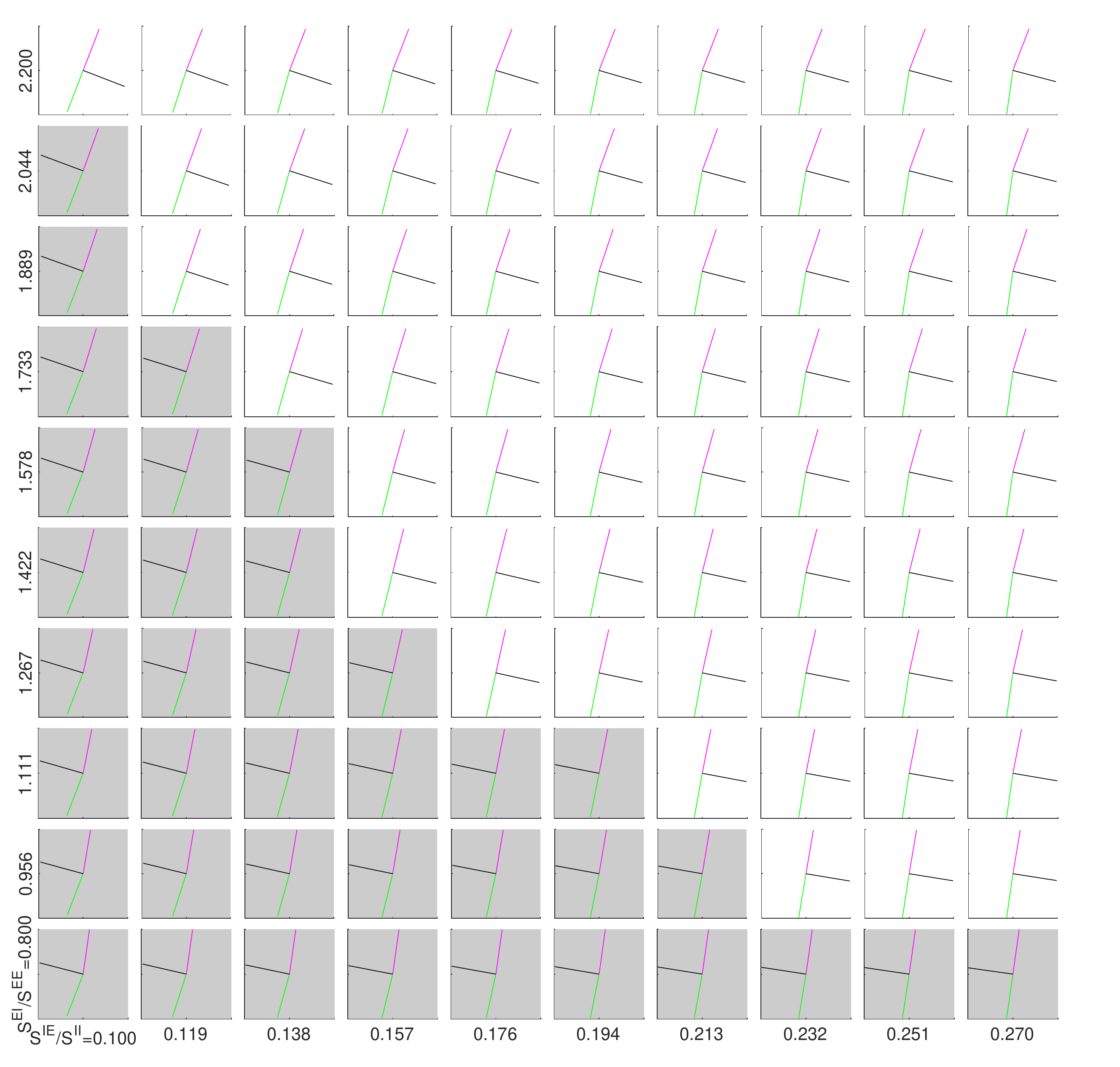}
    \caption{Solvability of Eq.~(\ref{Mthd:MFMat}) on the inhibition
      plane.  Here, $S^{EE} = 0.024$ and $S^{II} = 0.120$, and we fix
      $\B{v} = (v_E,v_I) = (0.55, 0.65)$ for all panels.  Panels with
      gray color corresponds to negative firing rates from the MF
      equations, and the boundary between gray and white panels roughly
      corresponds to where ${\rm det}(I-\BM(\B{v}))$ changes sign.}
    \label{FigS2:Singularity}
  \end{center}
\end{figure}
%%%%%%%%%%%%%%%%%%%%%%%%%%%%%%%%%%%%%%%%%%%%%%%%%%%%%%%%%%% FigS4

In Fig.~\ref{FigS2:Singularity}, we extend this picture to the
inhibition plane.  For each choice of ${S^{EI}}/{S^{EE}}$ and
${S^{IE}}/{S^{II}}$, we compute $\BM(\B{v})$ by assuming $\B{v}
= (v_E,v_I) = (0.55, 0.65)$.  Each panel shows the two (normalized)
column vectors $\B\alpha_1$ and $\B\alpha_2$, along with a black line
marking the region $S(\B\alpha_1,\B\alpha_2)$.  Observe that as
${S^{EI}}/{S^{EE}}$ and ${S^{IE}}/{S^{II}}$ decrease,
$\B\alpha_1$ and $\B\alpha_2$ first become linearly dependent (so that
$I-\BM(\B{v})$ becomes singular), then changing orientation and
resulting in negative firing rate estimates (gray panels).

Two final remarks.  First, the boundary between gray and white panels
corresponds roughly to where ${\rm det}(I-\BM(\B{v}))$ changes sign.
This is also where explosive dynamics occur in our {\em network}
simulations due to low suppression index $\mbox{SI}_E$ or very low
E-firing rates (caused by high external input to I-cells; more below).
This suggests that parameters where the MF+v algorithm fails to give
biologically meaningful estimates are also parameters where the network
model itself fails to give biologically meaningful results.

Second, concerning the narrow wedge of $\B{s}$-values in the first
quadrant that yield negative firing rate estimates, i.e., the set
bounded by $\B\alpha_2$ (oblique purple vector in the first quadrant)
and the vertical axis in Fig.~\ref{FigS2:Singularity_2}AB.  The same
pattern also occur in the upper part of Fig.~\ref{FigS2:Singularity}:
These correspond to when the external inputs to I-cells are too high.
Though the corresponding sets of $\B{s}$-values are small, they can have
a significant impact.  For example, for high-$S^{I\rm lgn}$ and/or
high-$F^{I\rm L6}$ regimes, this can lead to large swaths of gray.  See
Figs.~\ref{Fig2: LGN} and \ref{Fig3: LGN-vs-L6}, and {\bf SI}
(Sect.~\ref{SI-AddFrMap}).

\heading{Implementation and design of MF+v.}  As explained in Methods
(see M2), our first attempt at a fixed point iteration did not result in
a convergent algorithm.  So we iterate until the firing rate estimates
stabilize to a narrow, nearly linear band in firing rate space, then
average a number of successive estimates to produce an estimate.

\begin{algorithm}%[b]%[H]
  \SetAlgoLined
  \KwResult{MF computed firing rates $\B{f}$, and mean voltages $\B{v}$.}
  Set parameters $S^{EE}$, $S^{I{\rm lgn}}$, $F^{E{\rm L6}}$, etc.\;
  $(\B{f}_0,\B{v}_0)\gets{\mbox{initial conditions}}$\;
  $M\gets\mbox{maximum number of training iterations}$\tcp*{see text for value}
  $t^{\rm LIF}\gets\mbox{integration time}$\tcp*{default $t^{\rm LIF}=20$s}
  $\varepsilon\gets\mbox{early termination tolerance}$\tcp*{default $\varepsilon=0.05$}
  $k\gets\mbox{early termination horizon}$\tcp*{default $k=15$}
  \tcp{Training loop}
  \For{$p\gets1$ \KwTo $M$}{
    $\B{v}_{p} = {\rm LIF}(\B{f}_{p-1}, t^{LIF})$\;
    $\B{f}_{p} = {\rm MF}(\B{v}_{p},\B{f}_{p-1})$\; 
    \uIf{$CV(\B{f}_p,\cdots,\B{f}_{p-k}) < \varepsilon$}{
      goto {\small {\tt FINALIZE}}\;
    }
  }
  ~~{\small {\tt FINALIZE:}}\\
  \uIf{$p<M$}{
    \tcp{Convergence criterion satisfied}
    \tcp{Compute a more accurate estimate via a longer time average}
    $\ell_1\gets\mbox{number of voltages to use in moving
      average}$\tcp*{default $\ell_1=10$}
    $\ell_2\gets\mbox{number of firing rates to use in final
      estimate}$\tcp*{default $\ell_2=50$}
    \For{$q\gets p+1$ \KwTo $p+\ell_2$}{
      $\B{v}_{q} = {\rm LIF}(\B{f}^*_{q-1}, t^{\rm LIF})$\;
      $\B{v}^*_{q} = {\rm
        mean}(\B{v}_{q-\ell_1},\cdots,\B{v}_{q})$\tcp*{moving average}
      $\B{f}^*_{q} = {\rm MF}(\B{v}^*_{q}, \B{f}^*_{q-1})$\;
    }
    return ${\rm mean}(\B{f}^{*}_{p+1},\cdots,\B{f}^{*}_{p+\ell_2})$\;
  }
  \uElse{
    \tcp{Unconverged}
    return {\small {\tt FAIL}}
  }
  \caption{\label{Mthd:Algm1} The MF+v method.}
\end{algorithm}

Algorithm~\ref{Mthd:Algm1} gives a precise summary and lists all other
hyperparameter values used.  A practical issue is that we need to check
the variance of the voltage and firing estimates to determine when to
stop iterating.  To make this efficient, instead of carrying out
accurate but expensive long-time average for every iteration, we use
shorter runs that may be noisy by themselves but can be averaged
together to produce accurate estimates.  We then use a small number of
consecutive iterations to check convergence during an initial,
``training'' phase, and when certain stopping criteria are satisfied, we
compute a more accurate estimate and output the result.

In this paper, the precise stopping criterion is based on comparing the
{\it coefficient of variation}
\begin{equation}
  CV(\B{v}_p,\cdots,\B{v}_{p-k}) = \frac{{\rm
      var}(\B{v}_p,\cdots,\B{v}_{p-k})^{\nicefrac12}}{{\rm
      mean}(\B{v}_p,\cdots,\B{v}_{p-k})}
\end{equation}
of the last $k$ voltage estimates is against a pre-specified tolerance
$\varepsilon$.  When the stopping criterion is satisfied, we use moving
averages of the voltages to compute a larger number of iterations, and use
these to estimate the firing rate.  For the network models studied in
this paper, we have found the estimates to be insensitive to the exact
choice of the maximum iteration number $M$.  We typically set $M$ in the
range 300--500.

\subsection{Miscellaneous information on MF+v}

Here we record some additional information that have affected our
decision to use MF+v in this paper.

\heading{Effects of refractory period and different kick sizes.}  The
mean voltages $\B{v}$ produced by the LIF equations (and hence
the MF-computed firing rates $\B{f}$) can depend on parameters in a
nontrivial way, making it difficult to estimate $\B{v}$ using analytical
methods.  Here, we illustrate the parameter dependence of LIF neurons
via two examples, using the parameters in Sect.~\ref{SI-Para}.  In both
examples, a pair of LIF neurons (one E and one I) are each presented
with Poissonian spike trains modeling L4 inputs, with input rates
$f^{in}_E$ and $f^{in}_I$, in addition to L6, LGN, and amb inputs.  The
two neurons are uncoupled and given independent inputs.  We denote the
resulting output rates $f^{out}_E$ and $f^{out}_I$.
 
%%%%%%%%%%%%%%%%%%%%%%%%%%%%%%%%%%%%%%%%%%%%%%%%%%%%%%%%%%% FigS5
\begin{figure}%[htbp]
   %% \captionsetup{type=figure} 
   {\bf A}\\
   \begin{center}
      \includegraphics*[bb=0in 0in 6.96in 4.2in,width=\textwidth]{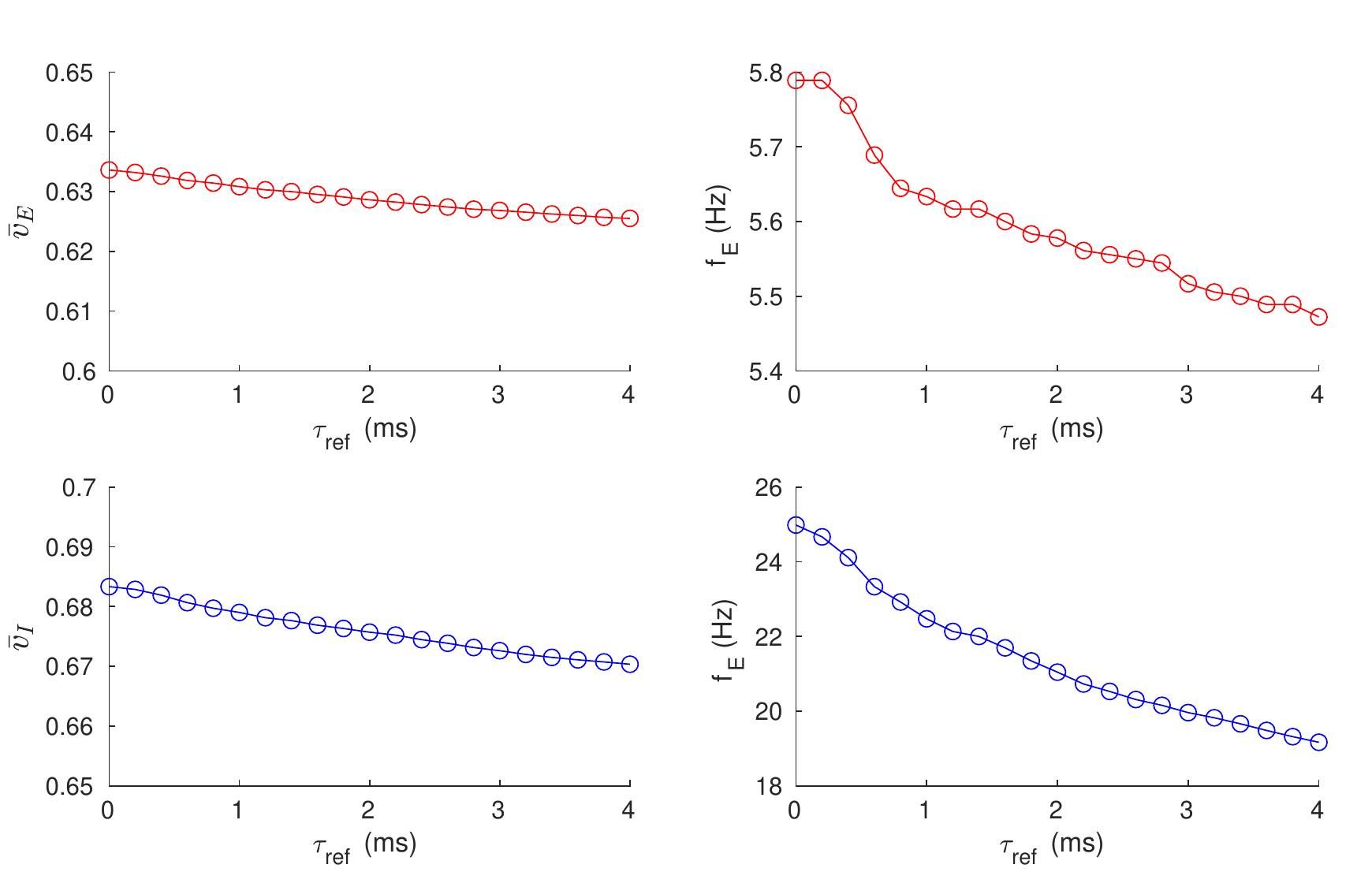}\\[1ex] 
   \end{center}
  {\bf B}\\
   \begin{center}
      \includegraphics*[bb=0in 0in 6.96in 4.2in,width=\textwidth]{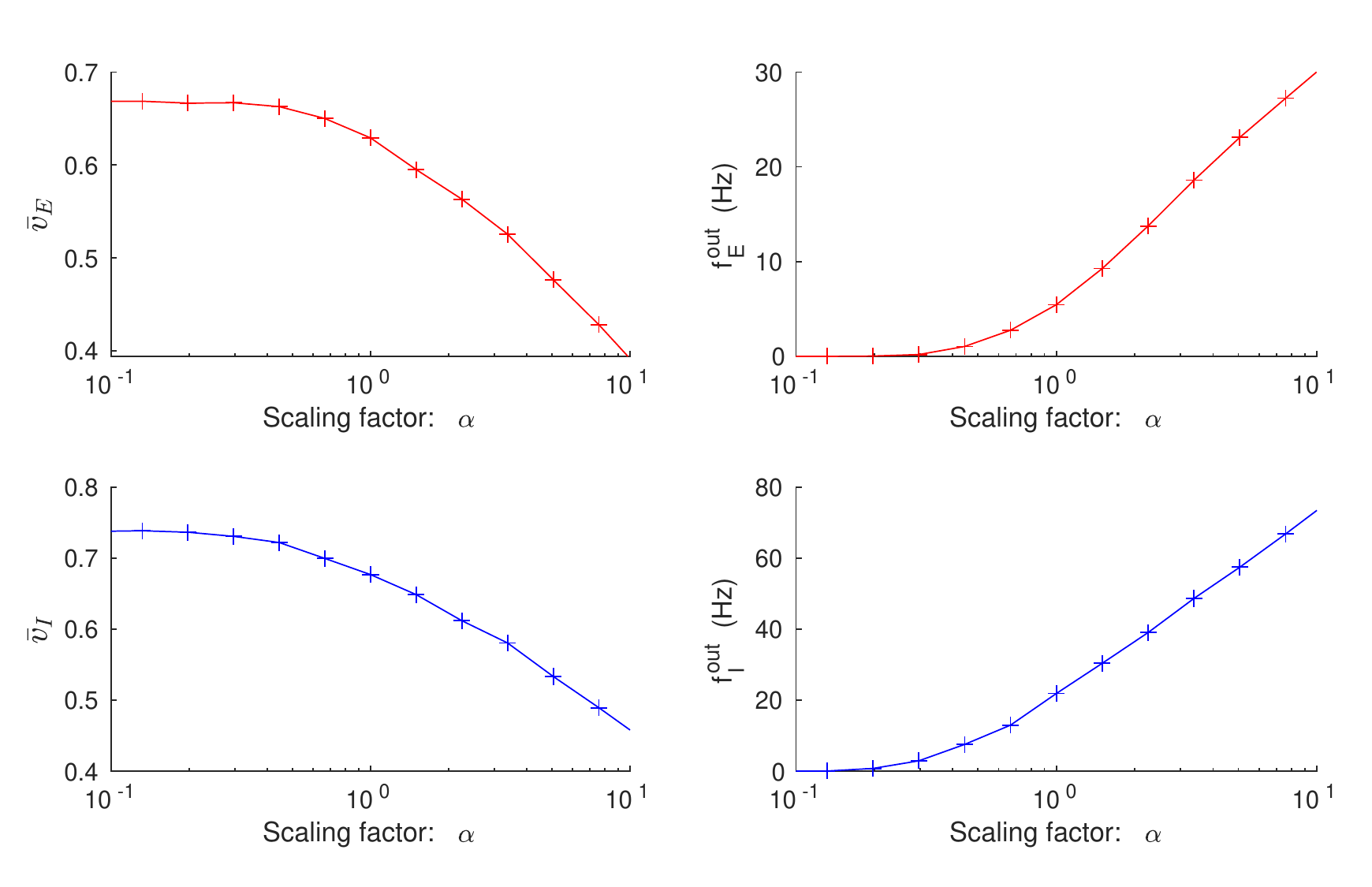}\\[1ex]       
   \end{center}
    \caption{Mean voltages and firing rates computed by MF+v with
      different \textbf{A.} refractory periods ($\tau_{\rm ref}$) and
      \textbf{B.} L4 excitatory kick sizes ($S^{EE}$ scaling factor
      $\alpha$). Here, $S^{EE}\times f^{in}_E = 0.13$, $S^{EI}\times
      f^{in}_I = 0.65$, $S^{IE}\times f^{in}_E = 0.10$, and
      $S^{II}\times f^{in}_I = 2.25$.}
    \label{FigS3: Analytical}
\end{figure}
%%%%%%%%%%%%%%%%%%%%%%%%%%%%%%%%%%%%%%%%%%%%%%%%%%%%%%%%%%% FigS5

Our first example concerns the refractory period $\tau_{\rm ref}$, which
can significantly impact neuronal activity because membrane conductances
and currents steadily decay during refractory periods; the larger the
$\tau_{\rm ref}$, the more conductance is ``missed'' by the neuron while
refractory.  For instance, for an E-cell with a 3 Hz firing rate, its
membrane potential stays unchanged for $3\times\tau_{\rm ref}$ in each
second.  This can be a non-negligible fraction of time, and the effect
is exacerbated by higher firing rates.  Fig.~\ref{FigS3: Analytical}A)
shows the mean voltages and firing rates as $\tau_{\rm ref}$ varies from
0 to 4 ms.  Though $f^{out}_E$ does not change much (5.5-5.8 Hz),
$f^{out}_I$ experiences a sharp change (19-25 Hz).

Our second example is motivated by the observation that for neurons in a
mean-driven regime (a limit often studied in theoretical analyses of
neuron models), their output rate depends on L4 kick sizes $S^{QQ'}$
($Q,Q'\in\{E,I\}$) only through the product $S^{QQ'}\times f^{in}_{Q'}$.
Thus, if we vary $S^{QQ'}$ while keeping $S^{QQ'}\times f^{in}_{Q'}$
constant, any variation in output rates ($\B{f}^{out}$) would be due to
fluctuations in the L4 inputs.  To test this, we perform the scaling
$S^{QQ'}\mapsto\alpha S^{QQ'}$ and $f^{in}_{Q'}\mapsto
f^{in}_{Q'}/\alpha$ for a range of scaling factors $\alpha$.  Other
parameters, including $(S^{Q{\rm lgn}},F^{Q{\rm lgn}})$ and $(S^{Q{\rm
    L6}},F^{Q{\rm lgn}})$ (which were previously indexed to $S^{QE}$)
are kept constant.  Fig.~\ref{FigS3: Analytical}B shows the results.
Observe that both $\B{v}$ and $\B{f}^{out}$ experience sharp changes
when $\alpha$ moves away from 1.  In particular, the firing rates are
almost 0 when $\alpha$ is small (low fluctuation), and unreasonably high
when $\alpha$ is large (high fluctuation).  These results suggest that
for the background regime studied here, MF+v (as is the network that it
models) operates in a fluctuation driven regime.

%%%%%%%%%%%%%%%%%%%%%%%%%%%%%%%%
\begin{figure}
  \begin{center}
    \begin{large}
      \begin{tikzcd}
        &\boxed{\mbox{Single LIF (E)}}\arrow[dashed,loop above,"f_E"]{r}\arrow[shift right=0.5ex,bend right=8,dashed,"~~~~~~f_E"]{dddd}\\
        \mbox{\small LGN}\arrow[dashed]{ru}\arrow[dashed]{rddd}\\
        \mbox{\small L6}\arrow[dashed]{ruu}\arrow[dashed]{rdd}&&\\
        \mbox{\small Ambient}\arrow[dashed]{rd}\arrow[dashed]{ruuu}\\
        &\boxed{\mbox{Single LIF (I)}}\arrow[dashed,loop below,"f_I~"]{r}\arrow[shift right=1.5ex,dashed, bend right=8,"f_I~~~~~~"]{uuuu}\\
      \end{tikzcd}
    \end{large}
  \end{center}
  \caption{The LIF-only algorithm.  We begin by choosing initial values
    $f^0_E$ and $f^0_I$.  In the first iteration, these values are used
    to drive a pair of LIF neurons for 20 seconds (biological time).
    The resulting firing rates (instead of the membrane voltages in
    MF+v) are then fed into the next iteration as the L4 input.  In the
    above, all dashed lines are modeled by Poisson processes.}
  \label{fig:lif-only-flowchart}
\end{figure}
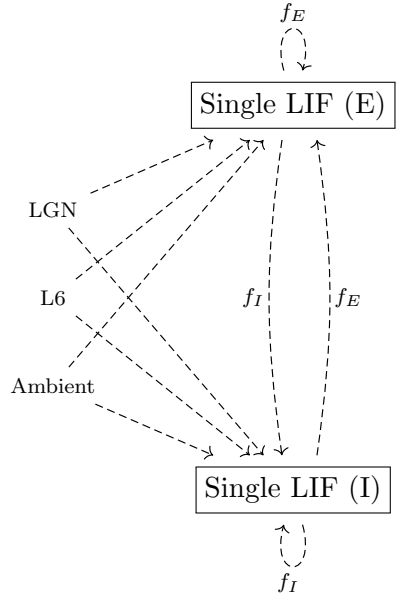

\begin{algorithm}%[htbp]
\SetAlgoLined
\KwResult{LIF-computed $\B{f}$.}
  Set parameters $S^{EE}$, $S^{I{\rm lgn}}$, $F^{E{\rm L6}}$, etc.\; 
  $\B{f}_0\gets{\mbox{initial conditions}}$\;
  $M\gets\mbox{maximum number of training iterations}$\tcp*{default $M=100$}
  $t^{\rm LIF}\gets\mbox{integration time}$\tcp*{default $t^{\rm LIF}=20$s}
  \tcp{Training loop}
  \For{$p\gets1$ \KwTo $M$}{
    $\B{f}_{p} = {\rm LIF}(\B{f}_{p-1}, t^{LIF})$\;
  }
  $\ell_1\gets\mbox{number of voltages to use in moving
    average}$\tcp*{default $\ell_1=10$}
  $\ell_2\gets\mbox{number of firing rates to use in final
    estimate}$\tcp*{default $\ell_2=50$}
  \For{$q\gets p+1$ \KwTo $p+\ell_2$}{
    $\B{f}_{q} = {\rm LIF}(\B{f}^*_{q-1}, t^{\rm LIF})$\;
    $\B{f}^*_{q} = {\rm mean}(\B{f}_{q-\ell_1},\cdots,\B{f}_{q})$\tcp*{moving average}
  }
  return ${\rm mean}(\B{f}^{*}_{p+1},\cdots,\B{f}^{*}_{p+\ell_2})$\;
 \caption{\label{Mthd:Algm2} The LIF-only algorithm.  We have removed
   the early termination criterion to simplify comparison with MF+v.}
\end{algorithm}
%%%%%%%%%%%%%%%%%%%%%%%%%%%%%%%%%%%%%%%%%%%%%%

\heading{Comparison to LIF-only.}  A natural alternative to MF+v is an
LIF-only method: We drive a pair of LIF models (one E and one I) with
Poissonian spike trains of rate $\B{f}$ as L4 input, along with Poisson
spike trains modeling inputs from LGN, L6, and amb, and look for values
of $\B{f}$ that lead to output rates equal to $\B{f}$.  That is, we look
for a self-consistent MF approximation without reference to
Eq.~(\ref{MF}).  This can be implemented by simply iterating LIF neurons
and feed $\ol{f}_p$ to iteration $p+1$ directly as the L4 E/I input.  A
schematic representation is in Fig.~\ref{fig:lif-only-flowchart}, with
details in Algorithm~\ref{Mthd:Algm2}.

\begin{figure}%[htbp]

     {\bf A}\\ %Contour maps for E and I firing rates\\
  \includegraphics*[bb=0in 3.2in 10.7in 5.9in,width=\textwidth]{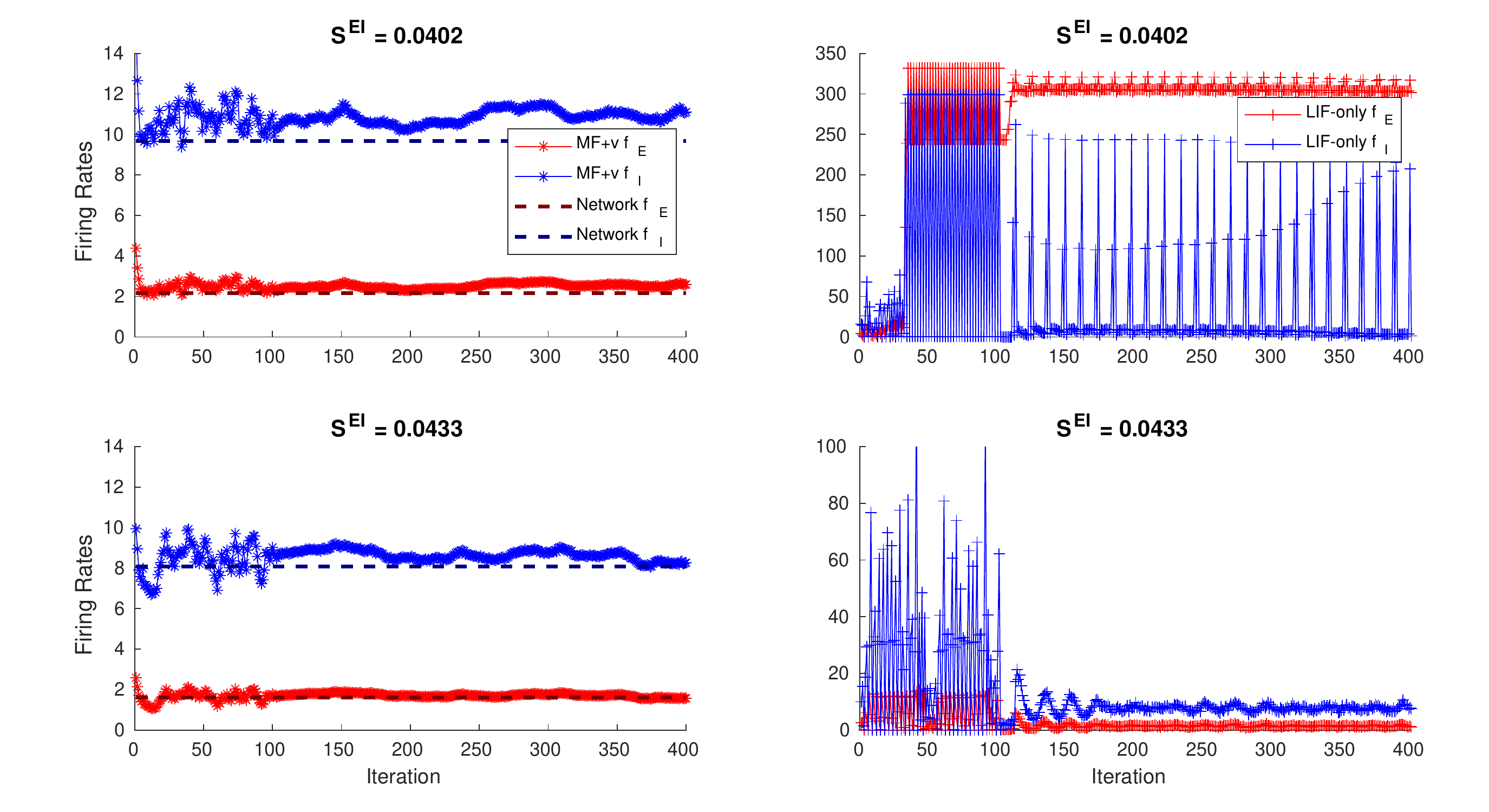}
     {\bf B}\\ %Network firing rates and MF predictions\\
  \includegraphics*[bb=0in 0.2in 10.7in 2.95in,width=\textwidth]{{FigureSLIFOnly}.pdf}\\
    \caption{Comparison of MF+v and LIF-only.  Both are trained for 100
      iterations with identical LIF neuron simulation time $t^{\rm LIF}
      = 20$s. {\bf A.} E/I firing rate trajectories from $\B{f}_p$
      (iteration 0-100) and $\B{f}^*_p$ (iteration 101-400), for $S^{EI}
      = 0.0402$.  Left: MF+v exhibits stable estimates for both
      parameter choices.  Right: LIF-only exhibit large oscillations.
      {\bf B.} Same as {\bf A}, but for $S^{EI} = 0.0433$.}

    \label{FigS1:LIF Only}
\end{figure}

We find that all else being equal, the LIF-only algorithm is much less
stable than MF+v.  To demonstrate this, we select two parameters from
Fig.~\ref{fig:flowchart}C ($S^{EI}\in \{0.0433,0.0402\}$) and compare a
simplified version of MF+v and LIF-only.  In these runs, to avoid
uncertainties associated with early termination, we train both
algorithms for $M=100$ iterations, then compute running averages.  For
MF+v, this means that in Algorithm~\ref{Mthd:Algm1}, we set $M=100$ and
$\varepsilon=0$, and always take the first branch after {\tt FINALIZE};
see Algorithm~\ref{Mthd:Algm2} for LIF-only.  Fig.~\ref{FigS1:LIF Only}
shows the results.  In the left panels, we plot the firing rates
$\B{f}_p$ for iterates $p\leq100$ and running averages $\B{f}^*_p$ for
$100<p\leq400$.  As can be seen, firing rates from MF+v
stabilizes quickly to network-computed rates, while LIF-only (right)
sometimes exhibits large oscillations.

%% For illustration purposes, we fix the number of training iterations
%% as 100 before going to the FINALIZE step (step 13 in
%% Algorithms~\ref{Mthd:Algm1} and \ref{Mthd:Algm2}) regardless whether
%% the stopping criterion has been met or not.

A potential explanation for the behavior of LIF-only is that (as we
noted in Methods Sect.~\ref{Mthd-MV}) one can obtain many more samples
of voltages per unit time than spikes.  Since the variance of firing
rate estimates is roughly inversely proportional to the number of
spikes, the single neuron rate estimates in the LIF-only algorithm are
far noisier at typical background firing rates.  We have also tested
other variants of LIF-only, such as averaging over ensembles of pairs of
LIF neurons.  However, the LIF-only method remains rather unstable (data
not shown).  Although LIF-only occasionally gives good predictions of
firing rates, it is far less reliable in comparison to MF+v.

%%%%%%%%%%%%%%%%%%%%%%%%%%%%%%%%%%%%%%%%%%%%%%%%%%%%%%%%%%% FigS2
%%%%%%%%%%%%%%%%%%%%%%%%%%%%%%%%%%%%%%%%%%%%%%%%%%%%%%%%%%% FigS2

%% \newpage
\subsection{Additional firing rate maps}
\label{SI-AddFrMap}

As mentioned in Results, here we show versions of Fig.~\ref{Fig2: LGN}
with different choices of parameters.

%% \newcommand{\AddFrMapsTopLabels}{\hspace{0.4in}{\scalebox{0.6}{$\mathbf{S^{Elgn}/S^{EE} = 1.5}$}}\hspace{0.45in}{\scalebox{0.6}{$\mathbf{S^{Elgn}/S^{EE} = 2.0}$}}\hspace{0.45in}{\scalebox{0.6}{$\mathbf{S^{Elgn}/S^{EE} = 2.5}$}}\hspace{0.45in}{\scalebox{0.6}{$\mathbf{S^{Elgn}/S^{EE} = 3.0}$}}}
%% \newcommand{\AddFrMapsLeftLabels}{\rotatebox{90}{\hspace{0.25in}{\scalebox{0.6}{$\mathbf{S^{Ilgn}/S^{Elgn}=2.5}$\hspace{0.7in}$\mathbf{S^{Ilgn}/S^{Elgn}=2.0}$\hspace{0.7in}$\mathbf{S^{Ilgn}/S^{Elgn}=1.5}$}}}}

%% \addtolength{\tabcolsep}{-6pt}

%%Below We start Group 1 of Rate Mapps: SEE = 0.021
%%%%%%%%%%%%%%%%%%%%%%%%%%%%%%%%%%%%%%%%%%%%%%%%%%%%%%%%%%% FigSAdd1-1
\begin{figure}[h!]
  \begin{center}
    %% \captionsetup{type=figure} 
    %% \resizebox{5.5.3in}{!}{\includegraphics[bb=0in 0in 14.94in 10.3in]{{FigureS3_SEE0.021_SII0.12_rIL6=750}.pdf}}
    \begin{center}
      \resizebox{6in}{!}{\includegraphics{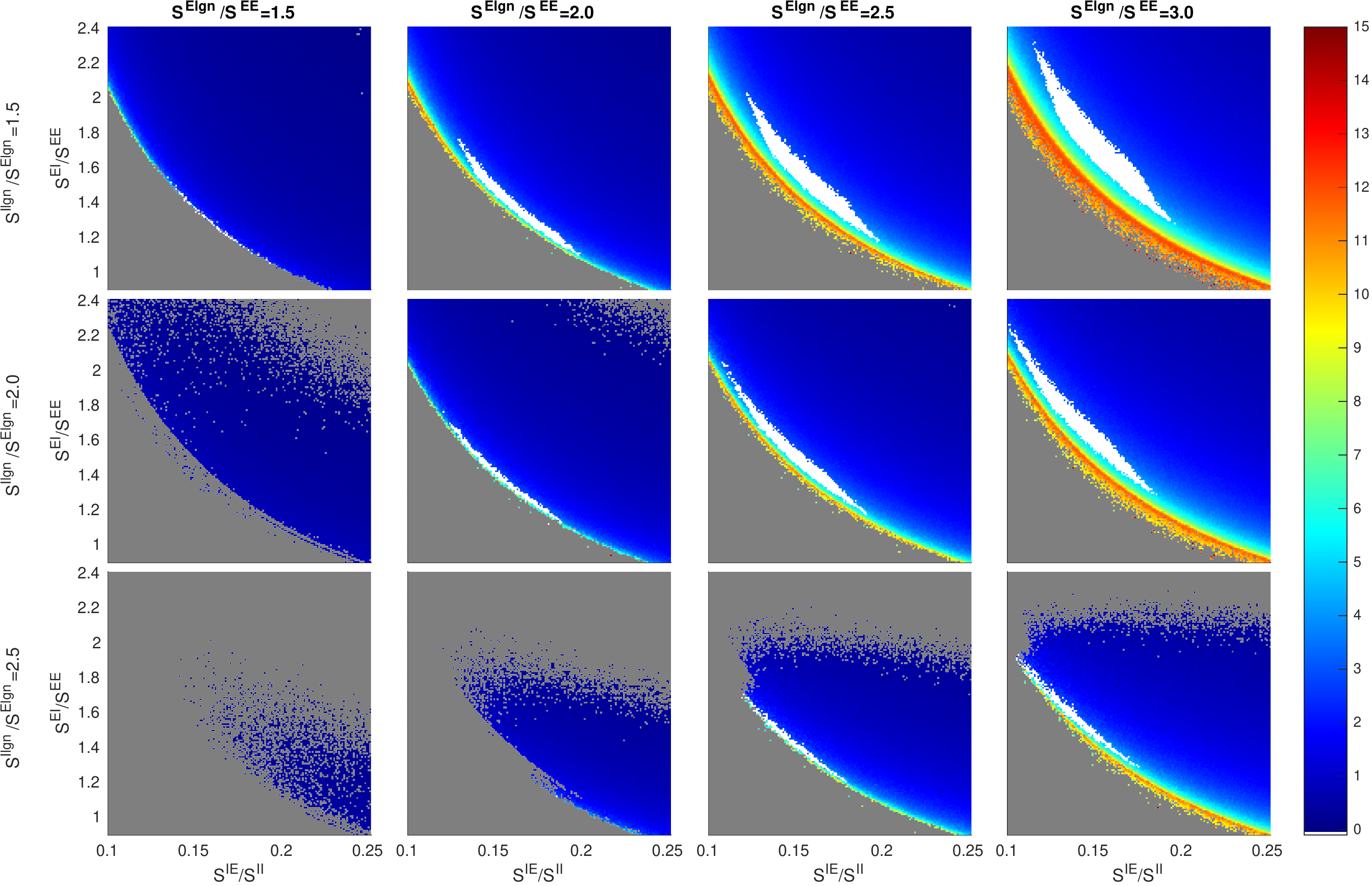}}
    \end{center}
    \caption{A version of Fig.~\ref{Fig2: LGN} with $S^{EE} = 0.021$, $S^{II} = 0.12$, and $F^{I\rm L6}/F^{E\rm L6} = 3$.}
    \label{FigSAdd1-1}
  \end{center}
\end{figure}
\vfill
%%%%%%%%%%%%%%%%%%%%%%%%%%%%%%%%%%%%%%%%%%%%%%%%%%%%%%%%%%% FigSAdd1-1
%%%%%%%%%%%%%%%%%%%%%%%%%%%%%%%%%%%%%%%%%%%%%%%%%%%%%%%%%%% FigSAdd1-2
\begin{figure}[h!]
  \begin{center}
    %% \captionsetup{type=figure} 
    \begin{center}
      \resizebox{6in}{!}{\includegraphics{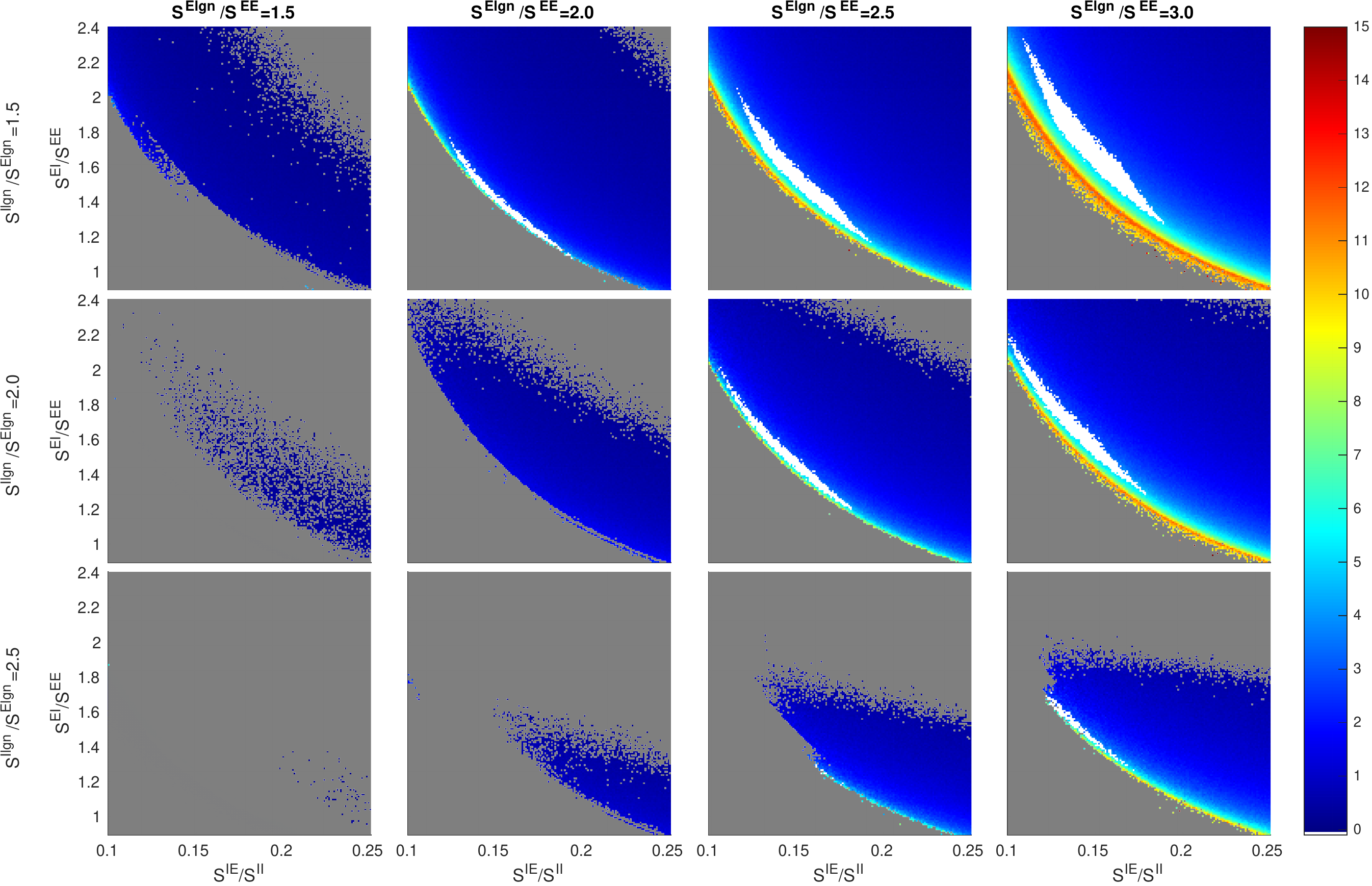}}
    \end{center}
    \caption{A version of Fig.~\ref{Fig2: LGN} with $S^{EE} = 0.021$, $S^{II} = 0.12$, and $F^{I\rm L6}/F^{E\rm L6} = 4.5$.}
    \label{FigSAdd1-2}
  \end{center}
\end{figure}
\vfill
%%%%%%%%%%%%%%%%%%%%%%%%%%%%%%%%%%%%%%%%%%%%%%%%%%%%%%%%%%% FigSAdd1-2
%%%%%%%%%%%%%%%%%%%%%%%%%%%%%%%%%%%%%%%%%%%%%%%%%%%%%%%%%%% FigSAdd1-3
\begin{figure}[h!]%[htbp]
  \begin{center}
    %% \captionsetup{type=figure} 
    \begin{center}
      \resizebox{6in}{!}{\includegraphics{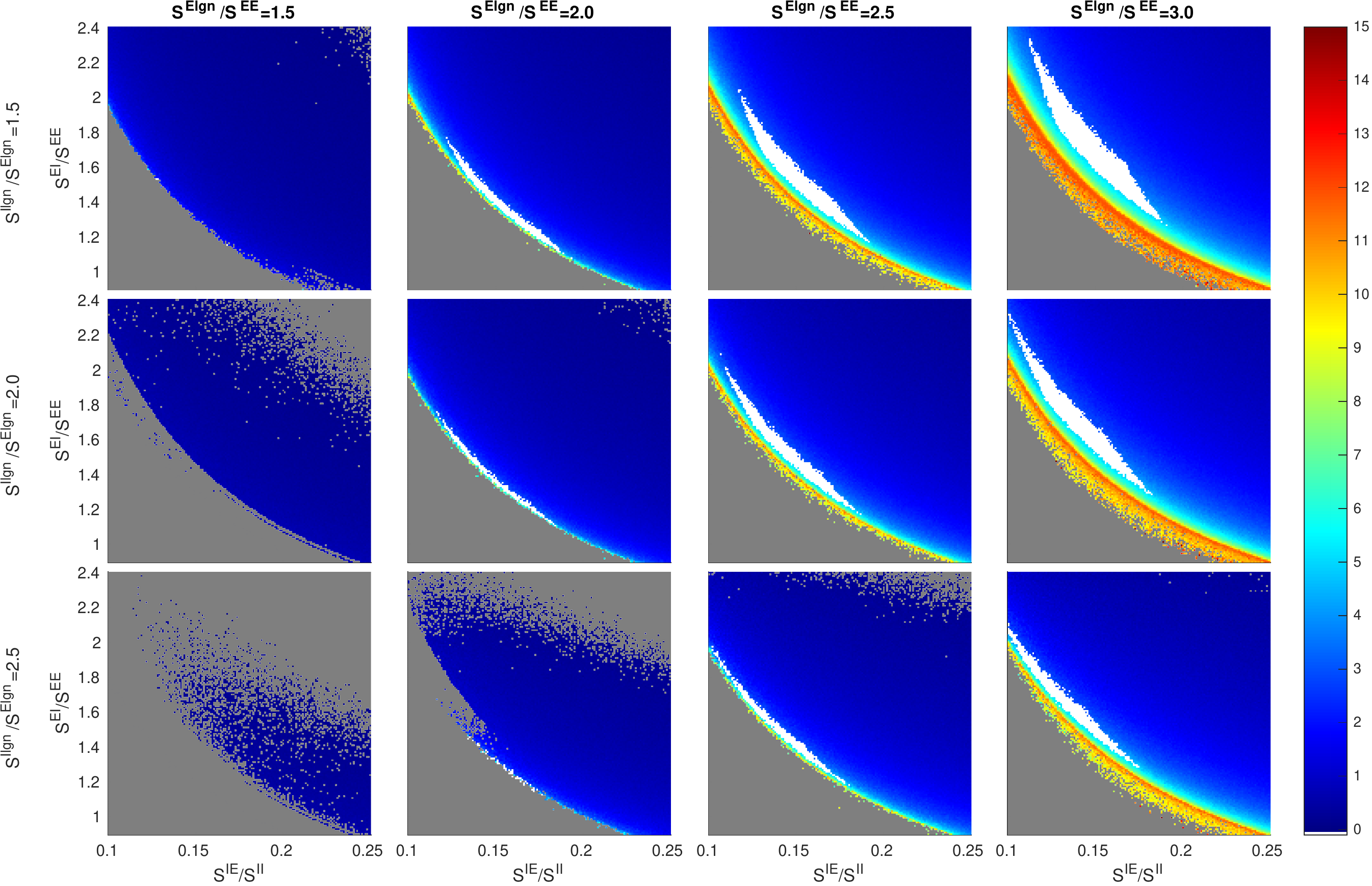}}
    \end{center}
    \caption{A version of Fig.~\ref{Fig2: LGN} with $S^{EE} = 0.021$, $S^{II} = 0.16$, and $F^{I\rm L6}/F^{E\rm L6} = 3$.}
    \label{FigSAdd1-3}
  \end{center}
\end{figure}
\vfill
%%%%%%%%%%%%%%%%%%%%%%%%%%%%%%%%%%%%%%%%%%%%%%%%%%%%%%%%%%% FigSAdd1-3
%%%%%%%%%%%%%%%%%%%%%%%%%%%%%%%%%%%%%%%%%%%%%%%%%%%%%%%%%%% FigSAdd1-4
\begin{figure}[h!]%[htbp]
  \begin{center}
    %% \captionsetup{type=figure} 
    \begin{center}
      \resizebox{6in}{!}{\includegraphics{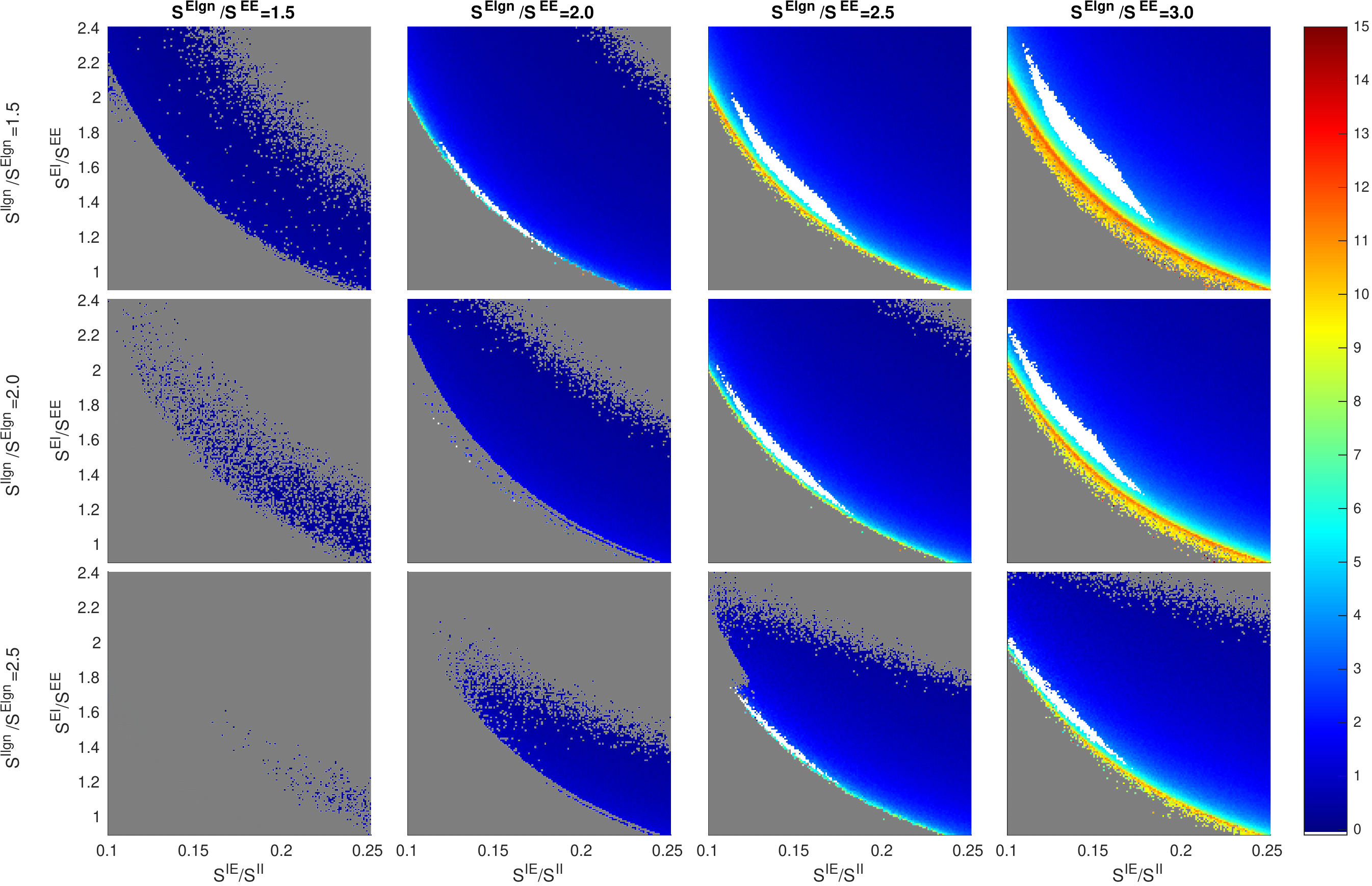}}
    \end{center}
    \caption{A version of Fig.~\ref{Fig2: LGN} with $S^{EE} = 0.021$, $S^{II} = 0.16$, and $F^{I\rm L6}/F^{E\rm L6} = 4.5$.}
    \label{FigSAdd1-4}
  \end{center}
\end{figure}
\vfill
%%%%%%%%%%%%%%%%%%%%%%%%%%%%%%%%%%%%%%%%%%%%%%%%%%%%%%%%%%% FigSAdd1-4
%%%%%%%%%%%%%%%%%%%%%%%%%%%%%%%%%%%%%%%%%%%%%%%%%%%%%%%%%%% FigSAdd1-5
\begin{figure}[h!]%[htbp]
  \begin{center}
    %% \captionsetup{type=figure} 
    \begin{center}
      \resizebox{6in}{!}{\includegraphics{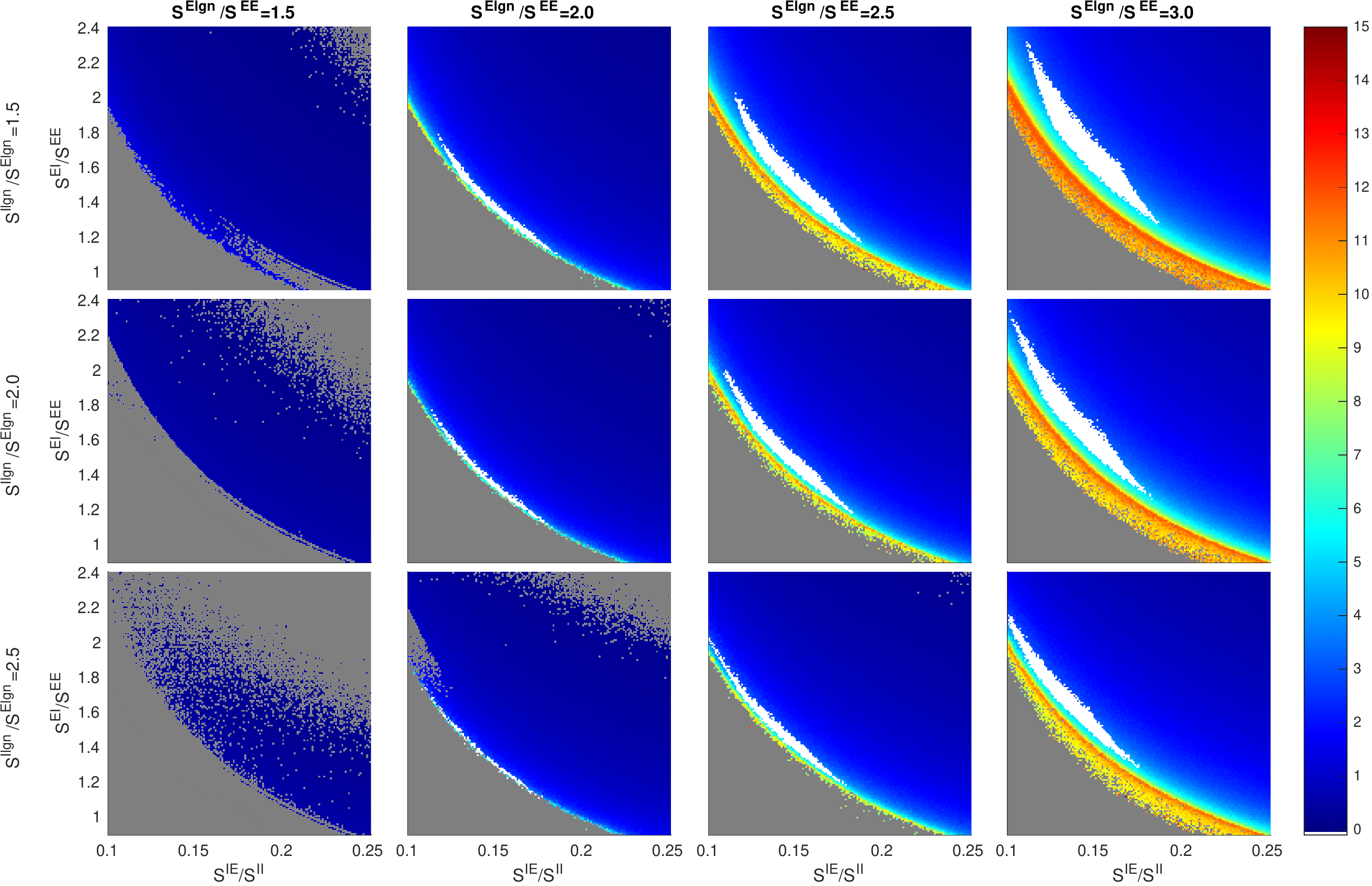}}
    \end{center}
    \caption{A version of Fig.~\ref{Fig2: LGN} with $S^{EE} = 0.021$, $S^{II} = 0.20$, and $F^{I\rm L6}/F^{E\rm L6} = 3$.}
    \label{FigSAdd1-5}
  \end{center}
\end{figure}
\vfill
%%%%%%%%%%%%%%%%%%%%%%%%%%%%%%%%%%%%%%%%%%%%%%%%%%%%%%%%%%% FigSAdd1-5
%%%%%%%%%%%%%%%%%%%%%%%%%%%%%%%%%%%%%%%%%%%%%%%%%%%%%%%%%%% FigSAdd1-6
\begin{figure}[h!]%[htbp]
  \begin{center}
    %% \captionsetup{type=figure} 
    \begin{center}
      \resizebox{6in}{!}{\includegraphics{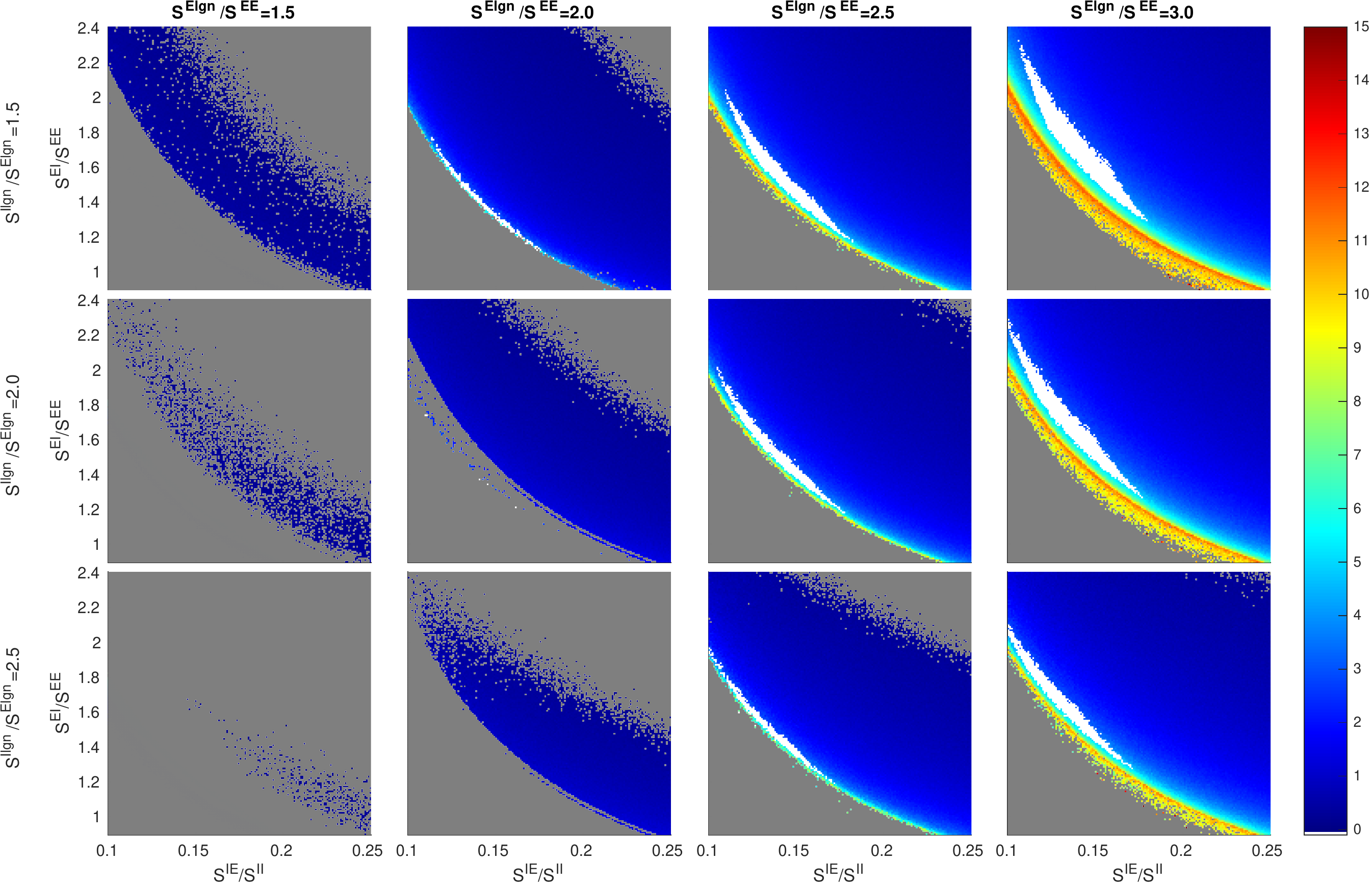}}
    \end{center}
    \caption{A version of Fig.~\ref{Fig2: LGN} with $S^{EE} = 0.021$, $S^{II} = 0.20$, and $F^{I\rm L6}/F^{E\rm L6} = 4.5$.}
    \label{FigSAdd1-6}
  \end{center}
\end{figure}
\vfill
%%%%%%%%%%%%%%%%%%%%%%%%%%%%%%%%%%%%%%%%%%%%%%%%%%%%%%%%%%% FigSAdd1-6

%%Below We start Group 2 of Rate Mapps: SEE = 0.024
%%%%%%%%%%%%%%%%%%%%%%%%%%%%%%%%%%%%%%%%%%%%%%%%%%%%%%%%%%% FigSAdd2-1
\begin{figure}[h!]%[htbp]
  \begin{center}
    %% \captionsetup{type=figure} 
    \begin{center}
      \resizebox{6in}{!}{\includegraphics{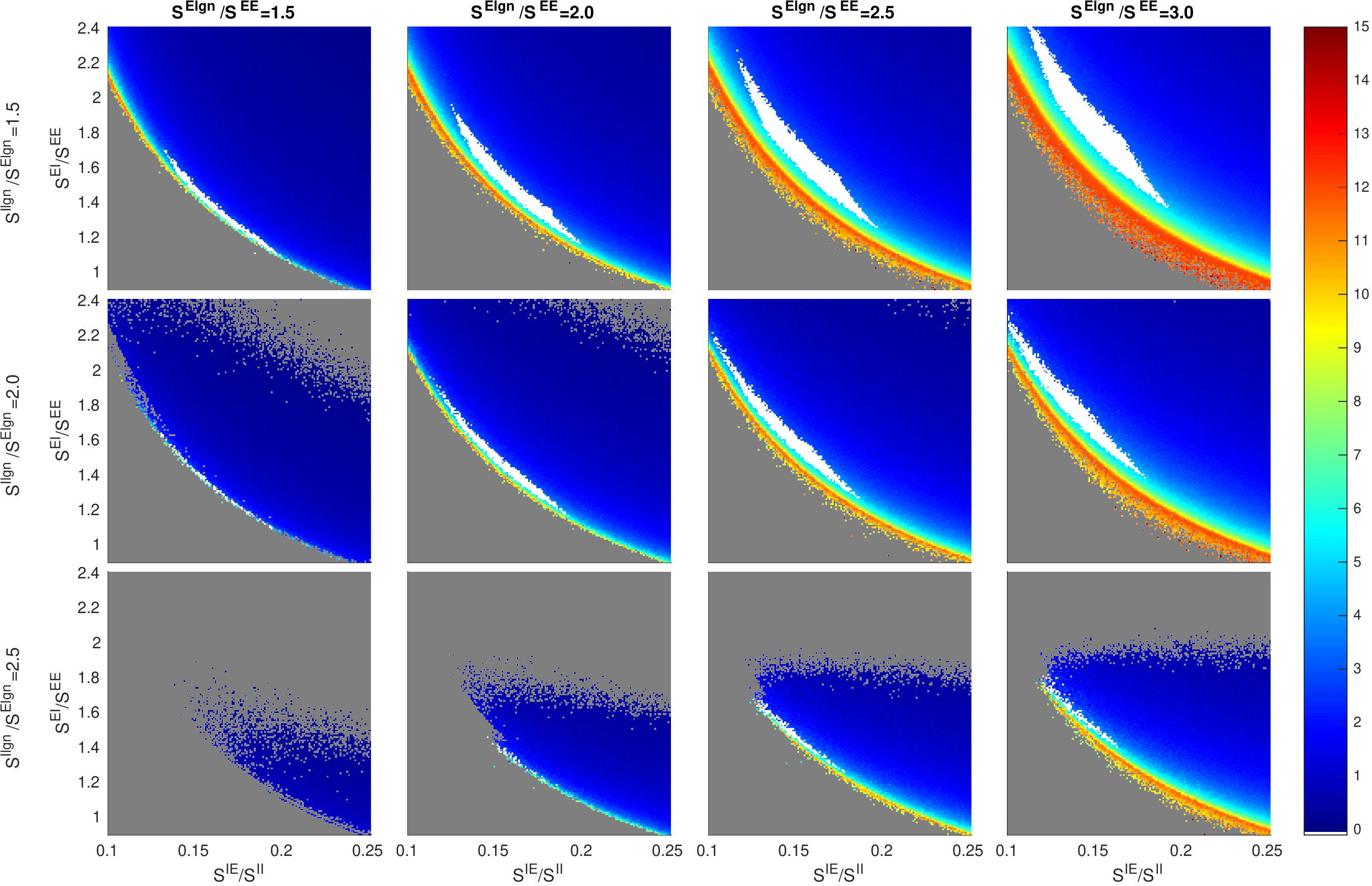}}
    \end{center}
    \caption{A version of Fig.~\ref{Fig2: LGN} with $S^{EE} = 0.024$, $S^{II} = 0.12$, and $F^{I\rm L6}/F^{E\rm L6} = 3$.}
    \label{FigSAdd2-1}
  \end{center}
\end{figure}
\vfill
%%%%%%%%%%%%%%%%%%%%%%%%%%%%%%%%%%%%%%%%%%%%%%%%%%%%%%%%%%% FigSAdd2-1
%%%%%%%%%%%%%%%%%%%%%%%%%%%%%%%%%%%%%%%%%%%%%%%%%%%%%%%%%%% FigSAdd2-2
\begin{figure}[h!]%[htbp]
  \begin{center}
    %% \captionsetup{type=figure} 
    \begin{center}
      \resizebox{6in}{!}{\includegraphics{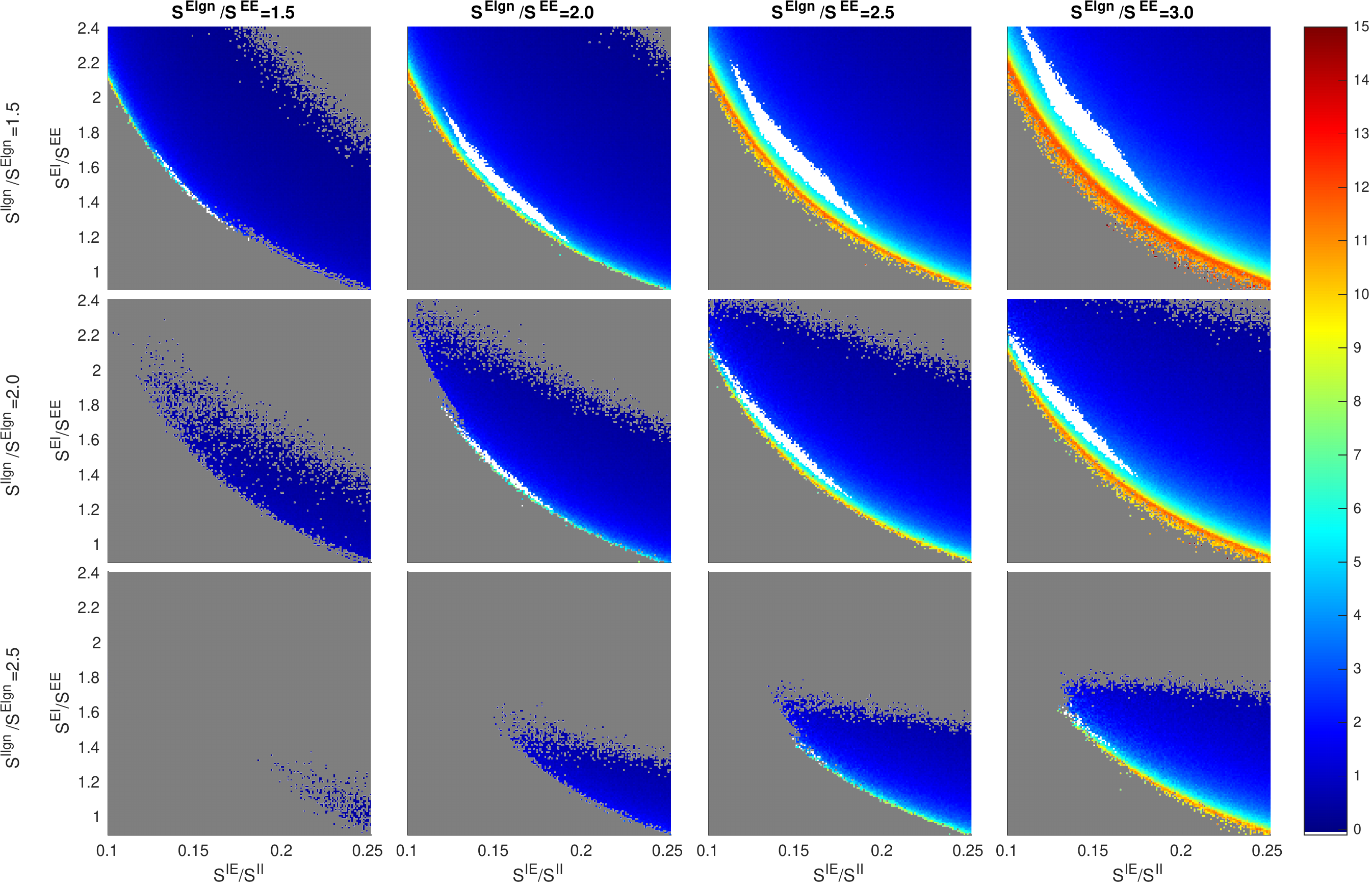}}
    \end{center}
    \caption{A version of Fig.~\ref{Fig2: LGN} with $S^{EE} = 0.024$, $S^{II} = 0.12$, and $F^{I\rm L6}/F^{E\rm L6} = 4.5$.}
    \label{FigSAdd2-2}
  \end{center}
\end{figure}
\vfill
%%%%%%%%%%%%%%%%%%%%%%%%%%%%%%%%%%%%%%%%%%%%%%%%%%%%%%%%%%% FigSAdd2-2
%%%%%%%%%%%%%%%%%%%%%%%%%%%%%%%%%%%%%%%%%%%%%%%%%%%%%%%%%%% FigSAdd2-3
\begin{figure}[h!]%[htbp]
  \begin{center}
    %% \captionsetup{type=figure} 
    \begin{center}
      \resizebox{6in}{!}{\includegraphics{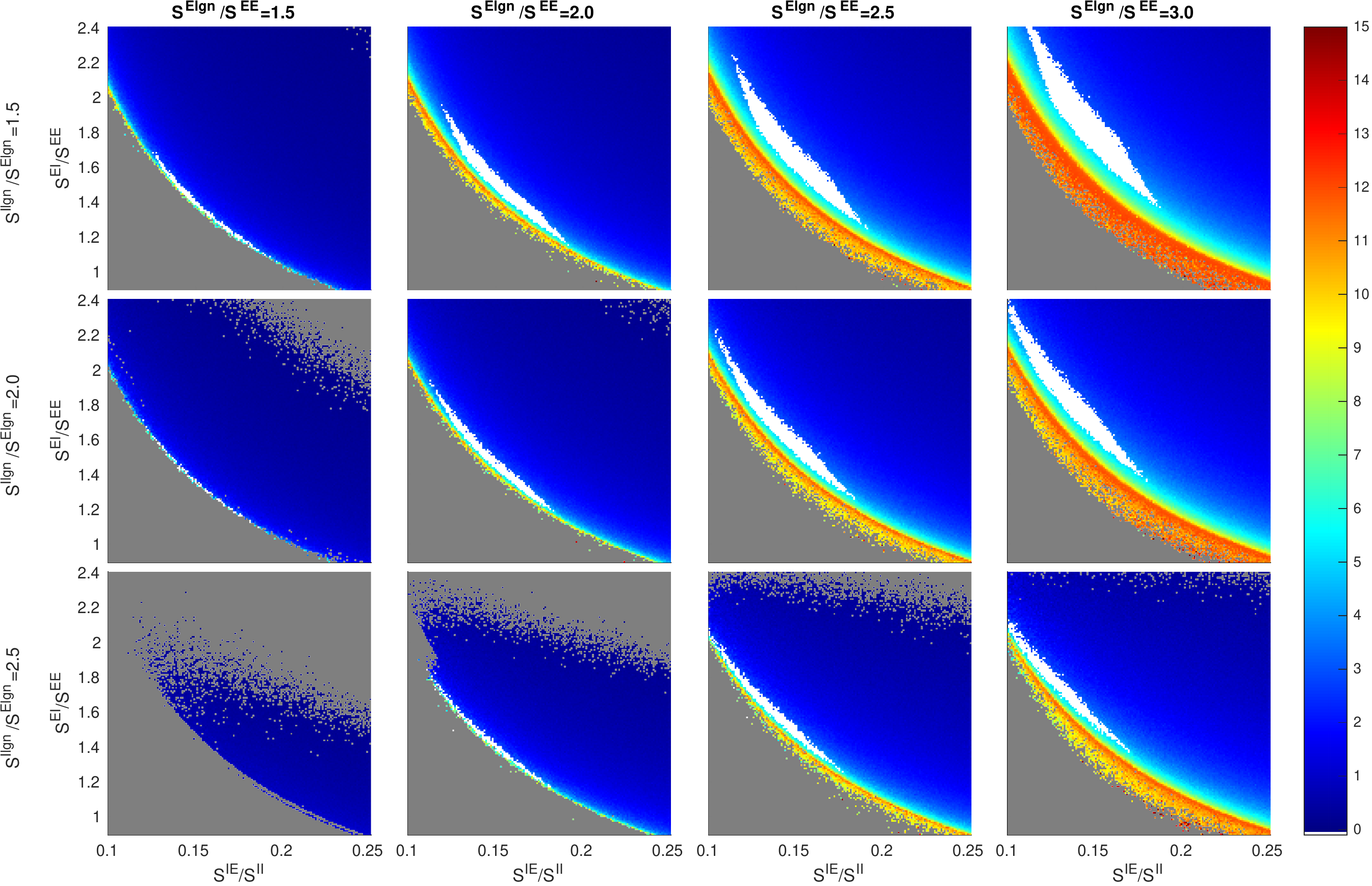}}
    \end{center}
    \caption{A version of Fig.~\ref{Fig2: LGN} with $S^{EE} = 0.024$, $S^{II} = 0.16$, and $F^{I\rm L6}/F^{E\rm L6} = 3$.}
    \label{FigSAdd2-3}
  \end{center}
\end{figure}
\vfill
%%%%%%%%%%%%%%%%%%%%%%%%%%%%%%%%%%%%%%%%%%%%%%%%%%%%%%%%%%% FigSAdd2-3
%%%%%%%%%%%%%%%%%%%%%%%%%%%%%%%%%%%%%%%%%%%%%%%%%%%%%%%%%%% FigSAdd2-4
\begin{figure}[h!]%[htbp]
  \begin{center}
    %% \captionsetup{type=figure} 
    \begin{center}
      \resizebox{6in}{!}{\includegraphics{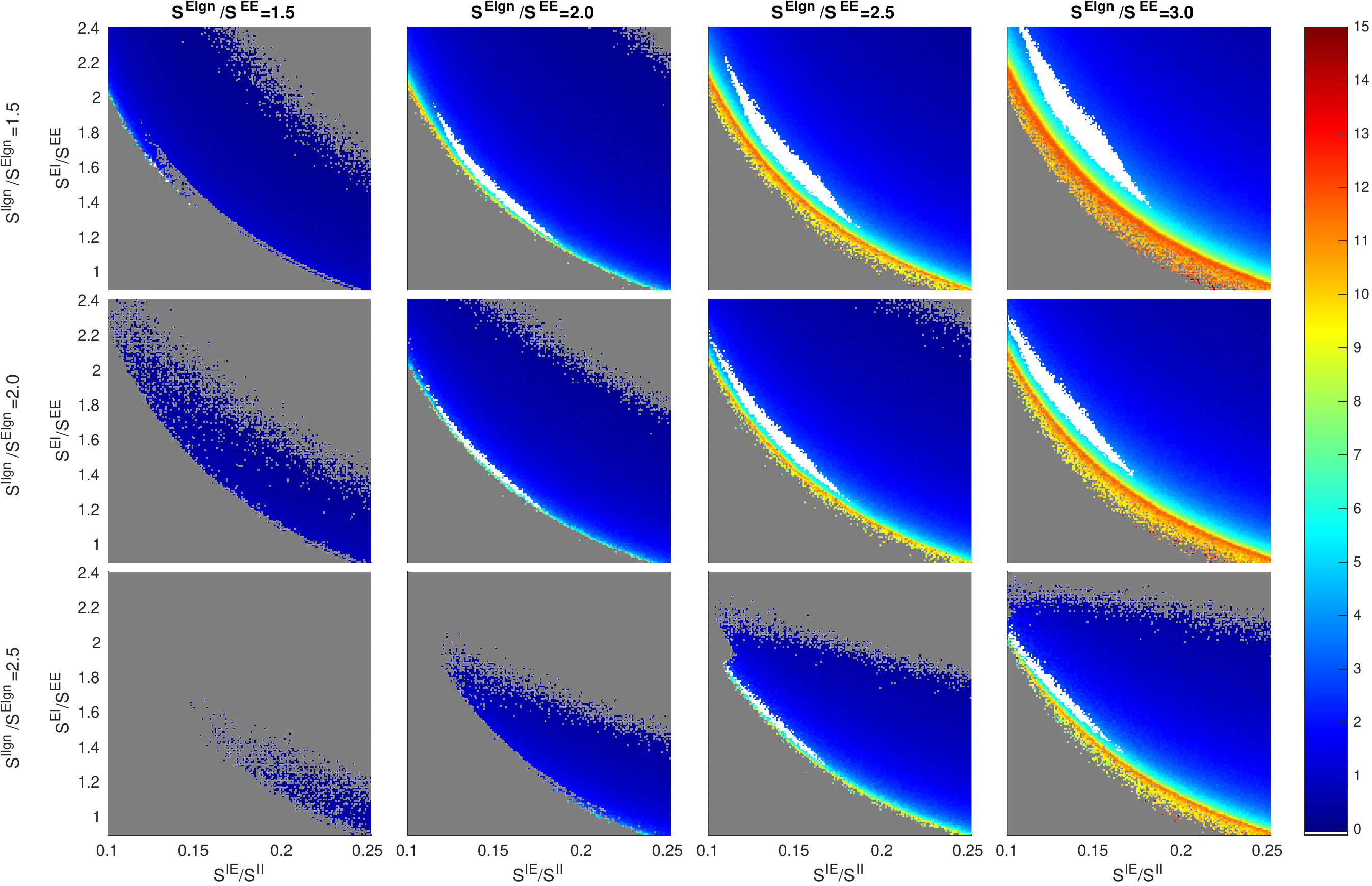}}
    \end{center}
    \caption{A version of Fig.~\ref{Fig2: LGN} with $S^{EE} = 0.024$, $S^{II} = 0.16$, and $F^{I\rm L6}/F^{E\rm L6} = 4.5$.}
    \label{FigSAdd2-4}
  \end{center}
\end{figure}
\vfill
%%%%%%%%%%%%%%%%%%%%%%%%%%%%%%%%%%%%%%%%%%%%%%%%%%%%%%%%%%% FigSAdd2-4
%%%%%%%%%%%%%%%%%%%%%%%%%%%%%%%%%%%%%%%%%%%%%%%%%%%%%%%%%%% FigSAdd2-5
\begin{figure}[h!]%[htbp]
  \begin{center}
    %% \captionsetup{type=figure} 
    \begin{center}
      \resizebox{6in}{!}{\includegraphics{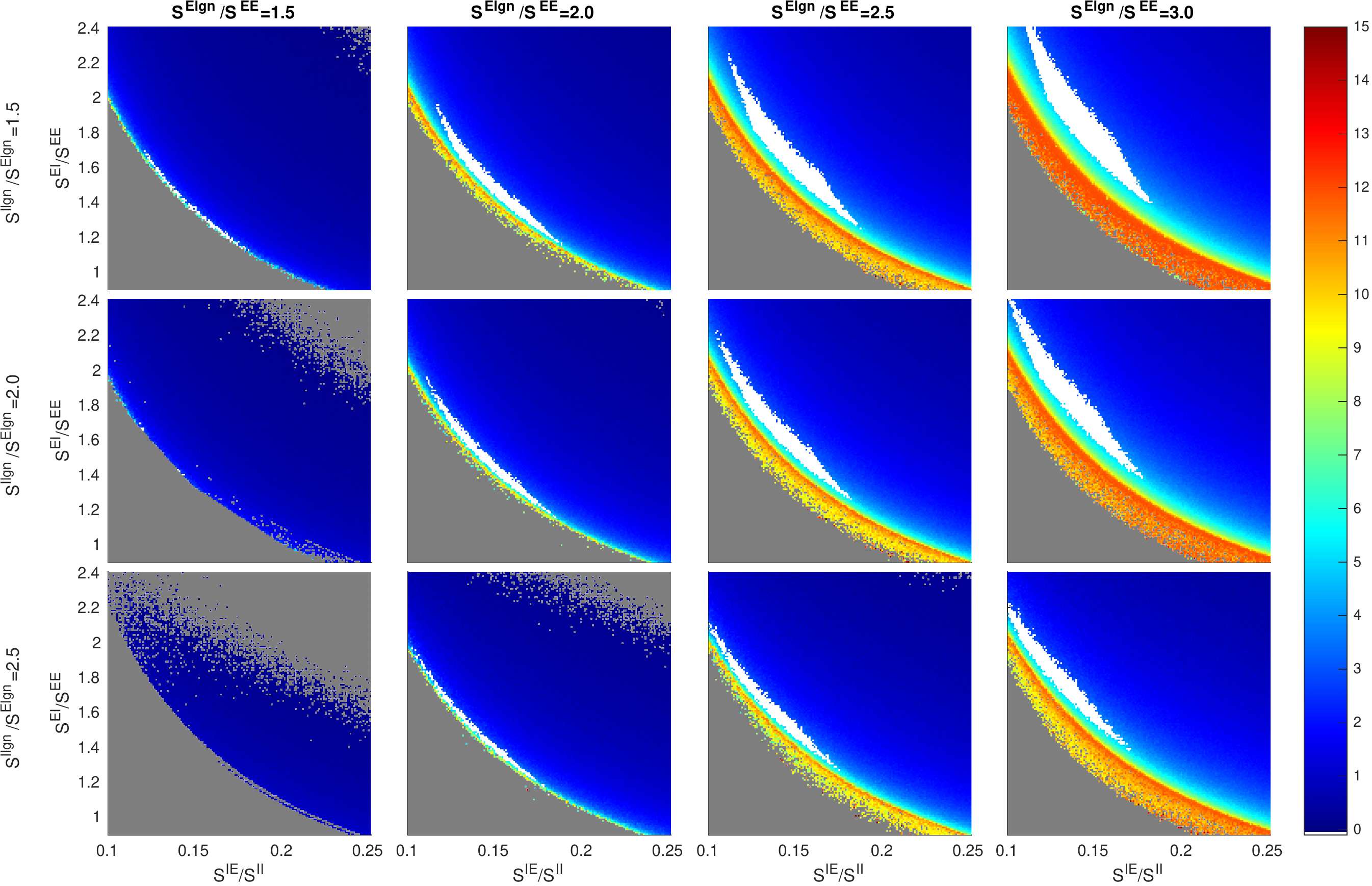}}
    \end{center}
    \caption{A version of Fig.~\ref{Fig2: LGN} with $S^{EE} = 0.024$, $S^{II} = 0.20$, and $F^{I\rm L6}/F^{E\rm L6} = 3$.}
    \label{FigSAdd2-5}
  \end{center}
\end{figure}
\vfill
%%%%%%%%%%%%%%%%%%%%%%%%%%%%%%%%%%%%%%%%%%%%%%%%%%%%%%%%%%% FigSAdd2-5
%%%%%%%%%%%%%%%%%%%%%%%%%%%%%%%%%%%%%%%%%%%%%%%%%%%%%%%%%%% FigSAdd2-6
\begin{figure}[h!]%[htbp]
  \begin{center}
    %% \captionsetup{type=figure} 
    \begin{center}
      \resizebox{6in}{!}{\includegraphics{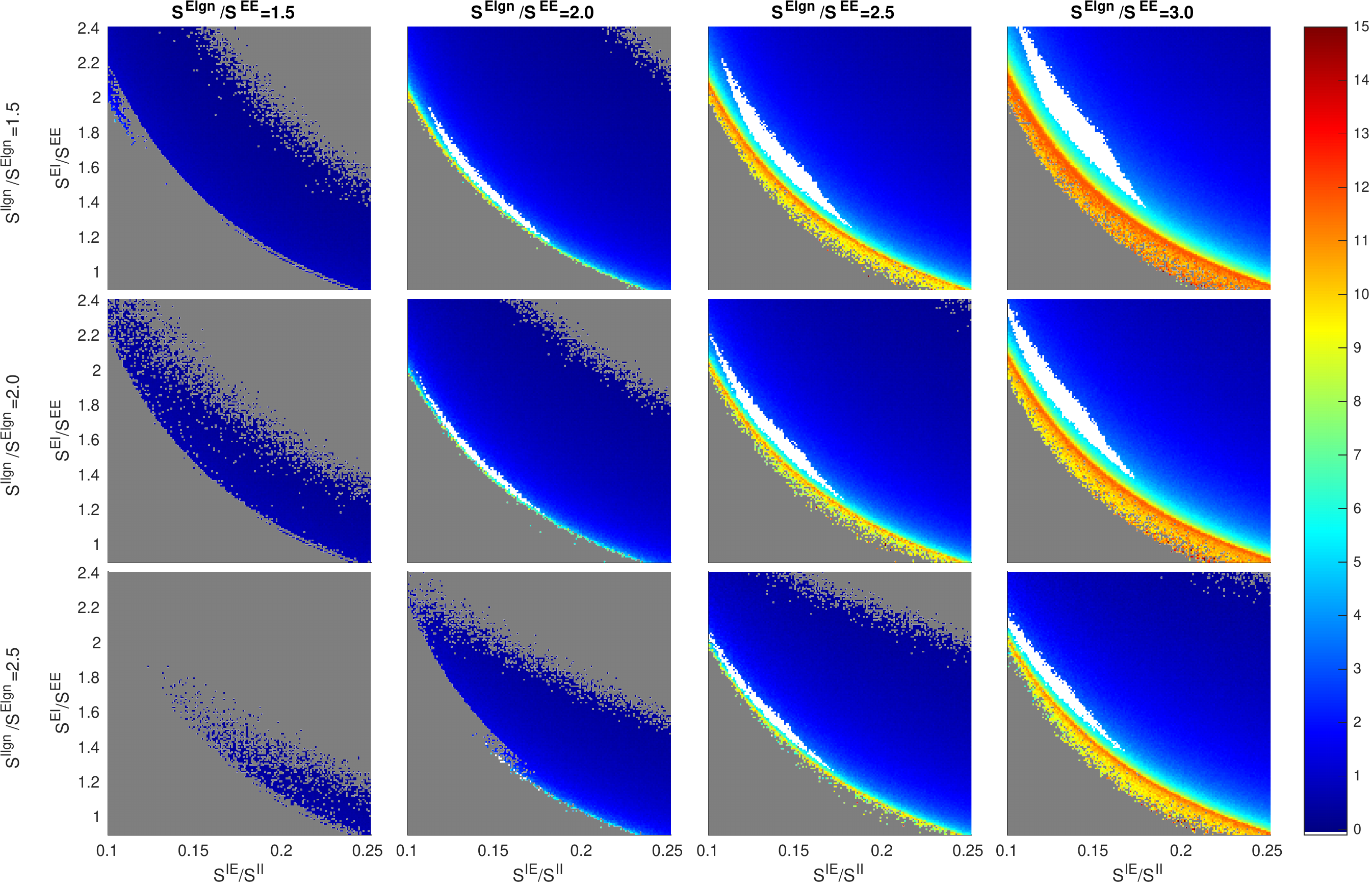}}
    \end{center}
    \caption{A version of Fig.~\ref{Fig2: LGN} with $S^{EE} = 0.024$, $S^{II} = 0.20$, and $F^{I\rm L6}/F^{E\rm L6} = 4.5$.}
    \label{FigSAdd2-6}
  \end{center}
\end{figure}
\vfill
%%%%%%%%%%%%%%%%%%%%%%%%%%%%%%%%%%%%%%%%%%%%%%%%%%%%%%%%%%% FigSAdd2-6

%%Below We start Group 3 of Rate Mapps: SEE = 0.027
%%%%%%%%%%%%%%%%%%%%%%%%%%%%%%%%%%%%%%%%%%%%%%%%%%%%%%%%%%% FigSAdd3-1
\begin{figure}[h!]%[htbp]
  \begin{center}
    %% \captionsetup{type=figure} 
    \begin{center}
      \resizebox{6in}{!}{\includegraphics{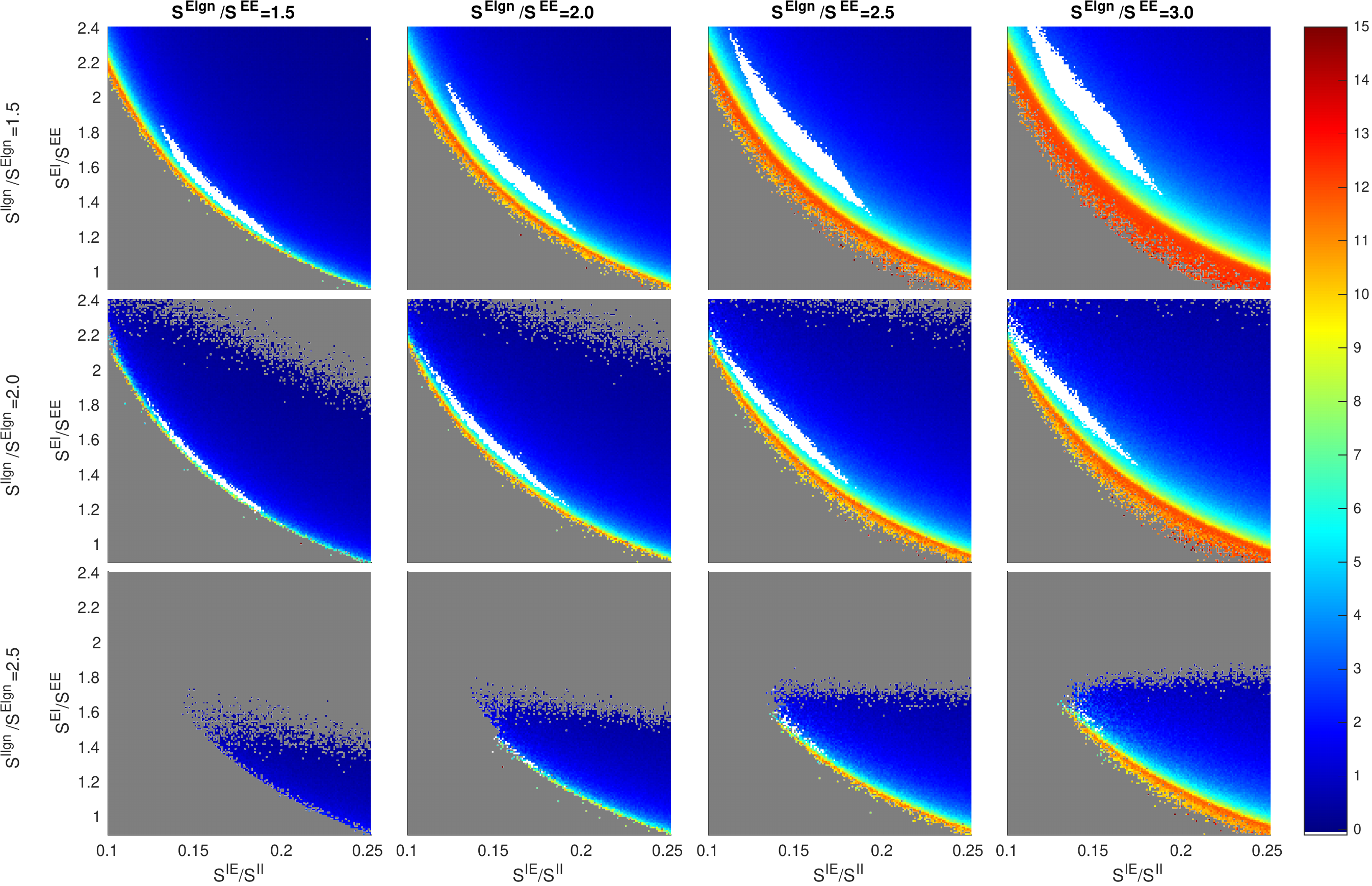}}
    \end{center}
    \caption{A version of Fig.~\ref{Fig2: LGN} with $S^{EE} = 0.027$, $S^{II} = 0.12$, and $F^{I\rm L6}/F^{E\rm L6} = 3$.}
    \label{FigSAdd3-1}
  \end{center}
\end{figure}
\vfill
%%%%%%%%%%%%%%%%%%%%%%%%%%%%%%%%%%%%%%%%%%%%%%%%%%%%%%%%%%% FigSAdd3-1
%%%%%%%%%%%%%%%%%%%%%%%%%%%%%%%%%%%%%%%%%%%%%%%%%%%%%%%%%%% FigSAdd3-2
\begin{figure}[h!]%[htbp]
  \begin{center}
    %% \captionsetup{type=figure} 
    \begin{center}
      \resizebox{6in}{!}{\includegraphics{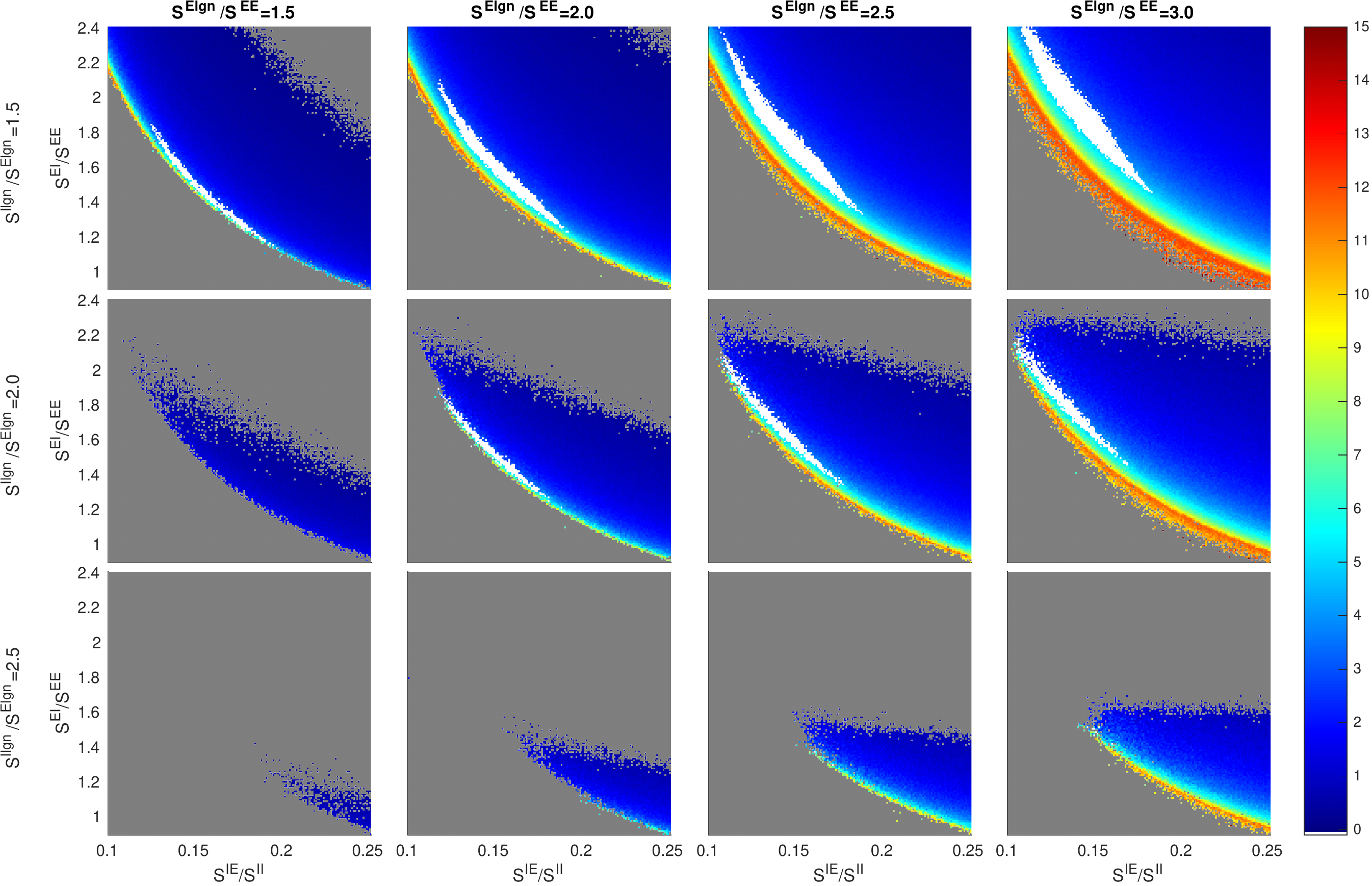}}
    \end{center}
    \caption{A version of Fig.~\ref{Fig2: LGN} with $S^{EE} = 0.027$, $S^{II} = 0.12$, and $F^{I\rm L6}/F^{E\rm L6} = 4.5$.}
    \label{FigSAdd3-2}
  \end{center}
\end{figure}
\vfill
%%%%%%%%%%%%%%%%%%%%%%%%%%%%%%%%%%%%%%%%%%%%%%%%%%%%%%%%%%% FigSAdd3-2
%%%%%%%%%%%%%%%%%%%%%%%%%%%%%%%%%%%%%%%%%%%%%%%%%%%%%%%%%%% FigSAdd3-3
\begin{figure}[h!]%[htbp]
  \begin{center}
    %% \captionsetup{type=figure} 
    \begin{center}
      \resizebox{6in}{!}{\includegraphics{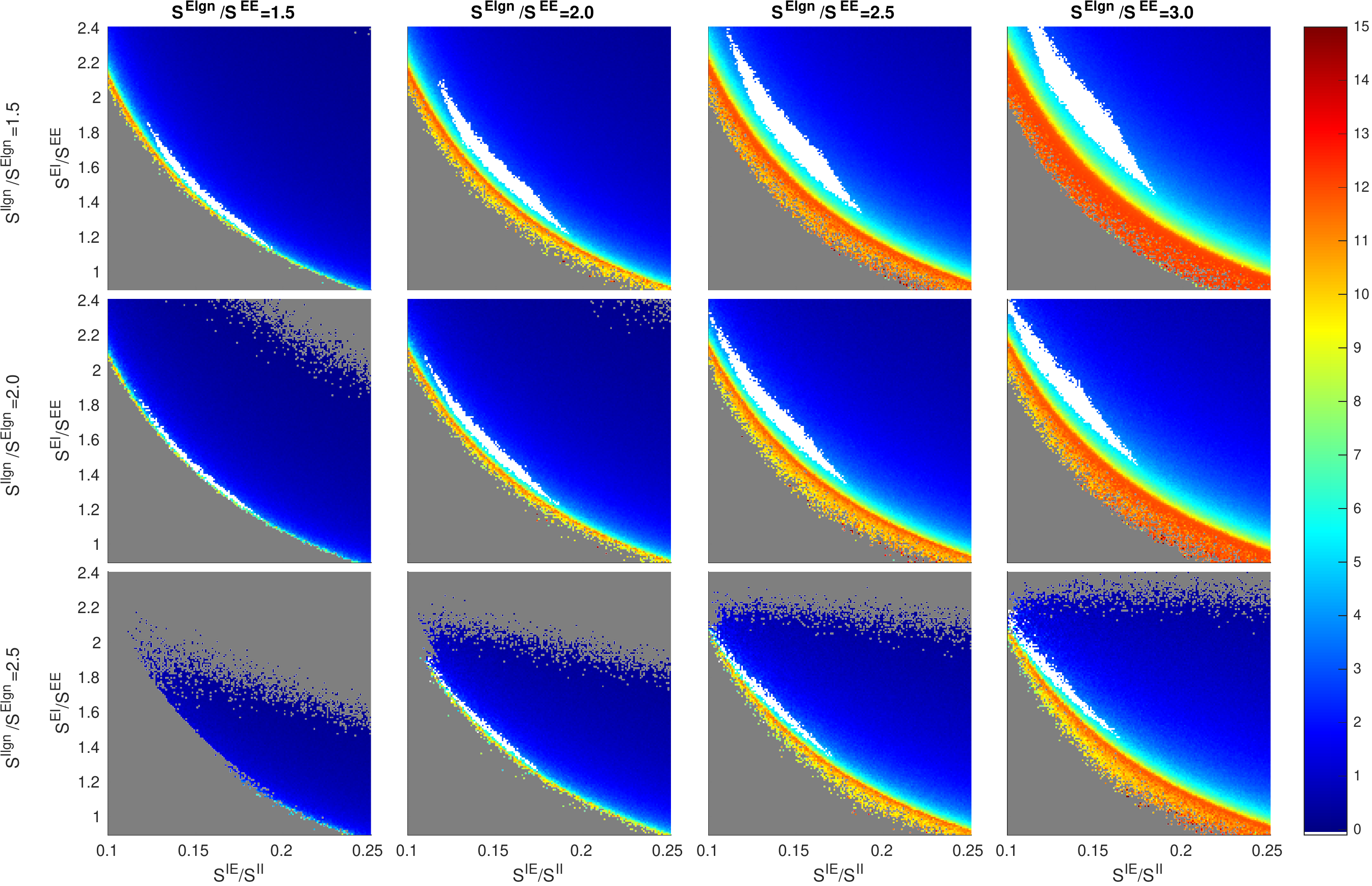}}
    \end{center}
    \caption{A version of Fig.~\ref{Fig2: LGN} with $S^{EE} = 0.027$, $S^{II} = 0.16$, and $F^{I\rm L6}/F^{E\rm L6} = 3$.}
    \label{FigSAdd3-3}
  \end{center}
\end{figure}
\vfill
%%%%%%%%%%%%%%%%%%%%%%%%%%%%%%%%%%%%%%%%%%%%%%%%%%%%%%%%%%% FigSAdd3-3
%%%%%%%%%%%%%%%%%%%%%%%%%%%%%%%%%%%%%%%%%%%%%%%%%%%%%%%%%%% FigSAdd3-4
\begin{figure}[h!]%[htbp]
  \begin{center}
    %% \captionsetup{type=figure} 
    \begin{center}
      \resizebox{6in}{!}{\includegraphics{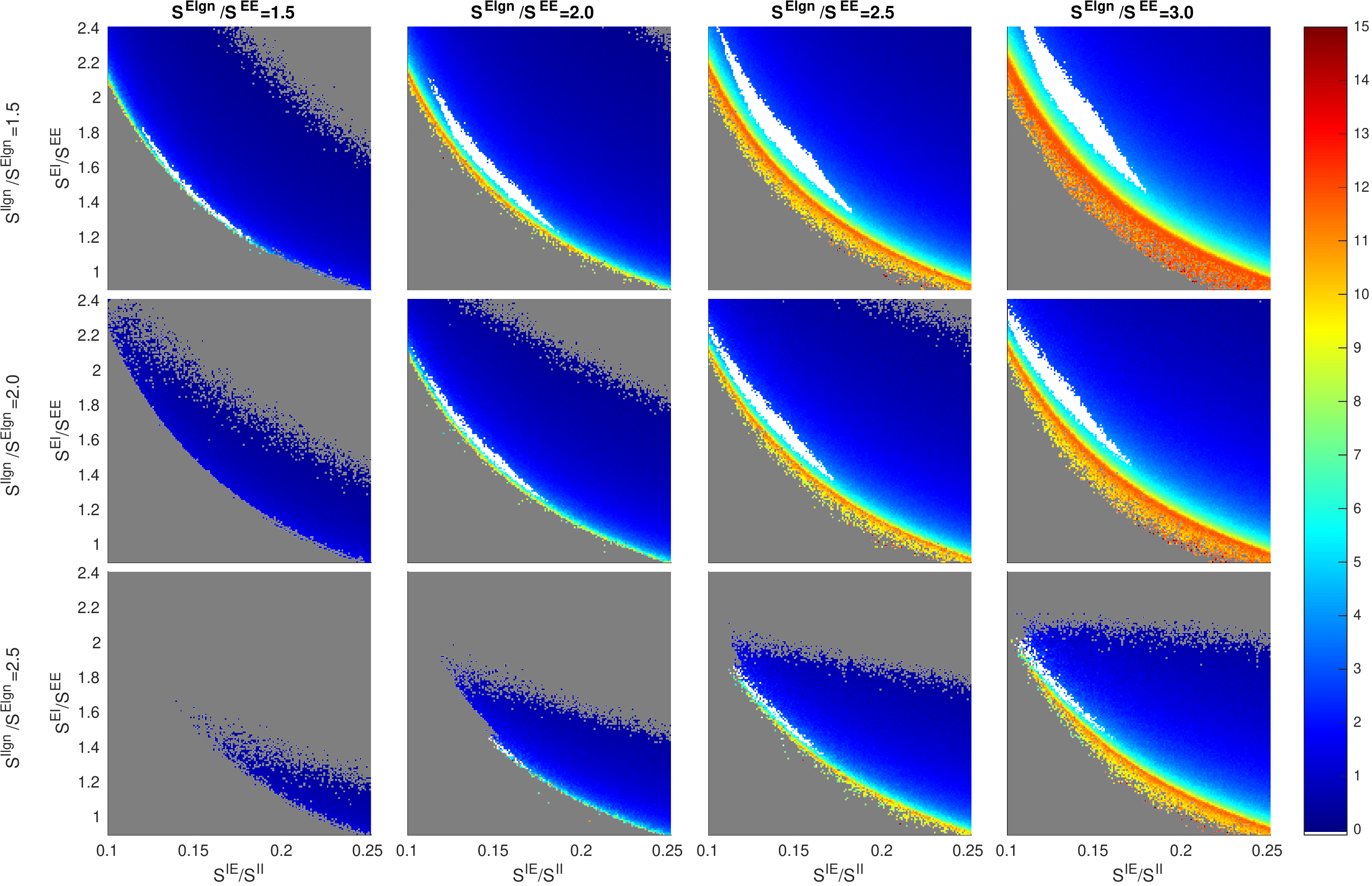}}
    \end{center}
    \caption{A version of Fig.~\ref{Fig2: LGN} with $S^{EE} = 0.027$, $S^{II} = 0.16$, and $F^{I\rm L6}/F^{E\rm L6} = 4.5$.}
    \label{FigSAdd3-4}
  \end{center}
\end{figure}
\vfill
%%%%%%%%%%%%%%%%%%%%%%%%%%%%%%%%%%%%%%%%%%%%%%%%%%%%%%%%%%% FigSAdd3-4
%%%%%%%%%%%%%%%%%%%%%%%%%%%%%%%%%%%%%%%%%%%%%%%%%%%%%%%%%%% FigSAdd3-5
\begin{figure}[h!]%[htbp]
  \begin{center}
    %% \captionsetup{type=figure} 
    \begin{center}
      \resizebox{6in}{!}{\includegraphics{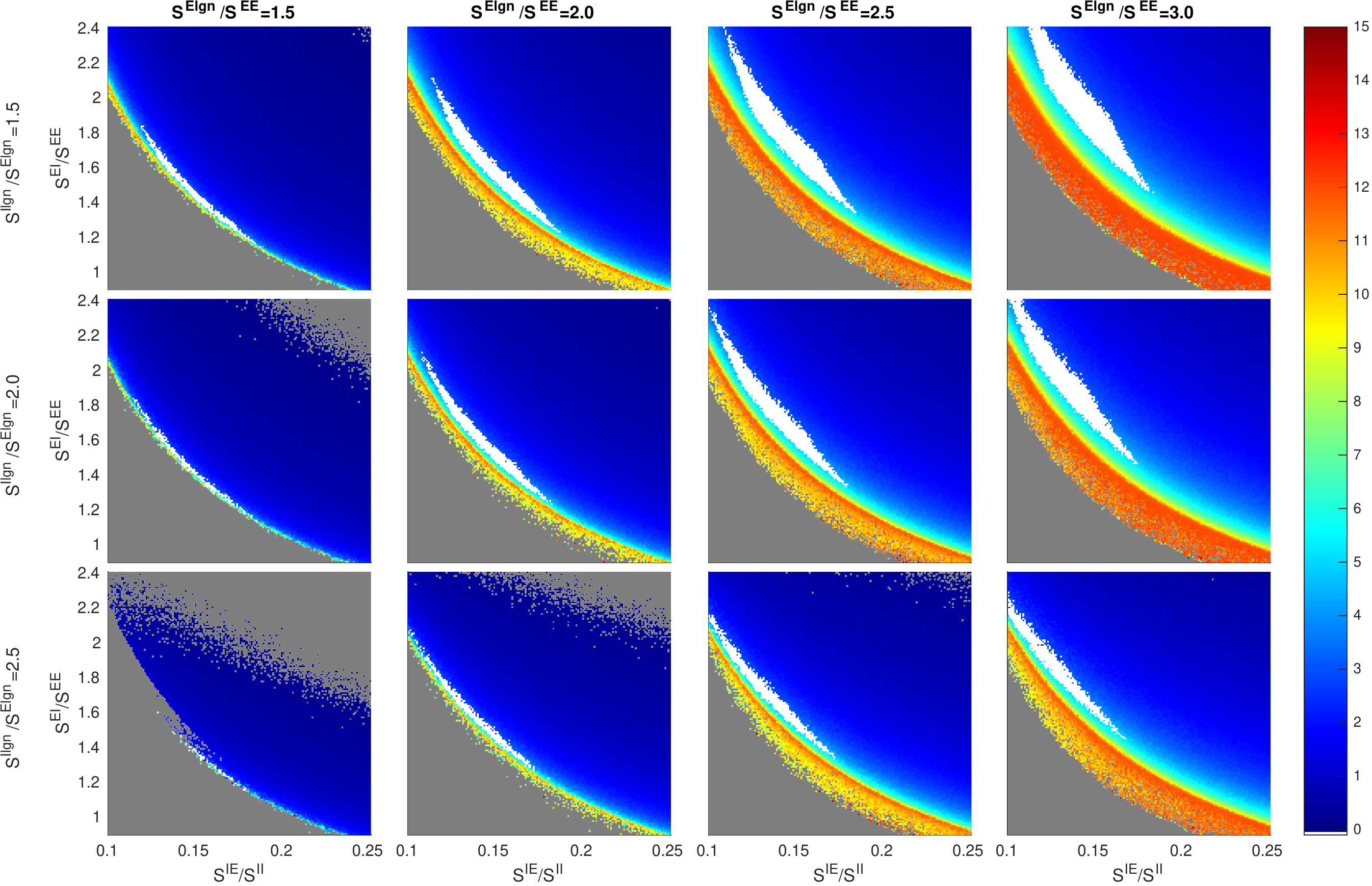}}
    \end{center}
    \caption{A version of Fig.~\ref{Fig2: LGN} with $S^{EE} = 0.027$, $S^{II} = 0.20$, and $F^{I\rm L6}/F^{E\rm L6} = 3$.}
    \label{FigSAdd3-5}
  \end{center}
\end{figure}
\vfill
%%%%%%%%%%%%%%%%%%%%%%%%%%%%%%%%%%%%%%%%%%%%%%%%%%%%%%%%%%% FigSAdd3-5
%%%%%%%%%%%%%%%%%%%%%%%%%%%%%%%%%%%%%%%%%%%%%%%%%%%%%%%%%%% FigSAdd3-6
\begin{figure}[h!]%[htbp]
  \begin{center}
    %% \captionsetup{type=figure} 
    \begin{center}
      \resizebox{6in}{!}{\includegraphics{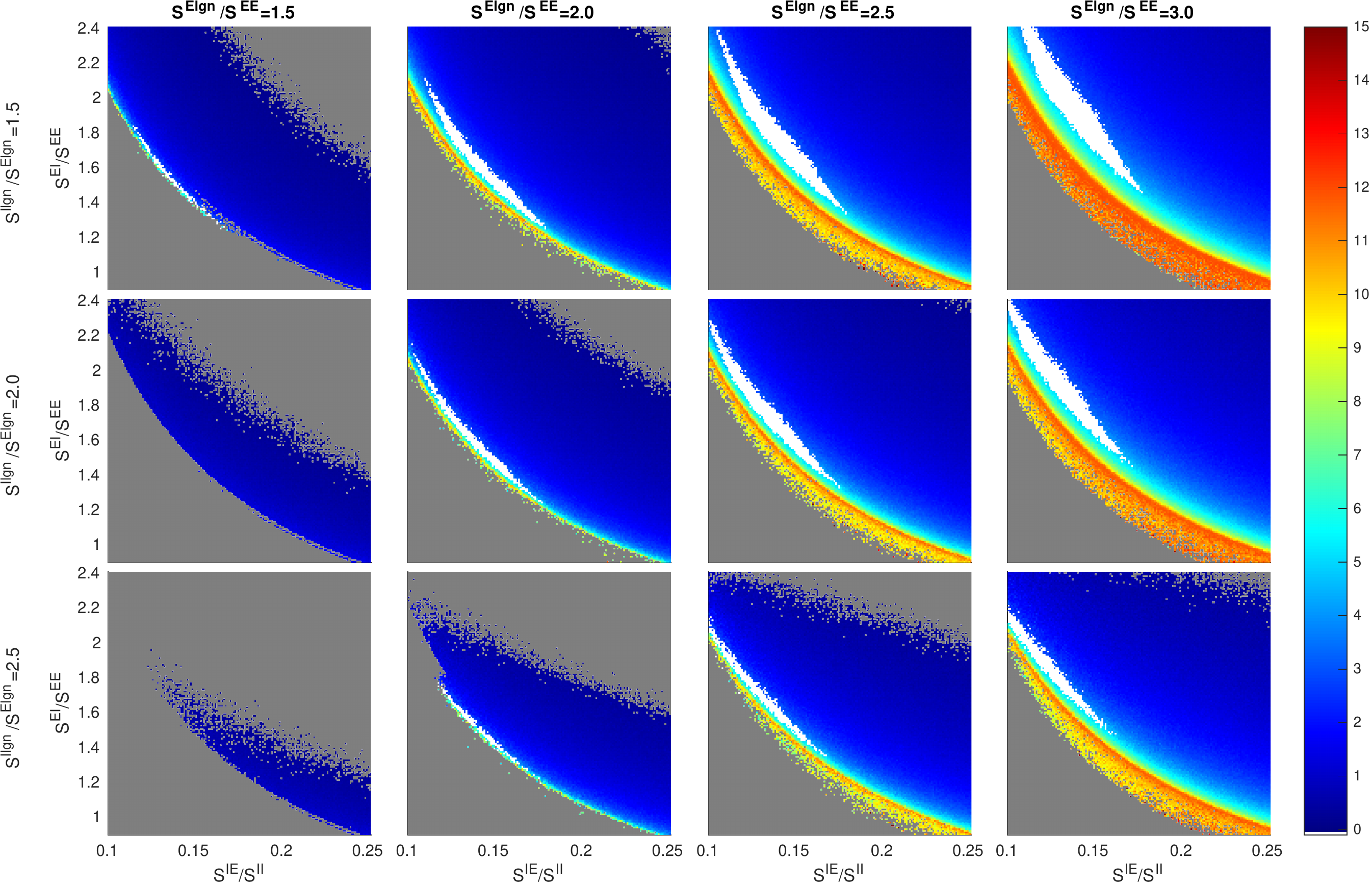}}
    \end{center}
    \caption{A version of Fig.~\ref{Fig2: LGN} with $S^{EE} = 0.027$, $S^{II} = 0.20$, and $F^{I\rm L6}/F^{E\rm L6} = 4.5$.}
    \label{FigSAdd3-6}
  \end{center}
\end{figure}
\vfill
%%%%%%%%%%%%%%%%%%%%%%%%%%%%%%%%%%%%%%%%%%%%%%%%%%%%%%%%%%% FigSAdd3-6
%%%%%%%%%%%%%%%%%%%%%%%%%%%% Supplemental %%%%%%%%%%%%%%%%%%%%%%%%%%%%%%%%%%%%%%%%%%%%
%% \addtolength{\tabcolsep}{6pt}

\end{document}